\documentclass[aps,prx,amsfonts,twocolumn,secnumarabic,nobalancelastpage,amsmath,amssymb,nofootinbib]{revtex4-1}

\usepackage{graphicx}
\usepackage[utf8]{inputenc}
\usepackage[T1]{fontenc}
\usepackage{dsfont}
\usepackage{tikz}
\usepackage{color}
\usepackage{aligned-overset}
\usepackage{hyperref}
\usepackage{xspace}

\newcommand{\bra}{\left\langle}
\newcommand{\ket}{\right\rangle}
\renewcommand{\vec}[1]{{\mathbf #1}}
\newcommand{\be}{\begin{equation}}
\newcommand{\ee}{\end{equation}}
\newcommand{\op}[1]{\operatorname{#1}}

\newlength{\diaglength}
\newcounter{diagcount}
\newcommand{\diag}[2]{
   \setlength{\diaglength}{#1mm}
\raisebox{1.2\diaglength}{\begin{tikzpicture}[scale=0.12]#2\end{tikzpicture}}
}
\newcommand{\particule}[1]{\draw [fill=white] (#1,0) circle (1) ;}
\newcommand{\ginc}[2]{\draw (#1,0) -- (#2,0) ;}
\newcommand{\identiquea}[2]{
   \draw (#1,0) -- (#1,-3) ;
   \draw (#1,-3) -- (#2,-3) ;
   \draw (#2,-3) -- (#2,0) ;
}
\newcommand{\identiqueb}[2]{
   \draw (#1,0) -- (#1,3) ;
   \draw (#1,3) -- (#2,3) ;
   \draw (#2,3) -- (#2,0) ;
}
\newcommand{\correldeux}[2]{
   \setcounter{diagcount}{#2}
   \addtocounter{diagcount}{-#1}
   \setcounter{diagcount}{\thediagcount}
   \draw [dashed] (#1,0) arc (180:0:\thediagcount * 0.5) ;
}

\newcommand{\gginc}[3]{\draw (#1,#3) -- (#2,#3) ;}
\newcommand{\pparticule}[2]{\draw [fill=white] (#1,#2) circle (1) ;}
\newcommand{\iidentique}[4]{\draw (#1,#2) -- (#3,#4) ;}
\newcommand{\iidentiquea}[2]{
   \draw (#1,3) -- (#1,6) ;
   \draw (#1,6) -- (#2,6) ;
   \draw (#2,6) -- (#2,3) ;
}
\newcommand{\iidentiqueb}[2]{
   \draw (#1,-3) -- (#1,-6) ;
   \draw (#1,-6) -- (#2,-6) ;
   \draw (#2,-6) -- (#2,-3) ;
}
\newcommand{\iidentiqueaa}[2]{
   \draw (#1,3) -- (#1,7) ;
   \draw (#1,7) -- (#2,7) ;
   \draw (#2,7) -- (#2,3) ;
}

\newcommand{\ccorreldeuxa}[2]{
   \setcounter{diagcount}{#2}
   \addtocounter{diagcount}{-#1}
   \setcounter{diagcount}{\thediagcount}
   \draw [dashed] (#1,3) arc (170:10:\thediagcount*0.5) ;
}

\newcommand{\ccorreldeuxd}[4]{\draw [dashed] (#1,#2) -- (#3,#4) ;}

\begin{document}

\title{Pseudo-gap and Localization of Light in Correlated Disordered Media}

\author         {R. Monsarrat}
\email          {romain.monsarrat@espci.psl.eu}
\author         {R. Pierrat}
\author         {A. Tourin}
\author         {A. Goetschy}
\email          {arthur.goetschy@espci.psl.eu}
\affiliation    {Institut Langevin, ESPCI Paris, PSL University, CNRS, 75005 Paris, France}
\date{\today}

\begin{abstract}
   Among the remarkable scattering properties of correlated disordered materials, the origin of pseudo-gaps and the
   formation of localized states are some of the most puzzling features. Fundamental differences between scalar and
   vector waves in both these aspects make their comprehension even more problematic. Here we present an in-depth and
   comprehensive analysis of the order-to-disorder transition in 2D resonant systems. We show with exact \textit{ab
   initio} numerical simulations in hyperuniform media that localization of 2D vector waves can occur in the presence of
   correlated disorder, in a regime of moderate density of scatterers. On the contrary, no signature of localization is
   found for white noise disorder. This is in striking contrast with scalar waves which localize at high density
   whatever the amount of correlation. For correlated materials, localization is associated with the formation of
   pseudo-gap in the density of states. We develop two complementary models to explain these observations.  The first
   one uses an effective photonic crystal-type framework and the second relies on a diagrammatic treatment of the
   multiple scattering sequences. We provide explicit theoretical evaluations of the density of states and localization
   length in good agreement with numerical simulations. In this way, we identify the microscopic processes at the origin
   of pseudo-gap formation and clarify the role of the density of states for wave localization in resonant correlated
   systems. 
\end{abstract}

\maketitle

\section{Introduction}

Light scattering in heterogeneous media is deeply affected by the existence of spatial correlations in the material
structure~\cite{vynck21}.  Transparency of the cornea~\cite{benedek71}, structural blue color of bird
feathers~\cite{jacucci20}, or strong anisotropic scattering in dense colloidal liquids~\cite{ochoa04} are representative
exemples of the impact of correlated disorder (CD) on optical properties. Remarkably, these effects can  be understood
in a simple manner from the dependence of the transport mean free path of light, $\ell_t$, on the pair correlation
function $h_2$ of the medium.  For monochromatic light of wavenumber $k=2\pi/\lambda$ propagating in a disordered medium
with number density $\rho$, the  inverse transport mean free path is $\ell_t^{-1}\sim \rho \int_0^{2k} dq F(q)S(q)$, where
$S(q)=1+\rho h_2(q)$ is the structure factor and $F(q)$ a function proportional to the form factor of individual
scatterers~\cite{fraden90, vynck21}. The interplay between local Mie resonance and non-local spatial correlations fully
determines the transmission and reflection spectra of the material: $T(\lambda)\sim \ell_t/L$ and
$R(\lambda)=1-T(\lambda)$  for a non-absorbing medium of thickness $L> \ell_t$. 

On the other hand, other fundamental properties of CD materials still withstand simple explanation. This is the case of
bandgaps or pseudo-gaps found  in the photonic density of states (DOS). Although pseudo-gaps are naturally expected in
photonic crystals with imperfections (as a result of partial gap filling by defect states)~\cite{john87, garcia11},
their presence in 2D and 3D materials without any periodicity remains puzzling~\cite{jin01, edagawa08, liew11}. Contrary
to the mean free path, there is no available theory predicting photonic bandgaps (PBGs) in these systems, and even no
consensus on the physical mechanisms at their origin~\cite{froufe17,vynck21}. The role of stealth hyperuniformity was
put forward in Ref.~\cite{florescu09}, but a recent numerical study showed that short-range order was actually
sufficient to observe similar PBG~\cite{froufe16}. The absence of a periodic lattice or long-range order makes the Bragg
interference scenario unreliable. It is thus tempting to describe PBG in CD materials as the result of a tight-binding
coupling between local resonances (such as Mie resonances), in a similar fashion to the computation of the electronic
DOS of amorphous semiconductors~\cite{weaire71, thorpe71}. The key role of local resonances in PBG formation has already
been emphasized for photonic crystals and photonic networks~\cite{lidorikis98, vynck09, rockstuhl06, sellers17}, as well as for dense media made of sub-wavelength resonators where propagation is essentially ballistic and exhibits a polaritonic dispersion
relation~\cite{lagendijk96}. However, a theoretical and comprehensive analysis for disordered media where multiple
scattering plays a prominent role is still missing.  

A second remarkable feature of CD materials is their ability to induce Anderson localization of light. In recent years,
localized states have been found numerically and experimentally in various amorphous correlated structures without
remnant periodicity~\cite{imagawa10, rechtsman11, conley14, haberko20}.  However, the physical origin of localization in
these materials remains unclear~\cite{vynck21}. Very recently, both pseudo-gap and localization of scalar waves have
been observed in 2D hyperfuniform materials, but no explanation has been provided for their
concomitance~\cite{froufe17,aubry20}.  In a first approach, localization can be thought to be related to the strong
modulation of $\ell_t$ mentioned above. Indeed, for scalar waves, reducing $\ell_t$ by a proper choice of the form and
structure factors is expected to favor wave localization both in 3D and 2D. A genuine localization transition is thought
to occur in 3D for $k\ell_t \sim 1$, whereas in 2D,  the commonly used expression of the localization length, $\xi \sim
\ell_t e^{\pi k\ell_t/2}$, quickly becomes larger than the system size and physically irrelevant for $k\ell_t \gg
1$~\cite{sheng06}. Such control of the localization length has been investigated numerically in Ref.~\cite{conley14}.
However, the use of previous 2D formula in CD materials presenting potentially strong local resonances is questionable
for two reasons. First, it ignores the strong modification of the energy velocity and diffusion coefficient of light in
resonant materials~\cite{lagendijk96}. Second, it fails to reveal the role of the DOS for localization, whereas
localized states have  been observed in the bands near the gap edge~\cite{hu08, rechtsman11, froufe17, aubry20}. John was the
first to highlight in a seminal work the crucial role of the DOS depletion for photon localization and the necessity to
modify the standard Ioffe-Regel criterion established initially for unperturbed DOS of free waves~\cite{john87,  john91,
john96}. His initial rigorous treatment considered slightly disordered photonic crystals~\cite{john87}, where
localization occurs in the vicinity of the original Bragg gap, and is, as such, inapplicable for CD materials where
local resonances are prominent and periodicity absent. However the general frameworks of the scaling
theory~\cite{abrahams79, john96}  and self-consistent theory of localization~\cite{vollhardt92}, which explicitly
involve the DOS, still apply, allowing to make an explicit connection between localization and DOS depletion.

In this work, we propose a thorough analysis of the photonic DOS and  localization signatures in 2D CD materials made of
resonant dipole scatterers. These materials allow an accurate representation of light scattering by atoms or local
inhomogeneities  supporting well-defined Mie-type resonances. They have been used in recent studies of DOS and
localization for both photonic crystals~\cite{antezza09_prl, perczel17, skipetrov20_epjb} and fully disordered
materials~\cite{goetschy11_pre, skipetrov14, bellando14, maximo15,  skipetrov16}.  Here we probe the continuous
transition from full disorder to complete order by tuning the degree of stealth hyperunifomity of the dipole pattern.
Importantly, our analysis emphasizes the differences between out-of plane and in-plane dipole excitations,  respectively
due to transverse magnetic (TM) and transverse electric (TE) field propagation. In the first case, the field exciting
the dipoles is perpendicular to the 2D plane, so that the scatterers behave as effective in-plane monopoles and wave
propagation reduces to a regular scalar problem. In the second case, the exciting field lies in the 2D plane, where the
dipoles can have different orientations: we refer to this situation as a vectorial light scattering problem.  Recently,
it has been demonstrated that the longitudinal coupling in 3D between fully disordered point scatterers indirectly opens
a new propagation channel at high density that prevents Anderson localization to occur~\cite{nieuwenhuizen94,
skipetrov14, skipetrov16_njp, vantiggelen21}. For in plane propagation in 2D disordered point dipole systems, this
longitudinal coupling is also present and as a result, no evidence of localized states has been reported so
far~\cite{maximo15}.  These recent observations have boosted the search for vector wave localization in amorphous
photonic materials with spatial correlations~\cite{vynck21, haberko20}.

In Sec.~\ref{dos}, we report the observation of localized states for TE (vector) waves in CD dipole point
patterns, that are clearly absent without spatial correlations. To our knowledge, such observation has not been reported
before. Remarkably, TE localization occurs at moderate density and disappears at high density, whereas TM (scalar) waves
are localized at high density only.  In addition, we show that localization is concomitant to the formation of
pseudo-gaps in the DOS. To elucidate these observations, we develop two complementary approaches, presented in
Sec.~\ref{crystal} and Sec.~\ref{dos_model}, respectively. The first one uses an effective photonic
crystal-type framework that expounds gap formation as the result of the superposition of different polaritonic
dispersion relations inside the effective Brillouin zone. This approach applies both for TE and TM waves and predicts
accurately the critical densities for which pseudo-gap  appear and disappear. In Sec.~\ref{dos_model}, we establish a general
expression of the DOS for an assembly of high-Q resonators, that turns out to be different from the commonly accepted
expression of the DOS. We then compute it theoretically for TE waves at moderate density using a diagrammatic expansion
including both spatial correlation and recurrent scattering. Good agreement is found with direct numerical simulation of
the DOS.  This approach allows us to identify microscopic scattering processes at the origin of the DOS depletion.
Finally, in Sec.~\ref{xi}, we express the equations of the self-consistent theory of localization in 2D in terms of the DOS
to reveal how the localization length $\xi$ explicitly depends on it. This prediction turns out to be in good agreement
with direct evaluation of $\xi$ from simulation. Hence, this work clarifies both the origin of pseudo-gap and
localization of vector waves in 2D correlated and locally resonant materials. 

\section{Density of states and localization of 2D vector waves}\label{dos}

\subsection{Resonances in finite-size hyperuniform materials}

\begin{figure*}
   \centering
   \includegraphics[width=.8\linewidth]{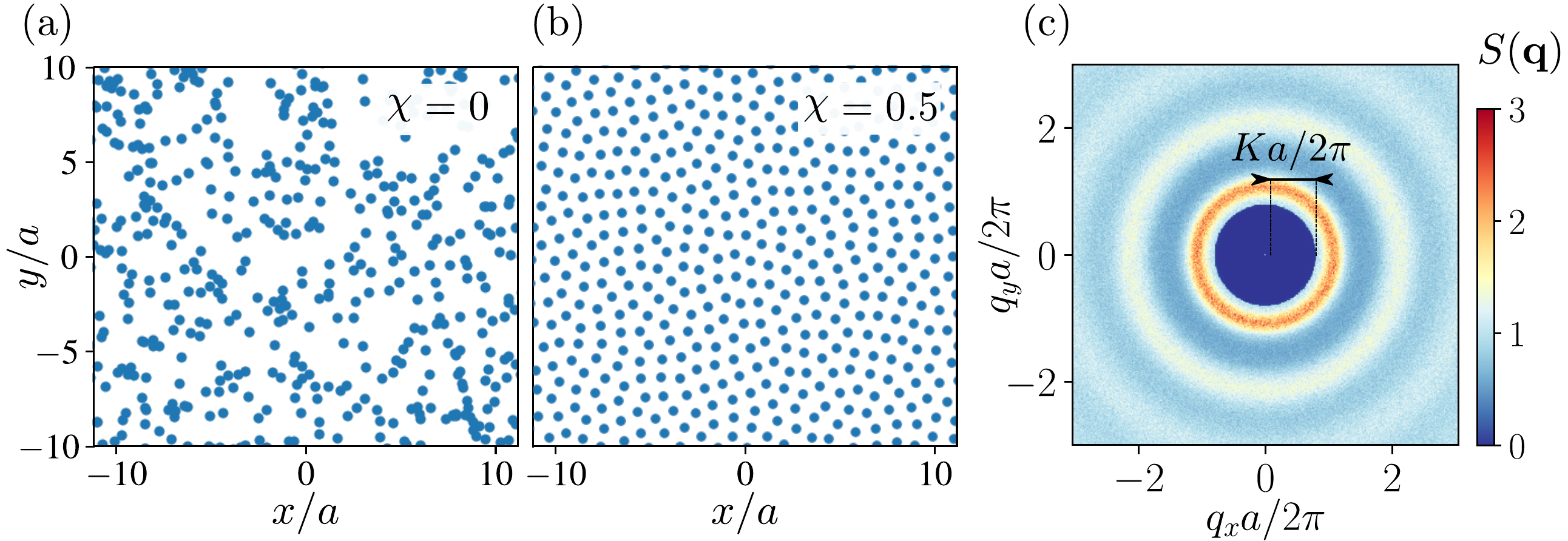}
   \caption{(a) and (b) Typical arrangements of scatterers corresponding to different degrees of spatial correlation.
   The continuous transition from white noise disorder (a), to strongly correlated material (b), and eventually to
   crystal, is probed by tuning the degree of stealth hyperuniformity $0\le\chi\lesssim 0.8$. (c) Isotropic structure
   factor of the correlated pattern shown in (b), which exhibits a dominant peak in $q\simeq 2\pi/a$ ($a=\rho^{-1/2}$ is
   the mean distance between scatterers). The procedure used to generate the pattern imposes $S(\vec{q})=0$ for $q<K$
   (see text for details).}
   \label{fig:pattern}
\end{figure*}

In this study, we consider light propagation at frequency $\omega=ck$ in an ensemble of $N$ identical resonant
scatterers characterized by their polarizability $\alpha(\omega)$. The field $\vec{E}_i$ exciting a scatterer $i$ is the
sum of the input external field $\vec{E}_0(\vec{r}_i)$ and the fields radiated by scatterers $j\neq i$:
\be
   \vec{E}_i=\vec{E}_0(\vec{r}_i)-k^2\alpha(\omega)\sum_{j\neq i}\vec{G}_0(\vec{r}_i, \vec{r}_j, \omega)\vec{E}_j.
   \label{EqExciting}
\ee
Here, $\vec{G}_0$ is the free space Green's function of the wave equation, propagating the field between different
scatterers. In 2D systems (invariant along direction $z$), $\vec{G}_0$ is a scalar for TM polarization and a $2\times 2$
matrix for TE polarization. Explicit expressions of $\vec{G}_0$ are given in App.~\ref{appendix:Green}. The derivation of coupled
equations~\eqref{EqExciting}, in a semi-classical~\cite{foldy45, lax51} or full quantum framework~\cite{lehmberg70,
goetschy11_thesis}, shows that $\alpha(\omega)$ also depends on the field polarization. Once the self-consistent
system~\eqref{EqExciting} is solved, the field at any position $\vec{r}$ is $\vec{E}(\vec{r})=\vec{E}_0(\vec{r})
-k^2\alpha(\omega) \sum_{i}\vec{G}_0(\vec{r}, \vec{r}_i, \omega)\vec{E}_i$. 

By definition, resonances of this scattering problem are solutions $\omega$ in the absence of external excitation
($\vec{E}_0=\vec{0}$). According to Eq.~\eqref{EqExciting} they satisfy the equation
\be
   \text{det}\left[\mathds{1}+k^2\alpha(\omega)\mathbb{G}_0(\omega)\right]=0,
   \label{EqDet1}
\ee
where $\mathbb{G}_0(\omega)$ is a $\beta N\times \beta N$  matrix ($\beta =1 $ for scalar waves and $\beta =2$ for
vector waves) with $i j$ element ($i \neq j $) equal to $\vec{G}_0(\vec{r}_i, \vec{r}_j, \omega)$ and $i i$ element equal to $\vec{0}$. 
Equation~\eqref{EqDet1} rigorously captures all the poles of the scattering operator, except those associated to
peculiar field solutions that are zero on each scatterer ($\vec{E}_i=0$ for all $i$). The latter, which behave as free
field without matter, may exist in crystals because of the periodicity~\cite{klugkist06, antezza09_pra} but are unlikely
for disordered materials. In the following, the polarizability of each scatterer is assumed to have a single resonance
at $\omega_0$ and a radiative decay rate $\Gamma_0$. For $\vert \omega-\omega_0\vert \ll \omega_0$, it takes the form
$\alpha(\omega)=4\beta\tilde{\alpha}(\omega)/k_0^2$, with 
\be
\label{EqDefAlpha}
   \tilde{\alpha}(\omega)=\frac{-\Gamma_0/2}{\omega-\omega_0+i \Gamma_0/2}.
\ee
The resonance condition~\eqref{EqDet1} is then conveniently expressed in terms of an effective Hamiltonian
$\mathcal{H}(\omega)$ as $\text{det}\left[\omega\mathds{1}-\mathcal{H}(\omega) \right] =0$, with
\be
   \mathcal{H}(\omega)=\left(\omega_0-i\frac{\Gamma_0}{2} \right)\mathds{1}-\frac{\Gamma_0}{2}\tilde{\mathbb{G}}_0(\omega),
   \label{EqHeff}
\ee
and $\tilde{\mathbb{G}}_0(\omega) =-4\beta(\omega/\omega_0)^2 \mathbb{G}_0(\omega)$. In this way, light-matter
interaction is entirely characterized by an effective potential proportional to the Green's matrix
$\tilde{\mathbb{G}}_0(\omega)$ \cite{akkermans08, antezza09_prl, goetschy11_thesis}. For $N\gg 1$ scatterers, the
frequency dependance of $\tilde{\mathbb{G}}_0(\omega)$ makes the search of resonances a cumbersome non-linear problem,
which can be treated exactly for periodic point patterns (see discussion in Sec.~\ref{crystal}), but not for the broad class of
hyperuniform patterns considered in this work. In the following, we will address the case of scattering resonators with
large quality factor ($Q=\omega_0/\Gamma_0 \gg 1$), for which the coupling term in Eq.~\eqref{EqHeff} can be treated as
a perturbation. This amounts to freeze the frequency of $\tilde{\mathbb{G}}_0(\omega)$ at the resonance frequency
$\omega_0$. This is an excellent approximation for light scattering in atomic systems, which  we expect to hold as well
for Mie-type resonators with $Q\gtrsim 10$. In this situation, there are $\beta N$ complex
resonances $\omega_n-i\Gamma_n/2$  given by
\begin{equation}
   \left\{\begin{aligned}
      \omega_n&=\omega_0-\frac{\Gamma_0}{2}\text{Re}\Lambda_n,
   \\
      \Gamma_n&=\Gamma_0\left(1+ \text{Im}\Lambda_n\right),
   \end{aligned}\right.
\end{equation}
where $\Lambda_n$ are the eigenvalues of the Green's matrix  $\tilde{\mathbb{G}}_0(\omega_0)$. For finite-size system,
$\Lambda_n$ occupy an extended domain in the complex plane, with $\text{Im}\Lambda_n>-1$~\cite{skipetrov11}.
$\text{Re}\Lambda_n$ represent collective Lamb shifts and $\text{Im}\Lambda_n$ collective decay rates. The distribution
of $\Lambda_n$ in the complex plane has been characterized theoretically in details for scalar and vector waves in 3D
fully disordered systems~\cite{skipetrov11, goetschy11_pre, goetschy11_epl, goetschy11_thesis}. In particular,  finite
mode life-times due to finite system size offer valuable information about localization properties of the corresponding
eigenstates~\cite{goetschy11_pre, skipetrov14, bellando14, skipetrov16}. 

\begin{figure*}
   \includegraphics[width=.85\linewidth]{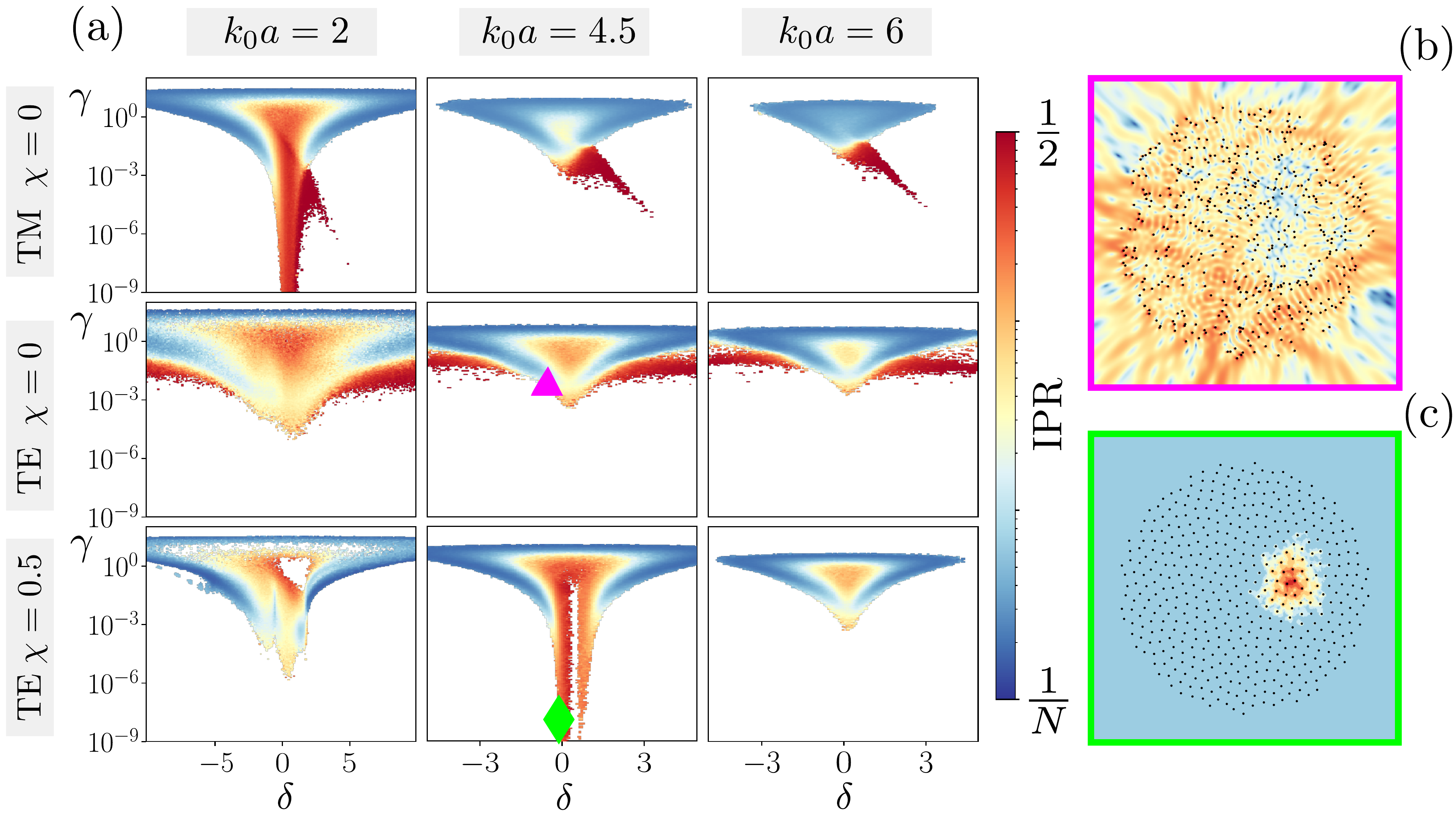}
   \caption{(a) Distribution of resonances $\omega_n-i\Gamma_n/2$ in the complex plane for different densities (columns)
   and degree of correlation or polarization (rows). Axes are labeled with normalized units
   ($\delta=2(\omega-\omega_0)/\Gamma_0$ stands for detuning and $\gamma=\Gamma/\Gamma_0$ for decay rate). The color
   refers to the inverse participation ratio of the corresponding eigenvectors in log scale. The system size is $k_0 R =
   55$ and the number of disorder realizations adjusted to have a number of eigenvalues of the order of $10^6$. (b) and
   (c) Typical spatial profiles in log scale of the intensity generated by exciting selectively eigenvectors associated
   to the eigenvalues marked with triangle (b) and diamond (c) in the complex plane.}
   \label{fig:spectra}
\end{figure*}

To probe the impact of the disorder-to-order transition on the  resonances of the system, we arranged the 2D scatterers
inside a disk of radius $R$ using stealthy hyperuniform (SHU) point patterns~\cite{torquato18, torquato15}.  The latter
are configurations $\{ \vec{r}_i \}$ which minimize to zero the structure factor
 $S(\vec{q})=1+\rho h_2(\vec{q})$ in a domain $\vert \vec{q} \vert < K$. In practice, they are
found by minimizing the two-body potential
\be
   U(\{ \vec{r}_i \})=\int_{q<K} d \vec{q} \,\tilde{S}(\vec{q}) \equiv \sum_{i,j}u(\vec{r}_i-\vec{r}_j),
\ee
where $\tilde{S}(\vec{q})=\sum_{i,j}e^{i\vec{q}.(\vec{r}_i-\vec{r}_j)}/N$, and $u(\Delta \vec{r})=2\pi K J_1(K\Delta
r)/\Delta r$ in the limit $R\to \infty$. We refer to App.~\ref{appendix:shu} for more details about the pattern
generation and the explicit link between $S(\vec{q})$ and  $\tilde{S}(\vec{q})$. The amount of spatial correlation is
controlled by the stealthiness parameter $\chi$ defined as the ratio between the number of constrained degrees of
freedom in reciprocal space ($NK^2a^2/8\pi$) and the total number of spatial degrees of freedoms ($2N$),
\be
   \chi=\frac{(Ka)^2}{16\pi},
\ee
where $a=\rho^{-1/2}=\sqrt{\pi/N}R$ is the mean distance between scatterers.  When $\chi$ is progressively increased
from $0$, the structure factor goes from a flat response $S(\vec{q})= 1$ for white noise disorder (see
Fig.~\ref{fig:pattern}\,(a) for a typical configuration) to a peaked profile that remains isotropic as long as $\chi
\lesssim 0.6$. Crystalline order is achieved for $\chi\simeq 0.7-0.8$ when $K$ coincides with the first Bragg peak of
the crystalline lattice. As an illustration, we show a typical disorder configuration for $\chi=0.5$ in
Fig.~\ref{fig:pattern}\,(b) and the corresponding isotropic structure factor in Fig.~\ref{fig:pattern}\,(c). Such isotropic
profile of $S(\vec{q})$, with a first dominant peak at $q\simeq 2\pi/a$, is very similar to the profile obtained with
hard disks at sufficiently large packing fraction~\cite{froufe16}. Since all our predictions for pseudo-gap and
localization established in the following are expressed in terms of $h_2(\vec{r})$ or $S(\vec{q})$, we expect them not to be specific to
SHU patterns, but rather generic for any type of correlated systems with similar structure factor. 

\subsection{Distribution of resonances in the complex plane and IPR}

\begin{figure*}
   \centering
   \includegraphics[width=0.85\linewidth]{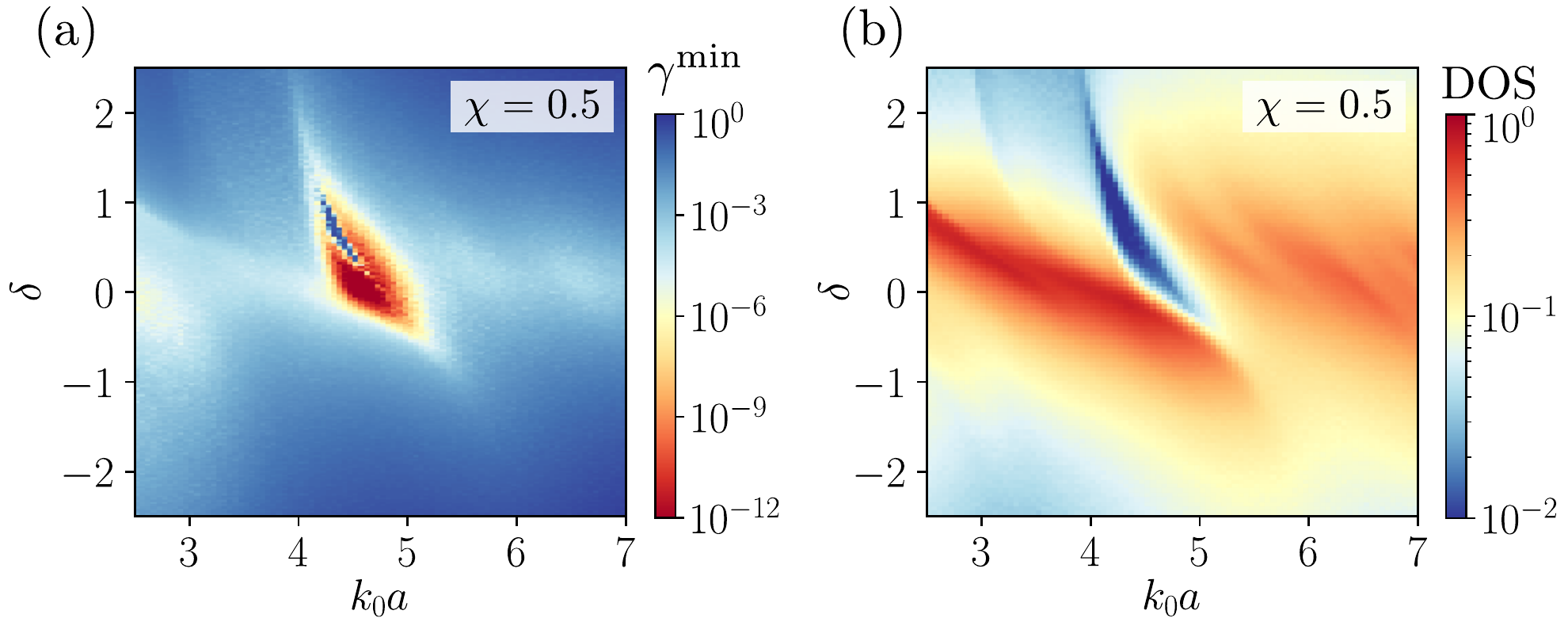}
   \caption{(a) Map of the smallest normalized decay rate $\gamma^\text{min}\equiv \left<\text{min}(\gamma)\right>$ in
   the phase space density-detuning. Small values (warm color) correspond to states localized in the bulk of the medium,
   associated to long lifetime. (b) Map of the normalized density of states [see Eqs.~\eqref{EqDOS1} and~\eqref{EqDOS2}
   for definition]. A pseudo-gap is found in the same phase-space domain as localized states in (a). For both maps (a)
   and (b), the system size is kept constant ($k_0 R = 55$) and the number of disorder realizations adjusted to have a
   number of eigenvalues of the order of $10^6$ for each $k_0a$.}
   \label{fig:phase_diagram}
\end{figure*}

The effect of correlated disorder on scattering can be apprehended by visualizing the repartition of resonances
$\omega_n-i\Gamma_n/2$ in the complex plane, as shown in Fig.~\ref{fig:spectra}\,(a).  For convenience, frequencies and
decay rates are expressed with normalized units, $\delta_n=2(\omega_n-\omega_0)/\Gamma_0$ and
$\gamma_n=\Gamma_n/\Gamma_0$. The degree of localization of the corresponding eigenstates $\psi_n$ (of components
$\psi_{n,i}$ at positions $\vec{r}_i$) is also shown, using the inverse participation ratio
$\operatorname{IPR}_n=\sum_i^{N}{\Vert \psi_{n,i} \Vert^4}/\sum_i^{N}{\Vert \psi_{n,i} \Vert^2}$ as color scale. We
first analyze the IPR map of scalar waves without correlation (first line), which will serves as reference for the
understanding of vector waves (second and third lines).  At low density ($k_0a=6$), resonances spread around the central
frequency $\omega_0$ with decay rates covering the range $\gamma\in [10^{-3}, 10] $, while most states are delocalized
(low IPR). The smallest decay rates stand for proximity resonances formed at $\delta>0$. They are due to localized
states on pairs of close scatterers ($\operatorname{IPR}_n\sim1/2$) and are unrelated to Anderson
localization~\cite{goetschy11_pre, goetschy11_epl, skipetrov14}. On the other hand, at large density  ($k_0a=2$), a long
tail of resonances with very small decay rates ($\gamma\in [10^{-15}, 10^{-6}] $) is clearly visible  near $\delta=0$.
It is associated to states with large IPR localized in the bulk of the scattering medium.  These signatures mark the
onset of Anderson localization, as already studied for scalar waves both in 3D~\cite{skipetrov14, skipetrov16,
bellando14} and 2D~\cite{maximo15}. Here localization is triggered by a reduction of the transport mean free path
$\ell_t$ at large density and small detuning~\cite{lagendijk96}. Approaches to localization in 3D and 2D are expected to
be different since a genuine transition occurs in 3D at $k_0a\simeq2.3$ corresponding to $k_0\ell_t \sim1$
~\cite{skipetrov18}, whereas in 2D, the critical density below which localization is visible ($\xi \lesssim R $) depends
on the system size $R$. However, the expected exponential dependence of $\xi$ on $k_0\ell_t $ in 2D (see Sec.~\ref{xi}
for a more precise statement) makes the condition $\xi \sim R $ accessible in realistic conditions for moderate values
of $k_0\ell_t$ only. 

For 2D vector waves without spatial correlation [second line of Fig.~\ref{fig:spectra}\,(a)], subradiant  proximity
resonances are still visible, covering now both the range of positive and negative detuning, in a similar fashion to the
3D vector case~\cite{goetschy11_thesis, skipetrov14}. More importantly, localization signatures previously found at high
density are now absent: decay rates near $\delta=0$ are orders of magnitude larger than in the scalar case and IPR
remain low. It has been shown numerically that such breakdown of localization at high density is due to near-field
contributions of the Green's tensor $\vec{G}_0$~\cite{maximo15}. Similar observations have also been reported previously
with 3D vector waves~\cite{skipetrov14, bellando14}. In 3D, near-field contributions present in the longitudinal part of
$\vec{G}_0$  become dominant precisely in the regime of high density where the Anderson transition takes place for
scalar wave. The longitudinal coupling between scatterers indirectly opens a channel for energy propagation that
prevents localization to occur~\cite{skipetrov14, vantiggelen21}. Should the same scenario apply in 2D, it would
indicate the existence of a transition between a delocalized phase at large density and a localized phase at low
density, albeit with an exponentially large localization length. In any case, no clear evidence of localization has been
reported for 2D vector waves in resonant dipolar systems yet~\cite{maximo15}.

Finally, for 2D vector waves with  substantial amount of spatial correlation [third line of
Fig.~\ref{fig:spectra}\,(a)], proximity resonances have completely disappeared because the short-range repulsion of the
pair potential $u(\Delta r)$ prevents scatterers to be too close to each other, as shown in Fig.~\ref{fig:pattern}\,(b).
Apart from this difference, the complex spectrum is left unchanged at low density (compare TM spectra at $k_0a=6$, for
$\chi=0$ and $\chi=0.5$). But remarkably, at intermediate density ($k_0a=4.5$), localization is established around
$\delta=0$ with states sharing small decay rates ($\gamma\in [10^{-12}, 10^{-6}] $) and strong confinement
($\operatorname{IPR}\gtrsim 0.1$). In Fig.~\ref{fig:spectra}\,(c), we show a representative example of the spatial
profile of localized states, with exponential shape bearing no resemblance to the delocalized profile found in the same
frequency range without correlation [see Fig.~\ref{fig:spectra}\,(b)]. A systematic study reveals that all states with
$\gamma \lesssim 10^{-6}$ are exponentially localized. We will provide a detailed study of the localization length $\xi$
in Sec.~\ref{xi}; typical distribution of $\xi$ can be found in App.~\ref{appendix:ls_xi_computation}. Increasing
further the density makes the localization signatures disappear again (see panel at $k_0a=2$). Hence, vector waves
exhibit localization when scalar waves do not, and \emph{vice versa}. 

\subsection{Localization and density of states}\label{SubSec:LocDos}

To cover exhaustively the properties of TE modes, we established the phase diagram of localization in the phase space
density-frequency. States that are exponentially localized in the bulk of the medium necessarily have small decay rates.
Hence, a good indicator of localization is  the smallest decay rate $\gamma$ for fixed density and frequency, averaged
over a large number of SHU configurations. It is more accurate than IPR, which can be large for states different from
those localized in the bulk (see discussion below).  A map of $\gamma^\text{min}\equiv \left<\text{min}(\gamma)\right>$,
where $\left< \dots \right>$ stands for ensemble average over different SHU configurations, is presented in
Fig.~\ref{fig:phase_diagram}\,(a) for $\chi=0.5$. The localized phase corresponds to the smallest values of
$\gamma^\text{min}$ marked with warm colors ($\gamma^\text{min}\lesssim10^{-6}$); it covers an intermediate range of
density corresponding to $k_0a\in [4.2, 5.2]$.  A systematic analysis for different degree of correlation $\chi$ reveals
that localization is preserved as long as $\chi \gtrsim 0.4$ with a localization island that is progressively submerged
as $\chi$ is reduced. Furthermore, a similar study for TM modes (not shown) establishes localization for $k_0a\lesssim
3$ for any $\chi$, confirming that vector and scalar waves exhibit localization in non-overlapping density ranges. 
 
\begin{figure}
   \centering
   \includegraphics[width=.8\linewidth]{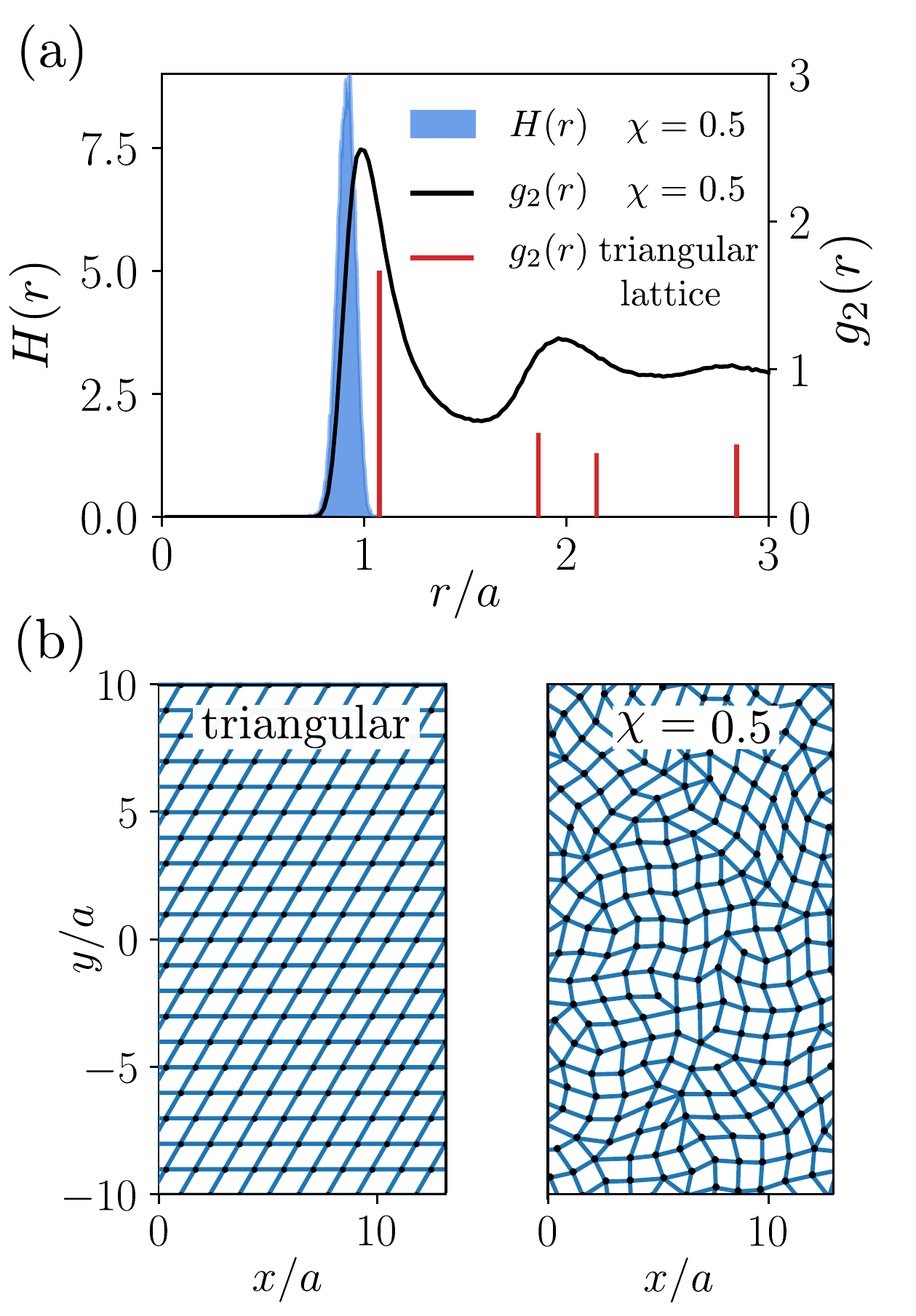}
   \caption{ (a) Pair correlation function $g_2(r)$ (left axis) and nearest neighbor distribution $H(r)$ (right axis) of
   a SHU configuration at $\chi=0.5$. The function $g_2(r)$ of a triangular lattice is superimposed for comparison.  (b)
   Connected point patterns of a triangular lattice (left) and a SHU configuration at $\chi=0.5$ (right).}
   \label{fig:triangulation}
\end{figure}

Although IPR maps in Fig.~\ref{fig:spectra}\,(a) give an indication of the spectral range covered by the resonances
of the system, they do not provide information about the DOS. The latter is defined as
\be
   p(\omega)=\frac{1}{\mathcal{A}}\left<\sum_{n=1}^{\beta N}\delta(\omega-\omega_n)\right>,
   \label{EqDOS1}
\ee
where $\mathcal{A}$ is the area occupied by the disordered sample. In the thermodynamic limit ($N \to \infty$,
$\mathcal{A}\to \infty$, at fixed density $\rho=N/\mathcal{A}$), almost all complex eigenvalues $\omega_n - i
\Gamma_n/2$ collapse on the real axis and the definition~\eqref{EqDOS1} coincides with the density of modes of the
Hamiltonian~\eqref{EqHeff}. The distribution $p(\omega)$ is related to the distribution of detuning
$\delta=2(\omega-\omega_0)/\Gamma_0$ as
\be
   p(\omega)=\frac{2\beta\rho}{\Gamma_0}p(\delta),
   \label{EqDOS2}
\ee
where $p(\delta)$ is normalized to unity ($\int d\delta \, p(\delta)=1$). 
 
By computing the DOS $p(\delta)$ for different values of density, we generated the map shown in
Fig.~\ref{fig:phase_diagram}\,(b). Remarkably, the DOS exhibits a strong depletion in the domain where localization is
observed in the map of $\gamma^{\text{min}}$ [Fig.~\ref{fig:phase_diagram}\,(a)]. A close comparison between the two
maps shows a correlation between weaker features as well: when the DOS is reduced or increased, so does
$\gamma^{\text{min}}$.  These observations are corroborated by similar finding for TM polarization: DOS is depleted for
$k_0 a \lesssim 3$ and $\delta \gtrsim 0$ in the presence of spatial correlation, in the same phase space domain where
localization occurs (not shown). Moreover, the value of the DOS in the depleted area of
Fig.~\ref{fig:phase_diagram}\,(b) is small but finite, marking the existence of a pseudo-gap. The value of $p(\delta)$
inside the pseudo-gap is independent of the system size (at fixed density), as can be expected for states localized in
the bulk of the medium. However, we also found a very thin part of the phase space where $p(\delta)$ decreases as the
medium gets larger. It corresponds to the slice of modes with moderated decay rates ($\gamma \gtrsim10^{-3}$) breaching
the localization island in Fig.~\ref{fig:phase_diagram}\,(a). The repartition of eigenvalues in the complex plane
represented in Fig.~\ref{fig:spectra}\,(b)  at $k_0a=4.5$ makes it also visible. Resonances inside this slice correspond
to states mostly located at the sample boundary, and thus potentially associated to relatively large IPR. They are an
artefact of the finite sample size used in simulations. In the thermodynamic limit, we expect the fraction of states in
this domain to vanish, revealing a real gap. Similar observations have been made for 3D disordered
crystals~\cite{skipetrov20_epjb, skipetrov20_prb}. Hence, our analysis indicates that localization of vector waves
occurs inside a broad pseudo-gap surrounding a thin gap identified in the map of $\gamma^{\text{min}}$. 

In the following sections, we will provide theoretical models that explain both the appearance of a pseudo-gap in the
DOS and the formation of localized states in the same phase space domain. In this regard, we will justify all the
differences discussed so far for TE and TM waves.  

\section{Effective photonic crystal model for the density of states}\label{crystal}

\begin{figure*}
   \centering
   \includegraphics[width=0.7\linewidth]{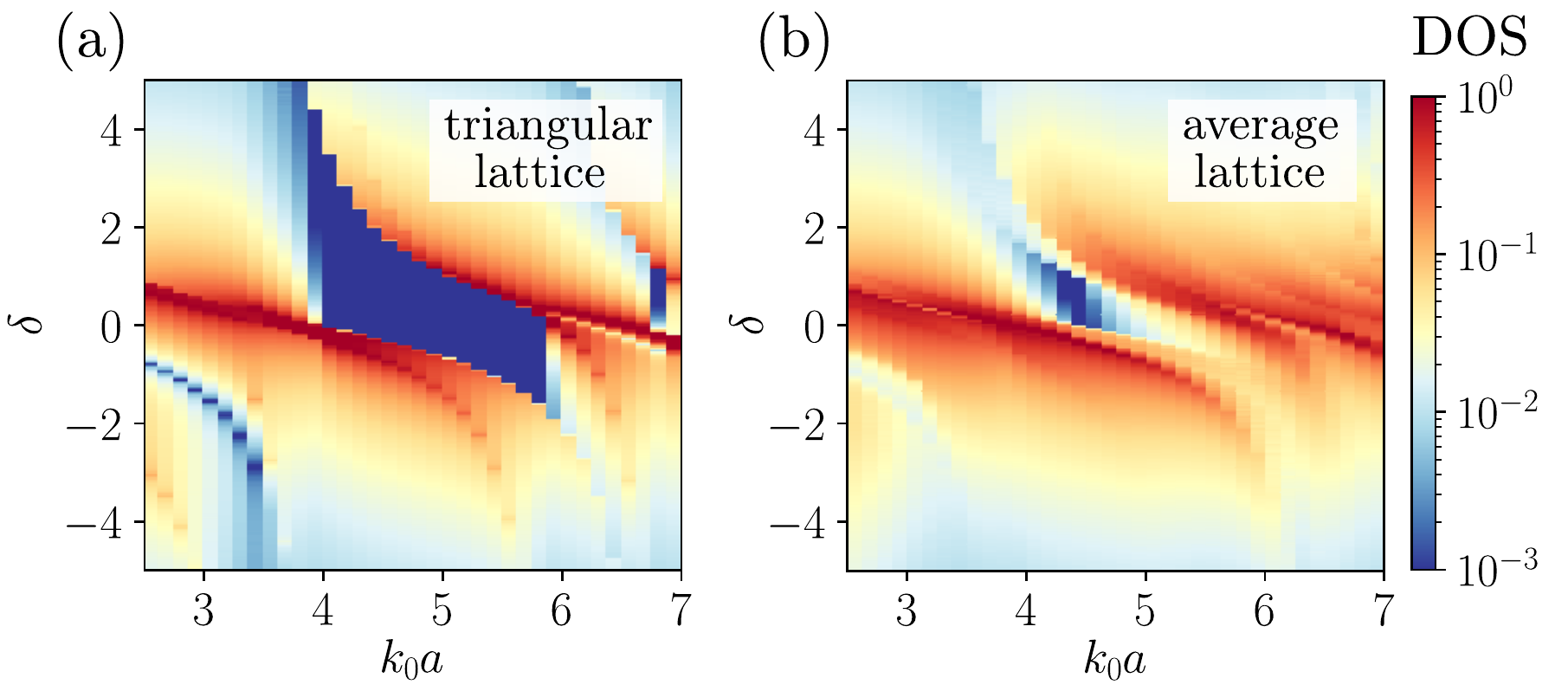}
   \caption{(a) Normalized DOS of TE waves propagating on a triangular lattice in the phase space density-detuning (b)
   Normalized DOS of TE waves, averaged over an ensemble of lattices with packing fraction superior to $0.75$ (see text
   for details).  The DOS in (b) reproduces with a good fidelity the DOS obtained with SHU configurations at $\chi=0.5$,
   shown in Fig.~\ref{fig:phase_diagram}(b). In particular, a pseudo-gap is preserved in the same range of $k_0a$ and
   $\delta$.}
   \label{fig:DOS_cry}
\end{figure*}

\begin{figure}[b!]
   \centering
   \includegraphics[width=.9\linewidth]{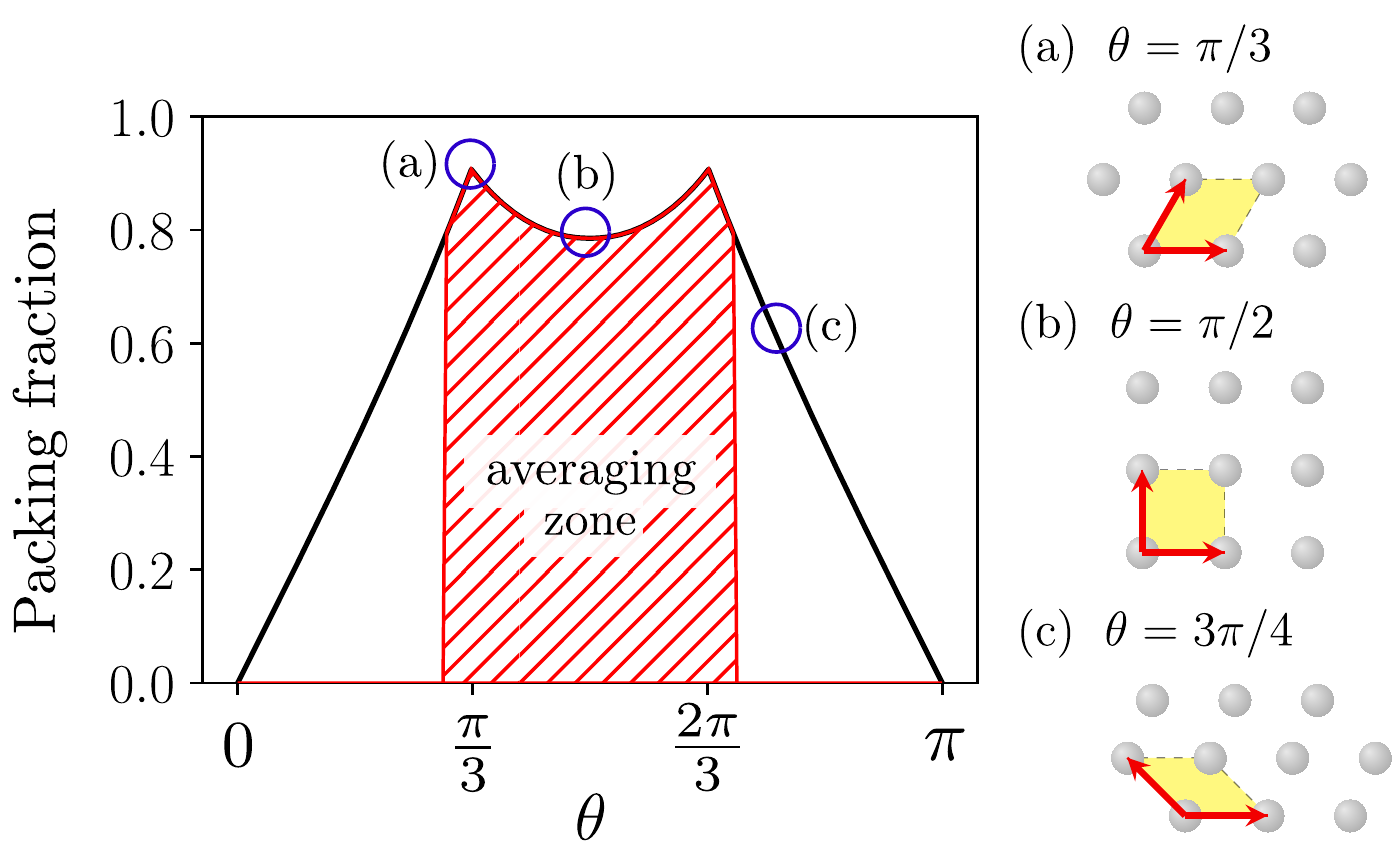}
   \caption{Packing fraction $f$ of oblique lattices with respect to the relative angle $\theta$ between the two basis
   vectors. The fraction $f$ is defined as the ratio between the area of packed non-interpenetrating disks placed on the
   lattice and the total area. Exemples of unit cells are shown on the right: triangular lattice (a) with $f=0.90$ ,
   square lattice (b) with $f=0.78$, and a less compact oblique lattice (c) with $f=0.60$. The dashed domain shows the range of
   angles used to obtain the average DOS of Fig.~\ref{fig:DOS_cry}\,(b).}
   \label{fig:compacity}
\end{figure}

\begin{figure*}
   \centering
   \includegraphics[width=.9\linewidth]{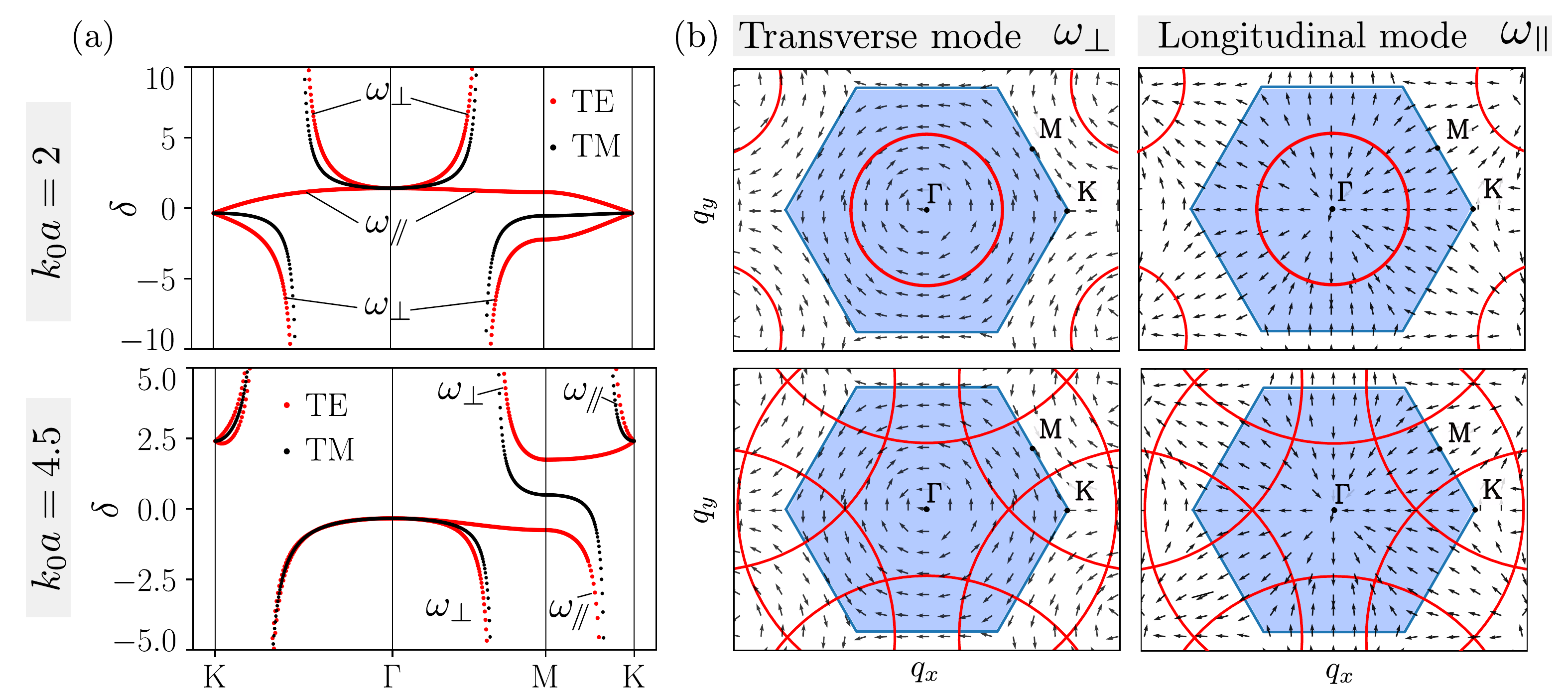}
   \caption{(a)  Photonic band structure of the triangular lattice at  high density ($k_0a=2$, top)  and moderate
   density ($k_0a=4.5$, bottom), for both TM (black) and TE (red) polarizations. For the TE case, bands are marked as
   transverse ($\omega_\perp$) or longitudinal ($\omega_\parallel$). (b) View of the first Brillouin zone (shaded blue),
   together with the light circles $\vert \vec{q} \vert =k_0$ and $\vert \vec{q} - \vec{Q} \vert =k_0$,  at high density
   (top) and moderate density (bottom).  Arrows represent the orientation $\vec{\epsilon}$ of the eigenstates
   corresponding to the transverse band ($\omega_\perp$, associated to $\vec{\epsilon} \perp \vec{q}$) and longitudinal
   band ($\omega_\parallel$, associated to $\vec{\epsilon} \parallel \vec{q}$) shown in (a).
  }
   \label{fig:crystal_band_struc}
\end{figure*}

\subsection{Density of states of crystals with large packing fraction} 
\label{sec:simu-crystal}

In the previous section, we established that a pseudo-gap in the DOS of TE modes forms at a relatively large degree of
spatial correlation ($\chi \gtrsim 0.4$). In this regime, SHU point patterns exhibit strong local order.  In
Fig.~\ref{fig:triangulation}\,(a), the distribution $H(r)$ of nearest neighbors at $\chi=0.5$ appears sharply peaked near $r\sim
a$ and the pair correlation function $g_2(\vec{r})=1+h_2(\vec{r})$ is isotropic, with strong oscillations associated to
the dominant peak of $S(\vec{q})$ shown in Fig.~\ref{fig:pattern}\,(c). Very similar results would be obtained with hard
disks of large packing fraction instead of SHU point patterns~\cite{froufe16, vynck21}. When the stealth parameter is
increased further and reach $\chi\simeq 0.7-0.8$, the patterns crystallize in lattices with large packing fraction, such
as the triangular lattice. To stress the similarities between SHU configurations at $\chi=0.5$ and the triangular
lattice, we connected points with lines to form a pattern of oblique unit cells, that preserve orientation locally. The
result is compared with the tiling of the lattice with primitive cells in Fig.~\ref{fig:triangulation}\,(b). In this way,
a disordered pattern can be seen as an assembly of small crystals, with orientation and lattice parameters that
fluctuate while preserving a large packing fraction. Earlier works noted that polycrystalline structures present
bandgaps similar to those found in perfect crystals as long as the crystal domains are sufficiently large~\cite{yang10,
froufe16}. It is thus instructive to compare the DOS of SHU resonant systems with the DOS of crystals made of
resonators. 
 
Using a method detailed in Sec.~\ref{sec:theory-crystal}, we first computed the map of DOS of TE modes for a triangular
lattice. The result presented in Fig.~\ref{fig:DOS_cry}\,(a) exhibits a full band gap for mostly positive detuning in a
broad range of density ($k_0a \in [3.8, 6]$). The comparison with Fig.~\ref{fig:phase_diagram}\,(b) reveals that this
gap covers a phase space domain that encompasses the domain of the pseudo-gap found at $\chi=0.5$. Performing a similar
analysis for different oblique lattices, we found that the size of the gap decreases with the packing fraction --- the
largest packing fraction and gap corresponding to the triangular lattice. In particular, a minimal gap exists for all
dense lattices, except for the square one (see Sec.~\ref{sec:theory-crystal} for explanation).  The DOS of an effective
crystal mimicking the SHU pattern can be obtained by averaging the DOS over the lattice parameter space restricted to
large packing fraction.  As the distance between scatterers is almost constant in SHU patterns, we considered different
lattices with basis vectors of equal length $a^*$ and different relative angle $\theta$. The range of spanned angles is
chosen to maintain a packing fraction larger than $0.78$, as illustrated in Fig.~\ref{fig:compacity}. In these
crystalline point patterns, the mean distance between points is $a=a^*\sqrt{\vert \text{sin}(\theta) \vert}$. We show in
Fig.~\ref{fig:DOS_cry}\,(b) the average DOS for a broad range of $k_0a$.  As expected, the gap of the triangular lattice
is now partially filled, revealing a strong depletion very similar to the pseudo-gap discussed in
Fig.~\ref{fig:phase_diagram}\,(b). This analysis corroborates the idea that SHU patterns behave as effective photonic
crystals, at least as far as their DOS is concerned. 

\subsection{Theoretical predictions for Bravais lattices}\label{sec:theory-crystal}

Previous considerations justify a deeper theoretical treatment of crystals with large packing fraction, such as the
triangular lattice. For infinite Bravais lattice, eigenstates of the Hamiltonian~\eqref{EqHeff} are Bloch modes that can
be labelled by a vector $\vec{q}$ in the first Brillouin zone~\cite{antezza09_prl}. The corresponding eigenvalues are
noted $\omega_\vec{q}$.  Using the periodicity of the lattice, the eigenvalue problem reduces to
$\text{det}\left[\omega_\vec{q}\mathds{1}-\mathcal{H}_\vec{q}(\omega_\vec{q}) \right] =0$, where the $\beta \times
\beta$ matrix $\mathcal{H}_\vec{q}(\omega_\vec{q})$ is
\begin{equation}
   \mathcal{H}_\vec{q}(\omega_\vec{q})=\left(\omega_0-i\frac{\Gamma_0}{2} \right)\mathds{1}-\frac{\Gamma_0}{2}\tilde{\mathbb{G}}_0(\vec{q},\omega_\vec{q}),
   \label{EqHamiltonQ}
\end{equation}
with $\tilde{\mathbb{G}}_0(\vec{q},\omega)=-4\beta(\omega/\omega_0)^2\sum_{\vec{R}\neq \vec{0}}\vec{G}_0(\vec{R},
\omega)e^{-i\vec{q}.\vec{R}}$. Since the free space Green's function $\vec{G}_0$ decays slowly in real space, it is not
possible to perform a nearest-neighbor-type approximation of  the summation involved in
$\tilde{\mathbb{G}}_0(\vec{q},\omega)$. Incidentally, this indicates that the tight-binding model used in
Refs.~\cite{weaire71, thorpe71} to compute the electronic DOS of amorphous semiconductors is not relevant for
Hamiltonian of the form~\eqref{EqHeff}.  Using Poisson's formula instead, we convert the sum over lattice positions
$\vec{R}$ into a sum over reciprocal lattice vectors $\vec{Q}$,
\begin{equation}
   \tilde{\mathbb{G}}_0(\vec{q},\omega)=-\frac{4\beta\omega^2}{\omega_0^2} \left[\rho\sum_{\vec{Q}} \vec{G}_0(\vec{q}-\vec{Q}, \omega) - \vec{G}_0(\vec{R}=\vec{0}, \omega)\right],
   \label{EqPoisson}
\end{equation}
where the Fourier transform of the Green's function is $\vec{G}_0(\vec{q},
\omega)=1/[(k^2-q^2)\mathds{1}+q^2\Delta^\parallel_\vec{q}\,\delta_{\beta,2}]$,  with $\Delta^\parallel_\vec{q}$ the
projector parallel to $\vec{q}$.  The formulation~\eqref{EqPoisson} is more appropriate for computation because
convergence is much faster in momentum space. The price to pay is the necessity to regularize $\vec{G}_0$ since the two
terms in Eq.~\eqref{EqPoisson} diverge, whereas their difference does not~\cite{antezza09_pra, perczel17}. This is done
by convoluting $\vec{G}_0$ in real space with a function of finite but small width that smears out the divergence of
$\text{Re}[\vec{G}_0(\vec{R}=\vec{0}, \omega)]$. The regularization of $\vec{G}_0$ as well as the numerical procedure
used to obtain, from Eqs.~\eqref{EqHamiltonQ} and~\eqref{EqPoisson}, the DOS shown in Fig.~\ref{fig:DOS_cry} are
detailed in App.~\ref{appendix:crystal}.

Dispersion relations, $\omega_\vec{q}$ versus $\vec{q}$, for the triangular lattice ($\theta=\pi/3$) are shown in
Fig.~\ref{fig:crystal_band_struc}\,(a), along the irreducible path $\Gamma \to M \to K \to \Gamma$ of the first Brillouin
zone. For consistency with previous treatment of SHU patterns, dispersion relations have been obtained, for both TM and TE polarizations,
by solving the eigenvalue equation $\text{det}\left[\omega_\vec{q}\mathds{1}-\mathcal{H}_\vec{q}(\omega_\vec{q}) \right]
=0$, with the large quality factor approximation  $\mathcal{H}_\vec{q}(\omega_\vec{q})\simeq
\mathcal{H}_\vec{q}(\omega_0)$. Let us first consider the high density regime ($k_0a=2$, top), which is the simplest to
analyze. TM polarization  gives a single solution $\omega^{0}_\vec{q}$, which exhibits a typical polaritonic dispersion
relation (black dots), while TE polarization gives two solutions, labeled  $\omega^\perp_\vec{q}$ and
$\omega^\parallel_\vec{q}$ (red dots). These labels refer to the orientation of the corresponding eigenvectors, that
can be either transverse or parallel to the wave vector $\vec{q}$ [see Fig.~\ref{fig:crystal_band_struc}\,(b) and (c)].
The band $\omega^\perp_\vec{q}$ presents a polaritonic dispersion while $\omega^\parallel_\vec{q}$ is mostly flat. These
different behaviors can be understood by considering the long-wavelength limit of Eq.~\eqref{EqPoisson}, where the sum
is dominated by the Fourier component $\vec{Q}=\vec{0}$:
\begin{multline}\label{eq::G0_q}
\tilde{\mathbb{G}}_0(\vec{q},\omega) \simeq \frac{4\beta}{(k_0a)^2}
  \left(
 \frac{k^2}{[q^2-k^2]\mathds{1}-q^2\Delta^\parallel_\vec{q}\,\delta_{\beta,2}}
+\frac{\mathds{1}}{2}\delta_{\beta,2}
 \right)
 \\
 -i\mathds{1}.
\end{multline}
Inserting this expression into Eq.~\eqref{EqHamiltonQ}, we find that the frequencies $\{\omega^{0}_\vec{q},
\omega^{\perp}_\vec{q}, \omega^{\parallel}_\vec{q}\}$ satisfy the following dispersion relations,
\begin{equation}
   \begin{aligned}
      \omega^{0}_\vec{q}&=\omega_0-\frac{\Gamma_0}{2}\frac{4}{(k_0a)^2}\frac{(\omega^{0}_\vec{q}/c)^2}{q^2-(\omega^{0}_\vec{q}/c)^2},
   \\
      \omega^{\perp}_\vec{q}&=\omega_0-\frac{\Gamma_0}{2}\frac{8}{(k_0a)^2}\left[\frac{(\omega^{\perp}_\vec{q}/c)^2}{q^2-(\omega^{\perp}_\vec{q}/c)^2}+\frac{1}{2}
   \right],
   \\
      \omega^{\parallel}_\vec{q}&=\omega_0+\frac{\Gamma_0}{2}\frac{4}{(k_0a)^2},
   \end{aligned}
   \label{EqDispersion}
\end{equation}
which can be interpreted as solutions of an effective medium wave equation $\left[ -\nabla\times \nabla \times
+k^2\epsilon(\omega)\right]E(\vec{r},\omega)=\vec{0}$. TM plane waves of vector $\vec{q}$ obey the dispersion relation
$q^2=k^2\epsilon(\omega)$, where the dielectric function, $\epsilon(\omega)=1+\rho \alpha_e(\omega)$, is expressed in
terms of an effective polarizability $\alpha_e(\omega)=4\beta\tilde{\alpha}_e(\omega)/k_0^2$, with
$\tilde{\alpha}_e(\omega)=-\Gamma_0/2(\omega-\omega_0)$.  In addition, longitudinal and transverse TE  solutions satisfy
$\epsilon(\omega)=0$ and $q^2=k^2\epsilon(\omega)$, respectively, with $\epsilon(\omega)=1+\rho
\alpha_e(\omega)/(1-\rho\alpha_e(\omega)/2)$. Hence, in the long-wavelength limit, the crystal rigorously behaves as an
homogeneous medium of effective polarizability  $\alpha_e(\omega)$. 

The large quality-factor approximation of Eqs.~\eqref{EqDispersion} amounts to replace $\omega_\vec{q}$ by $\omega_0$ in
the right-hand sides. These solutions, independent of the geometry of the Bravais lattice, are in good agreement with
exact results shown in Fig.~\ref{fig:crystal_band_struc}\,(a) at $k_0a=2$. In particular, $\omega^{0}_\vec{q}$ and
$\omega^{\perp}_\vec{q}$ diverge when they cross the light circle at $\vert \vec{q}\vert=k_0$, whereas
$\omega^{\parallel}_\vec{q}$ does not. As a general guiding rule, divergence occurs when the polarization of the
eigenstate -- shown in Fig.~\ref{fig:crystal_band_struc}\,(b) and (c) for TE waves --  is tangential to light circle in
the plane $(q_x, q_y)$.  Because of the polaritonic dispersion, the TM solution predicts the existence of a gap at high
density, in the frequency range $\delta\in [0, 4/(k_0a)^2]$. This prediction matches with observations reported in
Sec.~\ref{dos} for TM waves in SHU materials. On the other hand, the TE long-wavelength approximation does not capture
the closing of the gap that occurs near the point $K$. The approximation~\eqref{EqDispersion} for
$\omega^{\parallel}_\vec{q}$ is flat, whereas the exact solution shown in Fig.~\ref{fig:crystal_band_struc}\,(a) bends
in the vicinity of $K$. The degeneracy between $\omega^{\perp}_\vec{q}$ and $\omega^{\parallel}_\vec{q}$ at point $K$ is
due to the equal contributions of the three adjacent Brillouin zones. By selecting the corresponding  components
[$\vec{Q}=\vec{0}$ and $\vec{Q}^{\pm}= 2\pi/a^*(1, \pm 1/\sqrt{3})$ for the triangular lattice] in
Eq.~\eqref{EqPoisson}, we find that $\tilde{\mathbb{G}}_0(\vec{q},\omega)$ is proportional to the identity at $K$,
making the distinction between longitudinal and transverse modes irrelevant.  The predicted absence of gap for TE waves
at high density is consistent with our findings in SHU materials [see Fig.~\ref{fig:phase_diagram}\,(b)].

\begin{figure}
   \centering
   \includegraphics[width=0.9\linewidth]{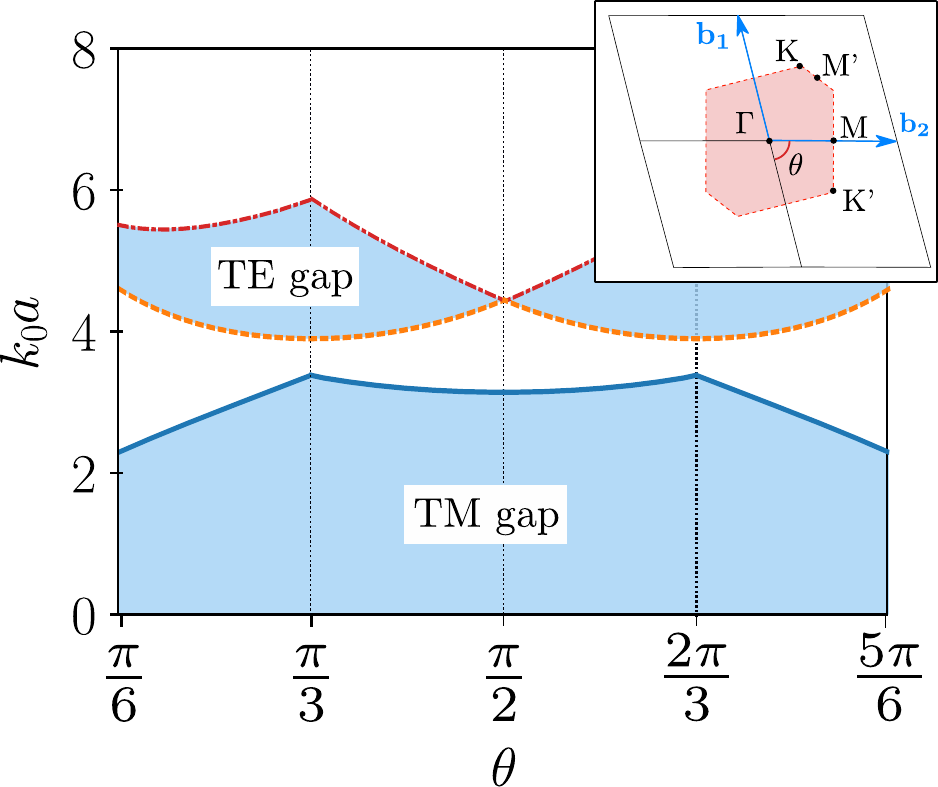}
   \caption{ Critical values $k_0a$ defining the gap regions as function of the direct lattice angle $\theta$. The TM
   gap disappears when the light circle reaches the closer inner edge of the first Brillouin zone (see inset), either in
   $M$ or $M'$ depending on $\theta$. It corresponds to the critical value $k_0a^*=\text{min}\left(\pi/\vert
   \text{sin}\theta\vert,\pi\sqrt{2-2\text{cos}\theta}/ \vert \text{sin}\theta\vert\right)$, where
   $a^*=a/\sqrt{\vert\text{sin}\theta}\vert$ is the distance between unit cells of the direct lattice. The  TE gap opens
   when the light circle reaches the outer edges in $K$ and $K'$, which corresponds to
   $k_0a^*=\pi/(1+\text{cos}\theta)\vert \text{sin}(\theta/2)\vert$; it closes  for  $k_0a^*=\text{min}\left(\pi/\vert
   \text{sin}(\theta/2)\vert,\pi \vert \text{sin}\theta\vert /\sqrt{5-4\text{cos}\theta} \right)$, when the light
   circles associated to two non-adjacent Brillouin zones first meet, either in $M$ or $M'$.  TM and TE gaps are found in
   non-overlapping  regimes of density. The range of density for the existence of the TE gap is maximal for the
   triangular lattice ($\theta=\pi/3$) and minimal for the square lattice ($\theta=\pi/2$).  }
   \label{fig:crystal_gap}
\end{figure}

By decreasing the density of scatterers, we reduce the size of the first Brillouin zone, which may become smaller than
the domain encompassed by the light circle centered in $\Gamma$. This is typically what is shown in
Fig.~\ref{fig:crystal_band_struc}\,(b) at $k_0a=4.5$. As a result, the light circles belonging to adjacent zones now
intersect the first Brillouin zone. The presence or absence of gap in this regime can be found by applying the general
guiding rule mentioned above. By definition, TM eigenstates are always transverse to the light circles, so that the TM
band diverges when crossing both the first circle along the path $\Gamma \to M$ and the second one along the path $M \to
K$ [see Fig.~\ref{fig:crystal_band_struc}\,(a)]. This precludes the possibility to observe a polaritonic gap for TM
polarization. On the other hand, TE eigenstates cannot be transverse to both light circles. Eigenstates associated to
$\omega^\perp_\vec{q}$ are transverse to the first circle, while those associated to $\omega^\parallel_\vec{q}$ are
transverse to the second one. Having a single polaritonic divergence in each of the two bands $\omega^\perp_\vec{q}$ and
$\omega^\parallel_\vec{q}$ allows the formation of a TE gap, while preserving the degeneracy at points $\Gamma$ and $K$.
The validity of this reasoning suggests that the sum in Eq.~\eqref{EqPoisson} is dominated by the component
$\vec{Q}=\vec{0}$ and its first neighbors on the reciprocal lattice. We checked numerically that a restriction of the
sum to these components  indeed reproduces  the band structure shown in Fig.~\ref{fig:crystal_band_struc}\,(a)
qualitatively, with a gap found at $k_0a=4.5$ for TE modes only. 

\begin{figure*}[t]
   \centering
   \includegraphics[angle=0,width=0.8\linewidth]{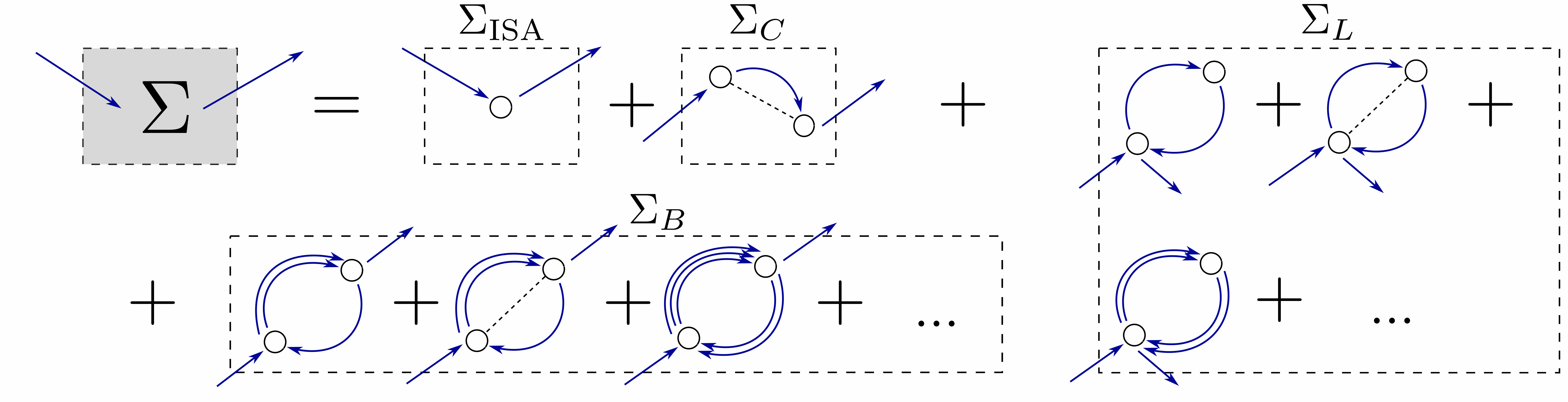}
   \caption{Pictorial representation of the scattering sequences included in the self-energy $\vec{\Sigma}$. Open
   circles represent scatterers, arrows account for field propagation through $\vec{G}_0$, and dashed line connecting
   scatterers stand for the spatial correlation function $h_2$. Diagrams are classified into four classes discussed in
   the text. The pseudo-gap found in the DOS $p(\delta)$ for strongly correlated systems is triggered by the destructive
   interference between the independent scattering contribution, $\vec{\Sigma}_{\text{ISA}}$, and the scattering loops
   involving two spatially correlated scatterers, $\vec{\Sigma}_L$.}
   \label{fig:rec_scat}
\end{figure*}

According to previous analysis, the TM gap observed at high density necessarily disappears when the light circle reaches
the inner edge of the first Brillouin zone (point M). This occurs at $k_0a^*=2\pi/\sqrt{3}$, where
$a^*=\sqrt{2}a/3^{1/4}$. On the contrary, a TE gap forms when the light circle reaches the outer edge (point K) and
disappears as soon as the circles associated to two non-adjacent Brillouin zones meet. For the triangular lattice, a gap
is thus found in the density window $k_0a^* \in [4\pi/3, 2\pi]$. For more general oblique lattices, such as those
considered in Sec.~\ref{sec:simu-crystal}, criteria for the gap formation in TM and TE polarization remain the same --
only the critical values of $k_0a$ defining the gap region are modified. We have calculated analytically and represented
in Fig.~\ref{fig:crystal_gap} these critical values as function of the angle $\theta$ between the two basis vectors of
the direct lattice. These predictions are in excellent agreement with direct evaluation of the band structure from
Eq.~\eqref{EqHamiltonQ}.  In particular, for TE polarization, a minimal gap exists for all dense lattices because of the
hexagonal shape of the Brillouin zone.  The only exception is the square lattice, which has a square Brillouin zone and
for which the conditions for gap opening and closing are the same. 

Our detailed examination of the band structure for arbitrary oblique lattice reveals that geometrical arguments
involving light circles and the shape of the Brillouin zone accurately capture the main features of the DOS at high or
moderate density. In particular, these arguments explain the robustness of the TE gap, even after averaging over an
ensemble of lattices mimicking SHU pattern, as shown in Fig.~\ref{fig:DOS_cry}\,(b). However, they hold for SHU patterns
with large degree of correlation $\chi$ only. In the following section, we aim at characterizing theoretically the
continuous evolution of $p(\delta)$ from white-noise disorder ($\chi=0$) to strongly correlated SHU patterns ($\chi\sim
0.5$). 

\section{Microscopic scattering model for the density of states}\label{dos_model}

Our objective is to evaluate the DOS of the effective Hamiltonian $\mathcal{H}(\omega_0)$ defined in Eq.~\eqref{EqHeff},
at finite density $\rho=N/\mathcal{A}$ and arbitrary $\chi$, in the limit of large system size ($N\to \infty,
\mathcal{A}\to \infty$). This DOS is related to the pdf of eigenvalues $\Lambda_n$ of the Green's matrix
$\tilde{\mathbb{G}}_0(\omega_0)$ through a simple rescaling. When increasing the area $\mathcal{A}$ occupied by the
scatterers, most of the eigenvalues $\Lambda_n$ that lie in the upper part of the complex plane
($\text{Im}\Lambda_n>-1$) progressively collapse towards the line $\text{Im}\Lambda =-1$, so that the imaginary part of
most of the eigenvalues of $\mathcal{H}(\omega_0)$ become vanishingly small while remaining negative. As a result, the
distribution of detuning $\delta=2(\omega-\omega_0)/\Gamma_0$ defined through Eqs.~\eqref{EqDOS1} and~\eqref{EqDOS2} can
be written in the form
\be
   p(\delta)=-\frac{1}{\pi\beta N}\text{Im} \left<\operatorname{Tr}\left[\frac{1}{(\delta +i)\mathds{1}
   + \tilde{\mathbb{G}}_0(\omega_0)}\right] \right>,
   \label{EqDOSTrace}
\ee 
where $\operatorname{Tr}$ stands for the trace of a matrix. The average trace involved in Eq.~\eqref{EqDOSTrace} may be evaluated theoretically by
generalizing the euclidean random matrix theory framework developed in Refs.~\onlinecite{skipetrov11, goetschy11_pre,
goetschy13_review} to correlated arrangement of scatterers. This is not the strategy adopted in the following, where we
wish to establish a clear connection with quantities commonly manipulated in mesoscopic physics theory, such as the
collective $\vec{T}$-operator or the self-energy $\vec{\Sigma}$~\cite{sheng06,akkermans07}.

The collective $\vec{T}$-operator in an ensemble of $N$ scatterers is the sum of all possible elastic scattering
sequences experienced by waves of frequency $\omega$~\cite{sheng06},
\begin{multline}
   \label{EqDefT}
   \vec{T}(\omega)=\sum_{l}\vec{t}_l+ \sum_{l}\sum_{m\neq l}\vec{t}_l\vec{G}_0(\omega)\vec{t}_m
\\
   +\sum_{l}\sum_{n\neq l}\sum_{m\neq n}\vec{t}_l\vec{G}_0(\omega)\vec{t}_n\vec{G}_0(\omega)\vec{t}_m
   + \dots,
\end{multline}
where $\vec{t}_l$ is the $\vec{t}$-operator of the point-like scatterer $l$ located in $\vec{r}_l$, of amplitude
$t(\omega)=-k^2\alpha(\omega)=-4\beta k^2\tilde{\alpha}(\omega)/k_0^2$. In the momentum representation, the
series~\eqref{EqDefT} is recast as
\be
   \label{EqDefT2}
   \vec{T}(\vec{q},\omega)=\frac{4\beta}{\mathcal{A}}\sum_{l=1}^N\sum_{m=1}^N\left[ \frac{1}{(\delta +i)\mathds{1} + \tilde{\mathbb{G}}_0(\omega_0)}  \right]_{lm} e^{-i\vec{q}.(\vec{r}_l-\vec{r}_m)}.
\ee
 This expression applies for resonators of large
quality factor described by the polarizability~\eqref{EqDefAlpha}. Since only the components $l=m$ contribute
significantly to Eq.~\eqref{EqDefT2} in the limit $q \to \infty$, we establish the following connection between
$p(\delta)$ in Eq.~\eqref{EqDOSTrace} and the average $\vec{T}$-operator:
\be
   p(\delta)=-\frac{1}{4\pi  \beta^2 \rho}\lim_{q\to \infty}\text{Im}\text{Tr}\left[\left<\vec{T}(\vec{q},\omega) \right> \right].
   \label{EqDOSasT}
\ee
Hence, we are left with the evaluation of $\left<\vec{T}(\vec{q},\omega) \right>$. Note that the distribution
$p(\delta)$ is expressed in terms of $\vec{T}(\omega)$ and not in terms of the full Green's operator
$\vec{G}(\omega)=\vec{G}_0(\omega)+\vec{G}_0(\omega)\vec{T}(\omega)\vec{G}_0(\omega)$, because it accounts for the
resonances of the complex system, defined as the poles of the scattering operator
$\vec{S}(\omega)=\mathds{1}+\vec{G}_0(\omega)\vec{T}(\omega)$. In particular, it does not reduce to the density of free
states in the absence of scatterers where $\vec{T}(\omega)=\vec{0}$.

The average collective $\vec{T}$-operator is conveniently expressed in terms of the self-energy $\vec{\Sigma}(\vec{q},\omega)$ as
\be
\label{DefSigma}
 \left<\vec{T}(\vec{q},\omega) \right>=\frac{\vec{\Sigma}(\vec{q},\omega)}{\mathds{1}-\vec{G_0}(\vec{q},\omega_0)\vec{\Sigma}(\vec{q},\omega)}.
\ee
By definition, $\vec{\Sigma}(\vec{q},\omega)$ is the sum of all possible irreducible scattering sequences contained in
the expansion of  $\left<\vec{T}(\vec{q},\omega) \right>$. In the following, we restrict our analysis to the regime of
moderate density ($k_0a\gtrsim 3$), where it is legitimate to perform an expansion of the self-energy  in density $\rho$.
By averaging Eq.~\eqref{EqDefT} and keeping all irreducible diagrams up to the second order in $\rho$, we get~\cite{vantiggelen94}
\begin{multline}
   \label{EqSigma}
   \vec{\Sigma}(\mathbf{q}, \omega) = \rho \, t(\omega) \mathds{1}
    + \rho^2t(\omega)^2
      \int \mathrm{d}\mathbf{r} \, h_2(\mathbf{r})\mathbf{G}_0(\mathbf{r},\omega_0) e^{-i \mathbf{q}\cdot \mathbf{r}} 
\\
   + \rho^2t(\omega)^3 \int \mathrm{d}\mathbf{r} \, \left[1+h_2(\mathbf{r})\right] \frac{
   \mathbf{G}_0^2(\mathbf{r},\omega_0)}{\mathds{1}-t(\omega)^2\mathbf{G}_0^2(\mathbf{r},\omega_0)}
\\
   + \rho^2t(\omega)^4
   \int \mathrm{d}\mathbf{r} \, \left[1+h_2(\mathbf{r})\right]
      \frac{\mathbf{G}_0^3(\mathbf{r},\omega_0)}{\mathds{1}-t(\omega)^2\mathbf{G}_0^2(\mathbf{r},\omega_0)} e^{-i \mathbf{q}\cdot \mathbf{r}}.
\end{multline}
The four terms of the expansion~\eqref{EqSigma} are represented schematically in Fig.~\ref{fig:rec_scat}. The first one,
$\vec{\Sigma}_{\text{ISA}}$, is the independent scattering approximation of the self-energy, which is the only relevant
term in the limit of dilute uncorrelated system. By keeping this first order in density in Eq.~\eqref{EqDOSasT}, we find
that the DOS $p(\delta)$ takes a Lorentzian profile,
\be
\label{EqDOSISA}
p(\delta)=\frac{1}{\pi}\frac{1}{1+\delta^2},
\ee
both for TM and TE waves. This result is in excellent agreement with numerical simulations at $\chi=0$ in the dilute
limit $k_0a\gg 1$ (results not shown). The second term in Eq.~\eqref{EqSigma}, noted $\Sigma_C$ in
Fig.~\ref{fig:rec_scat}, corresponds to scattering by two correlated scatterers, while the two last terms, $\Sigma_L$
and $\Sigma_B$, are recurrent scattering contributions involving pairs of scatterers; $\Sigma_L$ stands for scattering
loops which end where they start, and $\Sigma_B$ represents boomerang-like sequences in which the last scatterer differs
from the first one. For TM waves, $\Sigma_C$ and  $\Sigma_B$ do not contribute to $p(\delta)$ because they vanish in the
limit $q\to \infty$. On the contrary, for TE waves, they contribute to $p(\delta)$ through the singularity
$\delta(\vec{r})\mathds{1}/2k_0^2$ of $\mathbf{G}_0(\mathbf{r},\omega_0)$. The corresponding weight of $\Sigma_C +
\Sigma_B$ is $-\rho^2t(\omega)^2\mathds{1}/2k_0^2$, irrespectively of the amplitude of the correlation $h_2$. This term,
which slightly red-shifts  $p(\delta)$ by an amount $\Delta\delta \simeq -4/(k_0a)^2$ in 2D, is the so-called
Lorentz-Lorenz correction to the DOS, discussed in Ref~\onlinecite{morice95, cherroret16} for 3D uncorrelated systems. The
only contribution in Eq.~\eqref{EqSigma} that gives a dependence of $p(\delta)$ on the degree of spatial correlation is
thus the loop term $\Sigma_L$. 

\begin{figure*}
   \centering
   \includegraphics[width=.9\linewidth]{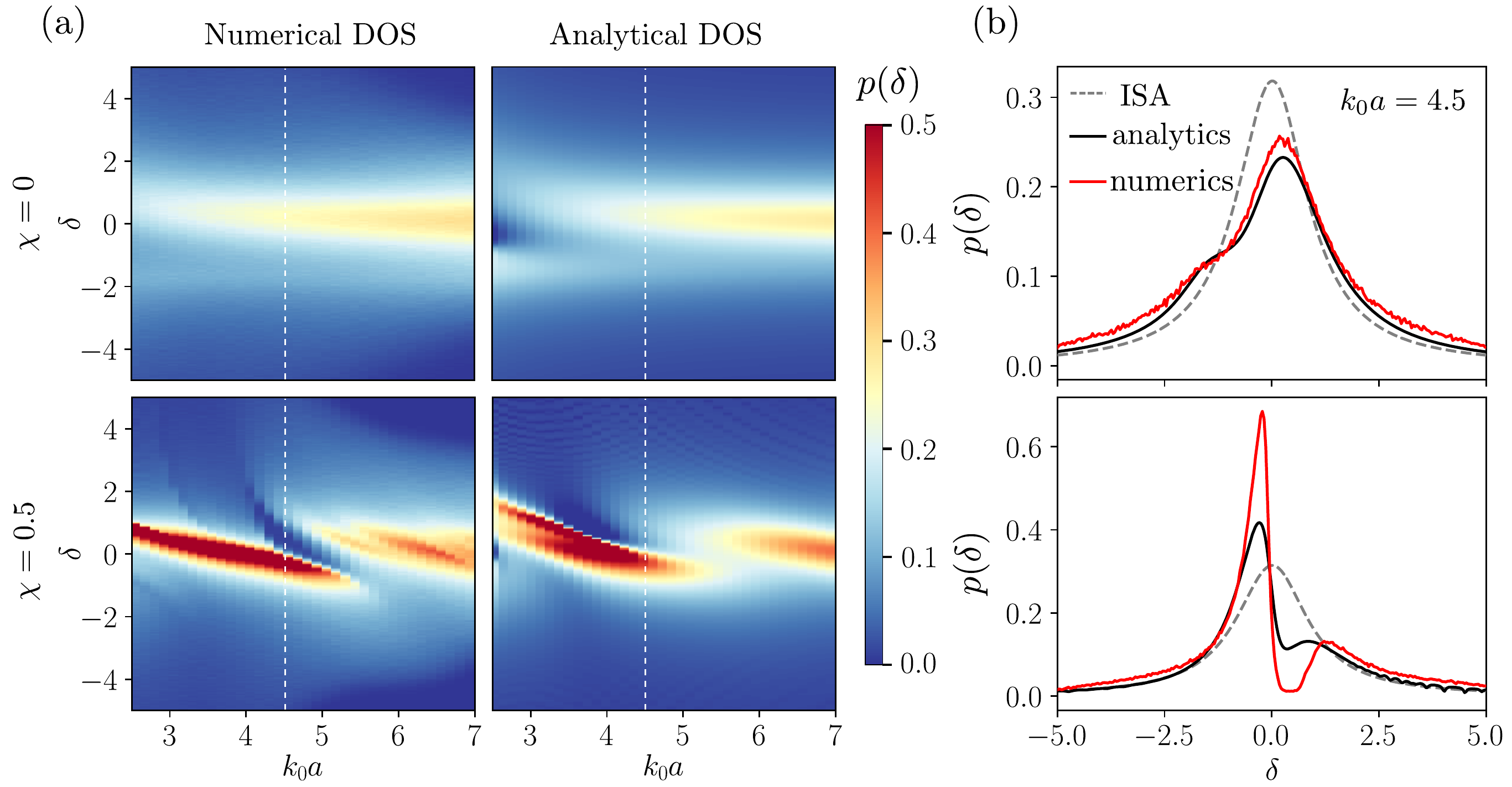}
   \caption{(a) Comparison between the numerical DOS $p(\delta)$ of TE waves (left) and the theoretical
   prediction~\eqref{EqDOSasT}, including recurrent scattering and spatial correlation at the second order in density
   (right). Good agreement is found for $k_0a\gtrsim 3$, without (top) and with (bottom) spatial correlation.  (b) Cuts
   along the line $k_0a=4.5$ of the maps shown in (a). The Lorentzian profile~\eqref{EqDOSISA} for dilute uncorrelated
   system is shown with a dashed line for reference. Numerical results shown in (a) and (b) have been obtained with a system size $k_0 R = 55$ and a number of disorder realizations adjusted to have $\sim10^6$ eigenvalues for each $k_0a$.
}
   \label{fig:DOSTheory}
\end{figure*}

We have represented in Fig.~\ref{fig:DOSTheory}(a) maps of the DOS $p(\delta)$ of TE waves in the regime of moderate
density $k_0a\in[2.5, 7]$, for two degrees of spatial correlation, $\chi=0$ (top) and $\chi=0.5$ (bottom). Numerical
distributions obtained from the diagonalization of the Hamiltonian~\eqref{EqHeff} (left pannels) are compared with the
theoretical prediction~\eqref{EqDOSasT}, evaluated at the second order in density with Eqs.~\eqref{DefSigma}
and~\eqref{EqSigma} (right pannels). Details regarding the explicit analytical computation of Eq.~\eqref{EqDOSasT} and
Eq.~\eqref{EqSigma} are given in App.~\ref{Appendix:Sigma}. Good agreement between numerics and theory is found over
a broad range of density and  detuning, all the way from uncorrelated system ($\chi=0$) to strongly correlated one
($\chi=0.5$). This confirms the validity of the connection established in Eq.~\eqref{EqDOSasT} between $p(\delta)$ and
the average collective $\vec{T}$-operator. 

Figure~\ref{fig:DOSTheory}\,(b) shows cuts of the two maps along the line $k_0a=4.5$. In the absence of spatial
correlation (top), $p(\delta)$ differs slightly from the Lorentzian profile~\eqref{EqDOSISA} found in the dilute limit
(dashed line). In particular, the loop correction $\vec{\Sigma}_L$ is responsible for an antisymetric contribution to
$p(\delta)$ through $\text{Im}[t(\omega)^3]$, which blue-shifts the maximum of $p(\delta)$ and creates a depletion at
$\delta<0$. On the contrary, for a large degree of correlation (bottom), $\vec{\Sigma}_L$ is responsible for a dip in
the DOS at $\delta>0$, which coincides with the pseudo-gap found in simulations. To understand the origin of this dip,
it is sufficient to consider the single loop approximation  $\vec{\Sigma}_L\simeq  \rho^2t(\omega)^3 \int
\mathrm{d}\mathbf{r} \, g_2(\mathbf{r}) \mathbf{G}_0^2(\mathbf{r},\omega_0) $. At $\chi=0.5$, the pair correlation
function $g_2(\mathbf{r})=1+h_2(\mathbf{r})$ is zero for $r\lesssim a$ and presents a dominant peak at $r\simeq a$.
As the result, most of the scatterers involved in a scattering loop are separated by a distance $a$. For $\delta \gtrsim
0 $ and $k_0a=4.5$, the phase accumulated along a scattering loop of length $a$ is opposite to the phase of single
scattering, so that  $\vec{\Sigma}_\text{ISA}$ and $\vec{\Sigma}_L$  are of opposite sign and interfere destructively.
The fact that the theory does not reproduce quantitatively the pseudo-gap is attributed to the density expansion of
$\vec{\Sigma}$ performed above. We expect the agreement with numerical DOS to get better when including scattering loops
made of more than two spatially correlated scatterers. 

Our theoretical treatment of the  DOS not only provides explicit expressions of $p(\delta)$ for any $\chi$, but also
indicates which microscopic scattering mechanisms contribute to the formation of the pseudo-gap in correlated materials,
and what should be the profile of the spatial correlation function to make the pseudo-gap prominent.  

\section{Localization length in correlated resonant systems}\label{xi}

We found in Sec.~\ref{SubSec:LocDos} that the formation of a pseudo-gap in the DOS $p(\delta)$ of correlated materials
is concomitant to a modification of the eigenstates of $\mathcal{H}(\omega_0)$. States near the band edges become
exponentially localized in the bulk of the medium, and, as a result, acquire very long life-times in finite-size
samples. Here we want to establish, in an explicit manner, the connection between $p(\delta)$ and the localization
properties of  $\mathcal{H}(\omega_0)$. To characterize the latter, it is custom to consider the spatial properties of
the intensity associated to the field operator
$\mathcal{G}(\omega)=1/[\omega\mathds{1}-\mathcal{H}(\omega_0)]$~\cite{sheng06}. The mean intensity is defined as
\be
   I^{\alpha \gamma}(\vec{R}, \Omega)=\left<\mathcal{G}^{\alpha \gamma}_{ij}\left(\omega^+ \right)\mathcal{G}^{\alpha \gamma}_{ij}\left(\omega^-\right)^* \right>,
\ee
where  $\omega^\pm = \omega\pm\Omega/2$, $(i, j)$ are positions indices with $\vec{r}_i=\vec{r}_j+\vec{R}$, and
$(\alpha, \gamma)$ refer to output and input polarization channels. Using Eq.~\eqref{EqDefT2}, we can also express the
intensity in terms of the collective $\vec{T}$-operator,
\be
   \label{EqTT}
   I^{\alpha \gamma}(\vec{R}, \Omega) \propto \left< \vec{T}^{\alpha \gamma}_{ij} (\omega^+)\vec{T}^{\alpha \gamma}_{ij}(\omega^-)^*\right>,
\ee
where $\vec{T}^{\alpha \gamma}_{ij}(\omega)=\left([(\delta +i)\mathds{1} + \tilde{\mathbb{G}}_0(\omega_0)]^{-1}\right)^{\alpha \gamma}_{ij}$. 

One common way to find the expression of the correlator ~\eqref{EqTT} is to write a Behte-Salper equation for $\left<
\vec{T} \vec{T}^\dagger  \right> $ and compute its irreducible vertex at a given order in density using the
expansion~\eqref{EqDefT}. In the long time limit ($\Omega\to 0$) and large scale limit ($R\gg \ell_t$),   $I^{\alpha
\gamma}(\vec{R}, \Omega)$ is dominated by the scalar mode of the Bethe-Salpeter equation, which is independent of the
polarization channels: $I^{\alpha \gamma}(\vec{R}, \Omega)\equiv I(\vec{R}, \Omega)$~\cite{akkermans08}. In addition, in
the regime of weak scattering ($k\ell_s \gg 1$, where $\ell_s$ is the scattering mean free path discussed below), the
mean intensity effectively obeys a diffusion equation. In the Fourier domain, it reads
\be
   \label{EqDiff}
   (DQ^2-i\Omega)I(\vec{Q}, \Omega)=0,
\ee
with $I(\vec{Q}, \Omega)$ the Fourier transform of $I(\vec{R}, \Omega)$. If localization corrections are ignored, the
diffusion coefficient  $D$ for 2D waves reduces to
\be
   \label{EqDiffCoeff}
   D_0=\frac{\ell_t v_E}{2},
\ee
where $\ell_t=\ell_s/(1-g)$ is the transport mean free path, $g$ the scattering anisotropy factor, and $v_E$ the energy
velocity. In an ensemble of scatterers with large quality factors, $v_E$ is much smaller than $c$, because of the long
time ($\sim\Gamma_0^{-1}$) spent for each scattering event:   $v_E\simeq\Gamma_0\ell_s$~\cite{lagendijk96}. The
result~\eqref{EqDiffCoeff} includes the  vectorial nature of the problem at hand, in the sense that the explicit
expressions of $\ell_s$ and $g$, which involve the self-energy and the irreducible vertex, depend on
it~\cite{cherroret16, vynck21}. But it neglects other genuinely new vector effects that may show up in the regime of
large density, where near field radiation  carried by the longitudinal part of the Green's tensor
$\vec{G}_0(\vec{q},\omega)$ becomes dominant. Specifically, it was put forward  recently that the interference between
the longitudinal and transverse waves contribute to the diffusive current and thus increase the value of the diffusion
coefficient~\cite{vantiggelen21}.  In the regime of moderate density we are interested in ($k_0a\gtrsim 4$), we checked
explicitly that removing the longitudinal part of  $\vec{G}_0(\vec{q},\omega)$ in the definition of the
Hamiltonian~\eqref{EqHeff} does not change our findings for TE waves: both pseudo-gap and localization signatures shown
in Fig.~\ref{fig:phase_diagram} are preserved quantitatively.

In infinite 2D systems, localization corrections to the result~\eqref{EqDiffCoeff} cannot be ignored. When a pair of
complex conjugated fields evolves along a diffusive path, there is a high probability to form a scattering loop in which
the two fields propagate along the same path but in opposite directions. Accounting for these loops (cooperons)
self-consistently leads to a profound modification of $D$ which becomes solution of the equation~\cite{vollhardt92}
\be
   \label{EqDRenom}
   D=D_0-DP_0,
\ee 
where $P_0$ is the return probability to the origin by diffusion,
\be
   P_0=\frac{1}{\pi p(\omega)}\int \frac{d\vec{Q}}{(2\pi)^2} \frac{1}{DQ^2-i\Omega}.
\ee 
This expression explicitly depends on the DOS $p(\omega)$ of the Hamiltonian~\eqref{EqHeff}, defined in
Eq.~\eqref{EqDOS1}. The integral that appears in the expression of $P_0$ diverges at large $Q$, because the diffusion
process breaks down at distance shorter than $\sim\ell_t$. By introducing a cutoff $q_\text{max}= 1/\ell_t$, we get 
\be
   P_0=\frac{1}{(2\pi)^2p(\omega) D} \text{ln}\left(1+\frac{iD}{\Omega \ell_t^2} \right).
\ee
In the stationary limit $\Omega \to 0^+$,  the solution $D$ of Eq.~\eqref{EqDRenom} is thus purely imaginary. It reads
\be
   \label{EqSolD}
   D=-i\Omega \xi^2,
\ee
 where $\xi$ is a characteristic length defined as
\be
   \xi=\ell_t e^{2\pi^2 p(\omega)D_0}.
   \label{EqXi1}
\ee
With the result~\eqref{EqSolD}, the stationary solution of Eq.~\eqref{EqDiff} becomes $I(\vec{R})\sim e^{-R/\xi}$, so
that $\xi$ can be interpreted as the localization length. By inserting the expressions  of $D_0$ and $p(\omega)$, given
by Eqs.~\eqref{EqDiffCoeff} and~\eqref{EqDOS2}, into the formal solution~\eqref{EqXi1}, we finally get
\be
   \xi=\ell_t e^{2\pi^2\beta \rho \ell_t \ell_s p(\delta)},
   \label{EqXi2}
\ee  
which applies both for TM ($\beta=1$) and TE ($\beta=2$) waves.
 
To analyze the main features of the result~\eqref{EqXi2}, we need to specify the expression of $\ell_s$. In the absence
of absorption, $\ell_s$ characterizes the exponential extinction of the ballistic light inside the disordered medium. At
second order in density, its inverse takes the form
\be
   \frac{1}{\ell_s}=\frac{- \text{Im}\left[\Sigma^\perp(q=k_0, \omega)\right]}{\omega_0/v_\varphi},
 \label{EqMpf}
\ee
where $v_\varphi=c\left[1+  \text{Re}\Sigma_{\text{ISA}}^\perp(\omega)\right]$ is the phase velocity and
$\Sigma^\perp(q=k_0, \omega)$ is the transverse part of $\vec{\Sigma}(\vec{q}, \omega)$ given in Eq.~\eqref{EqSigma},
evaluated on the shell $\vert \vec{q} \vert = k_0$.  Its explicit analytical expression is given in
App.~\ref{Appendix:Sigma}. The result~\eqref{EqMpf} is based on the on-shell approximation that breaks down at large
density. In that case, the dielectric constant of the effective medium is non-local and the extinction of the ballistic
component is no more exponential. We checked numerically that this occurs in our systems for $k_0a \lesssim 4.6$.

The comparison of  Eqs.~\eqref{EqDOSasT} and~\eqref{EqMpf} shows that  $p(\delta)$ and $\ell_s$, whose product appears
in the expression~\eqref{EqXi2} of the localization length,  are both expressed in terms of $\vec{\Sigma}(\vec{q},
\omega)$. The first one however depends on both the transverse and longitudinal parts of $\vec{\Sigma}(\vec{q},
\omega)$, evaluated at $q\to \infty$ instead of $q=k_0$. As a consequence, the term $\vec{\Sigma}_B(\vec{q}, \omega)$
contributes to $\ell_s$ but not to $p(\delta)$.  The only case where the product $\rho\ell_s p(\delta)$ is independent
of detuning and density is the one of dilute systems ($k_0a\gg 1$), for which $\vec{\Sigma}(\vec{q}, \omega)\simeq
\Sigma_{\text{ISA}}(\omega)\mathds{1}$ is independent of $\vec{q}$. Then, Eq.~\eqref{EqXi2} reduces to
\be
\xi\simeq\ell_t e^{\frac{\pi k_0\ell_t}{2}},
  \label{EqXi3}
\ee
which is the commonly used expression for the localization length in 2D~\cite{vollhardt92, conley14}. We stress that the
result~\eqref{EqXi3} is not accurate as soon as spatial correlation and recurrent scattering become important. In
particular, for TE waves at $\chi=0.5$, we find that the frequency profile of $p(\delta)$ is strongly asymmetric [see
Fig.~\eqref{fig:DOSTheory}(b)], whereas the theoretical  prediction~\eqref{EqMpf} gives an almost symmetric profile of
$\ell_s$ vs $\delta$ (result not shown). Furthermore, the dependence of  $p(\delta)$ on $k_0a$ is strongly non-monotonic
[see Fig.~\eqref{fig:DOSTheory}(a)],  whereas Eq.~\eqref{EqMpf} yields to a monotonic increase of $\ell_s$ with $k_0a$.
In this situation, the product $\rho\ell_s p(\delta)$ is not constant and the approximation~\eqref{EqXi3} of
Eq.~\eqref{EqXi2} does not hold.

\begin{figure}
   \centering
   \includegraphics[width=.9\linewidth]{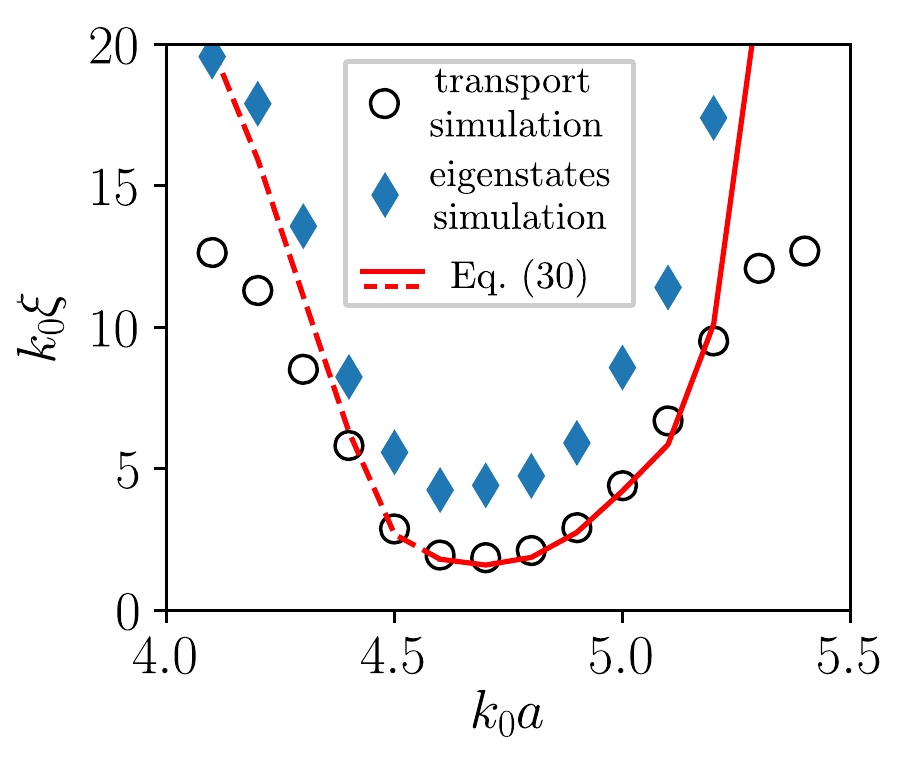}
   \caption{Comparison between the localization length $\xi$ obtained from transport simulation (open circles) and
   eigenstate profile (filled diamonds),  with the formula~\eqref{EqXi2} (solid and dashed line),  in a  ensemble of
   correlated resonant dipoles ($\chi=0.5$, $\delta=0$), illuminated with TE polarization. The localization length is
   strongly reduced when the DOS $p(\delta)$ is depleted.  Equation~\eqref{EqXi2} has been evaluated with numerical
   values of $p(\delta)$ and $\ell_s$.  For $k_0a<4.6$, reliable values of $\ell_s$ cannot be found, which is attributed
   to non-locality of the effective dielectric constant (see text for details); the value of  $\ell_s$ obtained at
   $k_0a=4.6$ is thus extrapolated  to get $\xi$  at lower $k_0a$ (dashed line). Details about transport simulation and
   eigenstate analysis are given in App.~\ref{appendix:ls_xi_computation}.}
   \label{fig:kxi_num}
\end{figure}

To test the validity on the prediction~\eqref{EqXi2}, and in particular its explicit dependance on $p(\delta)$, we first
computed numerically the values of $\ell_s$ and $\xi$ in the TE polarization at $\chi=0.5$. For this purpose, we
illuminated a disordered slab of thickness $L$ made of resonant dipoles and studied the dependence on $L$ of the mean
field intensity measured in transmission,  $\vert \left<\vec{E}(L) \right> \vert^2$, as well as the mean of the
logarithm of the intensity, $ \left< \text{ln}\left[\vert \vec{E}(L) \vert^2\right] \right> $.  Details of the
calculation are given in App.~\ref{appendix:ls_xi_computation}. We also checked that the values of $\xi$ found in this
way are compatible with a direct fit of the exponential profile of the localized eigenstates of $\mathcal{H}(\omega_0)$;
typical profile of the eigenstates and distribution of $\xi $ are shown in App.~\ref{appendix:ls_xi_computation}.
Second, we estimated $\ell_t$ by computing theoretically the irreducible vertex in the presence of spatial correlation
at  second order in density. This amounts to generalize the results of Ref.~\onlinecite{cherroret16} to 2D correlated
systems. After a lengthy calculation presented in App.~\ref{Appendix:Anisotropy}, we found that the scattering
anisotropy factor $g$ is of the order of $\sim 0.1$ for $k_0a \in [4, 5.5]$, so that $\ell_t\simeq 1.1 \, \ell_s$. In
Fig.~\ref{fig:kxi_num}, we compare the numerical values of $\xi$ with the formula~\eqref{EqXi2} evaluated with the
numerical values of $\ell_s$ and $p(\delta)$, in the range of $k_0a$ where $p(\delta)$ exhibits a pseudo-gap. We find
that $\xi$ is strongly reduced in this density window, confirming the critical dependence of $\xi$ on $p(\delta)$.
Hence, formula~\eqref{EqXi2} justifies quantitatively why localized states form when the DOS is depleted. 
 
\section{Conclusion}

We have demonstrated in this work that spatial correlation in disordered systems profoundly modify the properties of 2D
vector waves, by inducing pseudo-gap in the DOS as well as spatial localization. To explain these results, we
established a general formula for the DOS, Eq~\eqref{EqDOSasT}, that we expressed at the second order in density in
terms of the pair correlation function $g_2(\vec{r})=h_2(\vec{r})+1$. This expression captures the impact of the
disorder-to-order transition on the DOS. Strong spatial correlations force the scatterers to be at a well defined
distance from each other, which allows for efficient destructive interference between elementary scattering processes
involving one and two scatterers, in the density window $k_0a\in [4.2, 5.2]$. Hyperuniformity ($ \lim_{\vec{q}\to 0}
S(\vec{q})=0$) is not a requisite condition for this process, which can be realized with hard disk systems as well.

We also developed a complementary model for the DOS  based on an effective-crystal representation. This model
captures the fundamental difference between scalar and vector waves in the regime of high density. At $k_0a\lesssim 3$,
longitudinal and transverse TE modes interact to prevent the formation of a polaritonic gap similar to the one formed by
TM modes. On the other hand, at  $k_0a\gtrsim 4$, transverse and longitudinal modes exhibit independent polaritonic
dispersions that result in a TE gap, whereas TM waves support a double polaritonic dispersion that closes the gap. By
analyzing the mechanisms responsible for these band structures, we were able to predict accurately the non-overlapping
regimes of density where TE and TM  gaps are observed in strongly correlated materials. 
 
Finally, we proved that the localization length $\xi$ in 2D correlated systems takes the general form~\eqref{EqXi2}.
This result explicitly shows that spatial correlations can deeply affect localization, not only by modifying the
scattering mean free path $\ell_s$ and the scattering anisotropy factor $g$ as found numerically in
Ref.~\onlinecite{conley14}, but also by depleting the DOS $p(\delta)$. In this way, we explained why it is so difficult to
observe localization of TE modes in uncorrelated media of finite size and why spatial correlations turn out to be a
powerful knob to reach localization. Our predictions for $p(\delta)$ and $\xi $ are supported quantitatively by
extensive numerical simulations of wave propagation in hyperuniform systems. 

By providing theoretical grounds as well microscopic mechanisms for the emergence of photonic gap and localization, we
clarify recent experimental and numerical findings that put forward the role of short-range order for these
processes~\cite{edagawa08, liew11, froufe16, froufe17, aubry20}. We also establish the possibility to induce
localization for 2D TM waves propagating through in-plane dipole excitations, which has not been explored experimentally
so far. Finally, our theoretical treatments of the DOS and localization can, in principle, be extended to 3D systems.
Although the possibility to propagate along a third dimension evidently leads to qualitative different behavior, the
role of spatial correlations and the microscopic mechanisms responsible for $p(\delta)$ and $\xi$ are formally the same
as in 2D.  Hence, we think that our work will help to identify the key structural ingredients that affect significantly
wave propagation in 2D and 3D strongly correlated materials.    

\begin{acknowledgments}
We acknowledge useful discussion with  Rémi Carminati and Fabrice Lemoult.
   This research was supported by the ANR project LILAS under reference ANR-16-CE24-0001-01, by the ANR project
   MARS\_light under reference ANR-19-CE30-0026 and by LABEX WIFI (Laboratory of Excellence within the French Program
   "Investments for the Future") under references ANR-10-LABX-24 and ANR-10-IDEX-0001-02 PSL*. 
\end{acknowledgments}

\appendix

\section{Electromagnetic Green function and Green matrix in 2D}\label{appendix:Green}

Throughout this article, we have considered a translationally invariant medium along one direction so that the usual
$3\times3$ Green tensor splits into two independent block elements, a scalar one in the TM polarization when the
excitation lies along the invariance direction, and a $2\times2$ block for the TE polarization with a field
perpendicular to it. For the sake of simplicity both operators will be denoted $\mathbf{G_0}$. For TM polarization,
$\mathbf{G_0}$ is such that
\begin{gather}
   \left[ \nabla^2 + k^2\right]\mathbf{G_0}(\vec{r}-\vec{r}',\omega) =  \delta(\vec{r}-\vec{r}'),
\\
   \mathbf{G_0}(\vec{R},\omega) = -\frac{i}{4} \op{H}_0^{(1)}(k_0 R),
\end{gather}
where $\op{H}_{\alpha}^{(1)}$ is the Hankel function of the first kind and order $\alpha$. Similarly in the TE polarization
the Green tensor $\mathbf{G_0}$ reads
\begin{gather}
   \left[ -\nabla\times \nabla \times +k^2\right]\mathbf{G_0}(\vec{r}-\vec{r}',\omega) = \mathds{1}
   \delta(\vec{r}-\vec{r}'),
\\
   \begin{split}
      \mathbf{G_0}(\vec{R},\omega) & = -\frac{i}{4}\operatorname{VP}
         \left[  \left(\mathds{1} - \frac{\vec{R}\otimes \vec{R}}{R^2} \right)\op{H}_0^{(1)}(k_0 R)  \right.
   \\
      & \left. -  \left(\mathds{1} - 2\frac{\vec{R}\otimes \vec{R}}{R^2} \right) \frac{\op{H}_1^{(1)}(k_0 R)}{k_0R}
         \right] + \frac{\delta(\vec{R})}{2k_0^2}\mathds{1}.
   \end{split}
   \label{eq:G0}
\end{gather}

\section{Hyperuniform pattern generation and properties}\label{appendix:shu}

\begin{figure}[h]
   \centering
   \includegraphics[width=0.8\linewidth]{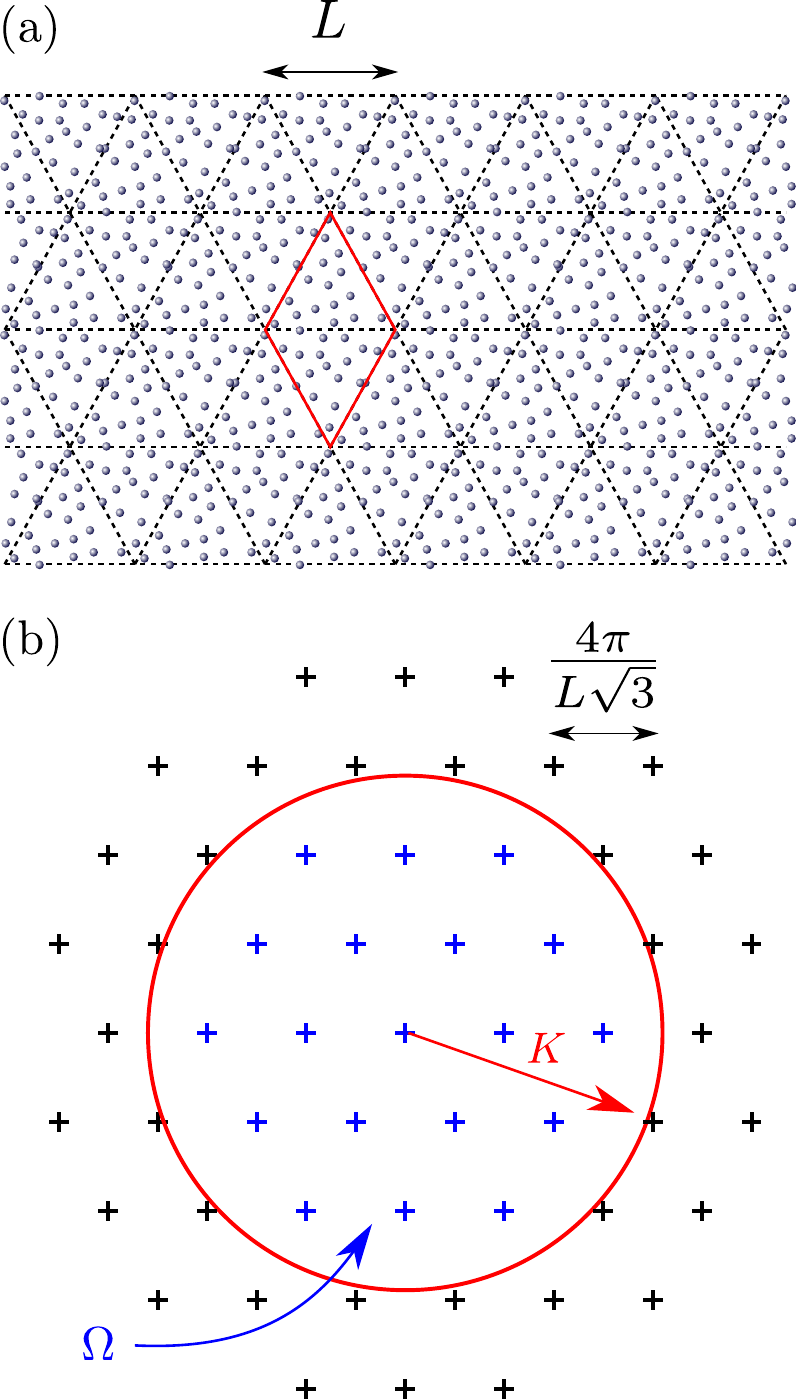}
   \caption{(a) Periodic system considered to generate stealth hyperuniform point patterns. (b) Ensemble $\Omega$ of
   points of the reciprocal space where the structure factor have to be optimized.}
   \label{fig:hyperuniform_generation}
\end{figure}

In this appendix, we focus on the process used to generate stealth hyperuniform point patterns. This process is
adapted from Refs.~\onlinecite{uche04,uche06,batten08,leseur16,froufe16}. SHU patterns are such that the
structure factor $\tilde{S}(\vec{q})$ vanishes in a domain $|\vec{q}|<K$.  Numerically, only a finite number of points can be
manipulated. To mimic an infinite medium, we thus divide the space into identical unitary cells. A square lattice is
often too restrictive since it forces the point pattern to crystallize into a square lattice at large $\chi$. For that
reason, we have chosen a triangular (or hexagonal) lattice [see Fig.~\ref{fig:hyperuniform_generation}\,(a)].

The induced periodicity implies that the structure factor vanishes for all $\vec{q}$ except the ones lying on the
reciprocal lattice. The problem of generating a SHU pattern reduces now to an optimization process to minimize the
structure factor for all $\vec{q}\in\Omega$ where $\Omega$ is the intersection of the reciprocal lattice and the domain
$|\vec{q}|<K$ [see Fig.~\ref{fig:hyperuniform_generation}\,(b)]. This can be translated into the determination of point
positions that minimizes the potential
\be
   U(\{ \vec{r}_j \})=\sum_{\vec{q}\in\Omega}\left|\sum_{j}\exp(i\vec{q}\cdot\vec{r}_j)\right|^2.
\ee
In practice, the first guess is a fully random configuration of points (Poisson point pattern) on which a conjugate
gradient method is applied which leads to a SHU point pattern. Depending on the intended use, the unitary cell structure
is then scaled and cropped to obtain a SHU point pattern inside a disk of radius $R$ and density $\rho=a^{-2}$.

Finally, we note that the structure factors $\tilde{S}(\vec{q})=\sum_{i,j}e^{i\vec{q}.(\vec{r}_i-\vec{r}_j)}/N$ and
$S(\vec{q})=1+\rho h_2(\vec{q})$ are linked through the relation
\be
   S(\vec{q})=\tilde{S}(\vec{q})-4\pi^2\rho\delta(\vec{q})
\ee
in an infinite medium (we have to remove the forward-scattering component). In a finite-sized medium, this relation
becomes~\cite{leseur16}
\be
   S(\vec{q})=\tilde{S}(\vec{q})-\frac{N-1}{\mathcal{A}^2}|\Theta(\vec{q})|^2
\ee
where $\mathcal{A}$ is the area occupied by the scattering medium and $\Theta(\vec{q})$ is the Fourier transform of the
function
\be
   \Theta(\vec{r})=\begin{cases}
      1 & \text{if $\vec{r}\in\mathcal{A}$,}
   \\
      0 & \text{otherwise.}
   \end{cases}
\ee

\section{Regularization of the crystalline Hamiltonian and large wavelength expansion}\label{appendix:crystal}

To construct the dispersion relation in infinite crystals, we recall the Hamiltonian from Eq.~\eqref{EqHamiltonQ} expressed
in term of Bloch waves
\be\label{EqHamiltonQ:app}
   \mathcal{H}_\vec{q}(\omega_\vec{q})=\left(\omega_0-i\frac{\Gamma_0}{2} \right)\mathds{1}-\frac{\Gamma_0}{2}\tilde{\mathbb{G}}_0(\vec{q},\omega_\vec{q})
\ee
with $\tilde{\mathbb{G}}_0(\vec{q},\omega)=-4\beta(\omega/\omega_0)^2\sum_{\vec{R}\neq \vec{0}}\vec{G}_0(\vec{R},
\omega)e^{-i\vec{q}\cdot\vec{R}}$. Following the procedure of Refs.~\onlinecite{perczel17,antezza09_pra}, we want to convert
the sum over the lattice into a sum over in reciprocal space.  Poisson's formula requires the evaluation of the Green's
function at its singularity.  This has been circumvented by regularizing the Green's function using a Gaussian cut-off
in momentum space, which smoothen the real space divergence at the origin.  Explicitly we used the
relation derived in Ref.~\onlinecite{antezza09_pra} given by
\begin{multline}
      \sum_{\vec{R}\neq \vec{0}} \mathbf{G}_0(\vec{R},\omega) e^{-i \vec{q}\cdot\vec{R}} \simeq
         e^{\frac{k_0^2b^2}{2}} \rho \sum_{\vec{Q}} \mathbf{G}_0^*(\vec{Q}-\vec{q},\omega)
   \\
      - \mathbf{G}_0^*(\vec{R}=\vec{0},\omega),\label{eq:regul1}
\end{multline}
where the regularized Green's function in momentum space takes the form
\begin{equation}\label{eq:regul2}
   \mathbf{G}_0^{*} (\vec{q},\omega) = \mathbf{G}_0(\vec{q},\omega) \frac{e^{-\frac{q^2b^2}{2}}}{2\pi b^2},
\end{equation}
where $b$ is the regularization parameter.
Using Eqs.~\eqref{eq:regul1} and \eqref{eq:regul2}, we can rewrite Eq.~\eqref{EqPoisson} as a sum of finite terms
\begin{multline}
\label{eq:Hq_reg}
   \tilde{\mathbb{G}}_0(\vec{q},\omega)=-\frac{4\beta\omega^2}{\omega_0^2} \,e^{\frac{k_0^2b^2}{2}} \left[\frac{1}{V_L}\sum_{\vec{Q}}
      \vec{G}^*_0(\vec{q}-\vec{Q}, \omega) 
   \right.
\\\left.\vphantom{\sum_{\vec{Q}}}
      - \vec{G}^*_0(\vec{R}=\vec{0}, \omega)\right],
\end{multline}
which becomes independent of $b$ for $b \ll a$. For the TM polarization,
the last term of Eq.~\eqref{eq:Hq_reg}  reads
\begin{align}
   \mathbf{G}_0^{*} (\vec{R}=\vec{0},\omega) & =-  \frac{i}{4}  \int \mathrm{d}\mathbf{r}\op{H}_0^{(1)}(k_0r) \frac{e^{-\frac{r^2}{2b^2}}}{2\pi b^2}
\notag
\\
   \underset{k_0b\ll 1} & \simeq -\frac{1}{\pi}\left[ \gamma + \ln\left(\frac{k_0^2b^2}{2} \right) \right] - \frac{i\mathds{1}}{4}.
\end{align}
Here the Hankel function has been approximated at small values by
\be
   \op{H}_0^{(1)}(x) \underset{x \ll 1} \simeq 1
+ \frac{i}{\pi} \left[2\gamma + \ln\left(\frac{x^2}{4}\right)\right],
\ee
where $\gamma$ is the Euler's constant. 
In the TE polarization, using
$\int_0^{2\pi} \mathrm{d}\theta \, \left(\mathds{1}-\hat{\vec{r}}\otimes\hat{\vec{r}}\right) = \pi \mathds{1} $ and
$\int_0^{2\pi} \mathrm{d}\theta \, \left(\mathds{1}-2\hat{\vec{r}}\otimes\hat{\vec{r}}\right) = 0$ where $\hat{\vec{r}}=\vec{r}/r$, we get
\begin{align}
   \mathbf{G}_0^{*} (\vec{R}=\vec{0},\omega) & = -\frac{i\pi\mathds{1}}{4} \int_0^{\infty} \mathrm{d}r\,r
      \op{H}_0^{(1)}(k_0 r) \frac{e^{-\frac{u^2}{2b^2}}}{2\pi b^2} + \frac{\mathds{1}}{4\pi k_0^2b^2}
\notag \\
   \underset{k_0b\ll 1} & \simeq -\frac{\mathds{1}}{2\pi}\left[ \gamma
      + \ln\left(\frac{k_0^2b^2}{2} \right) \right] \notag
\\ & \hphantom{\underset{k_0b\ll 1}{\approx}}
      - \mathds{1}\left(\frac{i}{8}-\frac{1}{4\pi k_0^2b^2} \right).
\end{align}

In the long wavelength limit ($qa\ll 1$), Eq.~\eqref{eq::G0_q} can be obtained from
Eq.~\eqref{eq:Hq_reg}, in the same fashion as in Ref.~\onlinecite{antezza09_pra} for 3D crystals. A more direct way is to go back to
Eq.~\eqref{EqHamiltonQ:app} and to compute $\sum_{\vec{R}\neq 0} \mathbf{G}(\vec{R}) e^{-i \vec{q}\cdot\vec{R}}$ in
the high density limit which gives
\begin{align}
   \sum_{\vec{R}\neq 0} \mathbf{G}(\vec{R}) e^{i \vec{q}\cdot\vec{R}} \underset{qa\ll1}{\simeq}& \frac{1}{a^2}
   \int \mathrm{d} \vec{R} \, \mathbf{G}(\vec{R})e^{-i \vec{q}\cdot\vec{R}} - \op{Im}[\mathbf{G}(\vec{R}=\vec{0})] \notag\\
   &- \frac{1}{a^2} \int_{\delta\mathcal{A}} \mathrm{d}\vec{R}\, \op{Re}[\mathbf{G}(\vec{R})] \notag\\
   \simeq& \frac{1}{a^2} \mathbf{G}(\vec{q}) + \mathds{1} \left(\frac{i}{4} - \frac{\delta_{\beta,2}}{2k_0^2a^2} \right) \\
   &+ \mathds{1}\frac{\delta\mathcal{A}}{a^2} \left[ \frac{2 \gamma -1 + \ln(\delta\mathcal{A}/2)}{8\pi} \right], \notag
\end{align}
where $\delta\mathcal{A}$ is a small surface enclosing the origin.
Taking the limit $\delta\mathcal{A} \rightarrow0$, we obtain Eq.~\eqref{eq::G0_q}.

For small enough $b/a$, the Hamitonian computed with Eq.~\eqref{eq:Hq_reg} is independent of the regularization
parameter $b$.  In practice, we set $b/a=0.01$, and truncate the sum over the reciprocal lattice by keeping $\vec{Q}$
which satisfy $\vert \vec{q} - \vec{Q}\vert \lesssim 7/b$. In order to generate dispersion relations such as shown in
Fig.~\ref{fig:crystal_gap}, the Hamiltonian is then diagonalized for $\vec{q}$ belonging to an irreducible path of the
first Brillouin zone. Moreover, the numerical density of states $p(\delta)$ shown  in Fig.~\ref{fig:DOS_cry} is obtained
by sampling an irreducible area of the first Brillouin zone and counting the number of resonances per unit frequency. 

\section{Computation of the self energy and $p(\delta)$}\label{Appendix:Sigma}

In order to evaluate Eq.~\eqref{EqDOSasT}, it is convenient to use the decomposition
$\vec{\Sigma}(\vec{q},\omega)=\Sigma^\perp(\vec{q},\omega)\vec{\Delta}^\perp_{\vec{q}}+\Sigma^\parallel(\vec{q},\omega)\vec{\Delta}^\parallel_{\vec{q}}$
and
$\vec{G_0}(\vec{q},\omega)=G_0^\perp(\vec{q},\omega)\vec{\Delta}^\perp_{\vec{q}}+G_0^\parallel(\vec{q},\omega)\vec{\Delta}^\parallel_{\vec{q}}$,
where $\vec{\Delta}^\perp_{\vec{q}}$ and $\vec{\Delta}^\parallel_{\vec{q}}$ are the projectors perpendicular and
parallel to $\vec{q}$. We get
\begin{multline}\label{eq:p_delta}
   p(\delta)=-\frac{1}{4\pi  \beta \rho}
      \lim_{q\to \infty}\text{Im}
       \left[ \frac{\Sigma^\perp(\vec{q},\omega)}{1-G_0^\perp(\vec{q},\omega)\Sigma^\perp(\vec{q},\omega)} \right.
\\
       + \left. \frac{\Sigma^\parallel(\vec{q},\omega)}{1-G_0^\parallel(\vec{q},\omega)\Sigma^\parallel(\vec{q},\omega)} \right].
\end{multline}
A first simplification comes from the fact that $G_0^{\perp}(\vec{q},\omega)\to 0$ when $q\to\infty$, reducing the first
term to the numerator $\Sigma^{\perp}$. Then, we take into account all scattering processes involving two different scatterers in the
computation of the transverse [$\Sigma^\perp(\vec{q},\omega)$] and longitudinal [$\Sigma^\parallel(\vec{q},\omega)$]
self-energies. This corresponds to a second order expansion in density and gives the following diagrams:
\begin{widetext}
   \begin{multline} \label{eq:sigma_2_diag}
      \Sigma = \overbrace{\vphantom{\Bigg/} \diag{0}{\particule{0}}}^{\displaystyle \Sigma_{\text{ISA}}} +
      \overbrace{\diag{0}{\ginc{0}{6}\correldeux{0}{6}\particule{0}\particule{6}}\vphantom{\Bigg/}}^{\displaystyle
      \Sigma_{C}}  +
      \overbrace{\diag{-2}{\ginc{0}{6}\ginc{6}{12}\identiquea{0}{12}\particule{0}\particule{6}\particule{12}} +
      \diag{-2}{\ginc{0}{6}\ginc{6}{12}\ginc{12}{18}\ginc{18}{24}\identiquea{0}{12}\identiquea{12}{24}\identiqueb{6}{18}\particule{0}\particule{6}\particule{12}\particule{18}\particule{24}}
      + \ldots \vphantom{\Bigg/} +
      \diag{-2}{\ginc{0}{6}\ginc{6}{12}\correldeux{0}{6}\identiquea{0}{12}\particule{0}\particule{6}\particule{12}}  +
      \diag{-2}{\correldeux{0}{6}\ginc{0}{6}\ginc{6}{12}\ginc{12}{18}\ginc{18}{24}\identiquea{0}{12}\identiquea{12}{24}\identiqueb{6}{18}\particule{0}\particule{6}\particule{12}\particule{18}\particule{24}}
      + \ldots\vphantom{\Bigg/}}^{\displaystyle \Sigma_{L}}
\\
      +
      \underbrace{\diag{-2}{\ginc{0}{6}\ginc{6}{12}\ginc{12}{18}\identiquea{0}{12}\identiqueb{6}{18}\particule{0}\particule{6}\particule{12}\particule{18}}
      +
      \diag{-2}{\ginc{0}{6}\ginc{6}{12}\ginc{12}{18}\ginc{18}{24}\ginc{24}{30}\identiquea{0}{12}\identiquea{12}{24}\identiqueb{6}{18}\identiqueb{18}{30}\particule{0}\particule{6}\particule{12}\particule{18}\particule{24}\particule{30}}
      + \ldots\vphantom{\Bigg/} +
      \diag{-2}{\ginc{0}{6}\ginc{6}{12}\ginc{12}{18}\correldeux{0}{6}\identiquea{0}{12}\identiqueb{6}{18}\particule{0}\particule{6}\particule{12}\particule{18}}
      +
      \diag{-2}{\correldeux{0}{6}\ginc{0}{6}\ginc{6}{12}\ginc{12}{18}\ginc{18}{24}\ginc{24}{30}\identiquea{0}{12}\identiquea{12}{24}\identiqueb{6}{18}\identiqueb{18}{30}\particule{0}\particule{6}\particule{12}\particule{18}\particule{24}\particule{30}}
      + \ldots \vphantom{\Bigg/}}_{\displaystyle \Sigma_{B}}.
   \end{multline}
\end{widetext}
This diagrammatic representation is similar to the one shown in Fig.~\ref{fig:rec_scat}. Circles represent scattering
events and horizontal solid lines are for the free space Green functions connecting these scattering events. Also,
disorder correlations are denoted by dashed curved lines and identical scatterers (recurrent scattering) are linked by
solid upper and lower lines. Eq.~\eqref{eq:sigma_2_diag} takes the compact form
\begin{multline}\label{eq:sigma_2}
   \vec{\Sigma}(\mathbf{q},\omega) =\vec{\Sigma}_{\text{ISA}}(\mathbf{q},\omega) + \vec{\Sigma}_{C}(\mathbf{q},\omega)
\\
   + \vec{\Sigma}_{L}(\mathbf{q},\omega)+\vec{\Sigma}_{B}(\mathbf{q},\omega),
\end{multline}
where
\begin{align}
   \vec{\Sigma}_{\text{ISA}}(\mathbf{q},\omega) & = \rho \, t(\omega) \mathds{1},
\\
   \vec{\Sigma}_{C}(\mathbf{q},\omega) & = \rho^2t(\omega)^2
      \int \mathrm{d}\mathbf{r} \, h_2(\mathbf{r}) \mathbf{G}_0(\mathbf{r},\omega_0) e^{-i \mathbf{q}\cdot
      \mathbf{r}},
\\
   \vec{\Sigma}_{L}(\mathbf{q},\omega) & = \rho^2t(\omega)^3
      \int \mathrm{d}\mathbf{r} \, \left[1+h_2(\mathbf{r})\right] \frac{
      \mathbf{G}_0^2(\mathbf{r},\omega_0)}{\mathds{1}-t(\omega)^2\mathbf{G}_0^2(\mathbf{r},\omega_0)},
      \label{eq:defSigmaL}
\\
   \vec{\Sigma}_{B}(\mathbf{q},\omega) & = \rho^2t(\omega)^4
      \int \mathrm{d}\mathbf{r} \, \left[1+h_2(\mathbf{r})\right]
      \frac{\mathbf{G}_0^3(\mathbf{r},\omega_0)e^{-i \mathbf{q}\cdot \mathbf{r}}}{\mathds{1}-t(\omega)^2\mathbf{G}_0^2(\mathbf{r},\omega_0)}.
\end{align}

Let us first consider $\vec{\Sigma}_C$. To compute this term, we expand the Green function given by Eq.~\eqref{eq:G0}
into its transverse and longitudinal projections in real space and its singular part as
\begin{equation}
\label{eq:decompoG_real}
   \mathbf{G}_0(\vec{r},\omega)=G^{t}(r,\omega) \mathbf{\Delta}_r^{\perp} + G^{l}(r,\omega)
      \mathbf{\Delta}_r^{\parallel}+\frac{\delta(\vec{R})}{2k_0^2}\mathds{1}
\end{equation}
where $\mathbf{\Delta}_r^{\perp}$ ($\mathbf{\Delta}_r^{\parallel}$) is the transverse (longitudinal) projector. Using the relations
\begin{align}\notag
   \int_0^{2\pi} \mathrm{d}\theta \, \mathbf{\Delta}_r^{\perp}e^{-i \vec{q}\cdot\vec{r}}
      &= 2\pi \left[\op{J}_0(qr) - \frac{\op{J}_1(qr)}{qr}\right] \mathbf{\Delta}_q^{\perp}
\\
      &+ 2\pi \frac{\op{J}_1(qr)}{qr} \mathbf{\Delta}_q^{\parallel}, \label{eq:int_proj}
\\\notag
   \int_0^{2\pi} \mathrm{d}\theta \, \mathbf{\Delta}_r^{\parallel}e^{-i \vec{q}\cdot\vec{r}}
      &= 2\pi \frac{\op{J}_1(qr)}{qr} \mathbf{\Delta}_q^{\perp}
\\
      &+ 2\pi \left[\op{J}_0(qr) - \frac{\op{J}_1(qr)}{qr}\right] \mathbf{\Delta}_q^{\parallel}, 
      \label{eq:int_proj2}
\end{align}
 where $\op{J}_0(x)$ and  $\op{J}_1(x)$ are  Bessel functions of the first kind, we find that $\vec{\Sigma}_C$ takes the form
\begin{equation}
   \vec{\Sigma}_{C}(\mathbf{q},\omega) =
      \Sigma_{C}^{\perp}(\mathbf{q},\omega)\mathbf{\Delta}_q^{\perp}
      +
      \Sigma_{C}^{\parallel}(\mathbf{q},\omega)\mathbf{\Delta}_q^{\parallel}
      +
      \Sigma_{C}^{\text{LL}}(\mathbf{q},\omega)\mathds{1},
\end{equation}
with
\begin{align}\notag
   \Sigma_{C}^{\perp}(\mathbf{q},\omega) & = 2\pi \rho^2 t(\omega)^2 \int_0^{\infty} \mathrm{d}r \, r \, h_2(r)
\\\times\label{eq::sigma_B2}
   &\left\{ G_0^{t}(r,\omega_0) \left[\op{J_0}(q r) - \frac{\op{J_1}(q r)}{q r}\right] + G_0^{l}(r,\omega_0) \frac{\op{J_1}(q r)}{q r}
   \right\},
\\\notag
   \Sigma_{C}^{\parallel}(\mathbf{q},\omega) & = 2\pi \rho^2 t(\omega)^2 \int_0^{\infty} \mathrm{d}r \, r \, h_2(r)
\\\times
   &\left\{ G_0^{t}(r,\omega_0) \frac{\op{J_1}(q u)}{q r} + G_0^{l}(r,\omega_0) \left[\op{J_0}(q r) - \frac{\op{J_1}(q u)}{q r}\right] \right\},
   \\
   \Sigma_{C}^{\text{LL}}(\vec{q},\omega) &= \frac{\rho^2t(\omega)^2h_2(0)}{2k_0^2}.
\end{align}
Here the last term $ \Sigma_{C}^{\text{LL}}(\vec{q})$ is due to the singular part of the Green function.

In the same way we can write the loop term $\vec{\Sigma}_L$ 
in the form
\begin{equation}
   \vec{\Sigma}_{L}(\mathbf{q},\omega) =
      \Sigma_{L}^{\perp}(\mathbf{q},\omega)\mathbf{\Delta}_q^{\perp}
      +
      \Sigma_{L}^{\parallel}(\mathbf{q},\omega)\mathbf{\Delta}_q^{\parallel}
      +
      \Sigma_{L}^{\text{LL}}(\mathbf{q},\omega)\mathds{1}.
\end{equation}
The first two terms follow from the decomposition
\begin{multline}
   \frac{\mathds{1}}{\mathds{1}-t(\omega)^2\mathbf{G}_0^2(\vec{r},\omega_0)} 
      = \sum_{j=0}^{\infty} \left[t(\omega)^2 \mathbf{G}_0(\vec{r},\omega_0)^2\right]^j
\\
   = \sum_{j=0}^{\infty} \left[t(\omega)^2 G^t(r,\omega_0)^2\right]^j \mathbf{\Delta}_r^{\perp} 
      + \left[t(\omega)^2 G^l(r,\omega_0)^2\right]^j \mathbf{\Delta}_r^{\parallel}
\\\label{eq:Gtl}
   = \frac{1}{1-t(\omega)^2G_0^{t}(r,\omega_0)^2} \mathbf{\Delta}_r^{\perp} 
      + \frac{1}{1-t(\omega)^2G_0^{l}(r,\omega_0)^2} \mathbf{\Delta}_r^{\parallel}.
\end{multline}
Taking the limit $\vec{q}\to\vec{0}$ in Eqs.~\eqref{eq:int_proj} and \eqref{eq:int_proj2}, we get
$\int_0^{2\pi} \mathrm{d}\theta \, \mathbf{\Delta}_r^{\perp}e^{-i\vec{k}\cdot\vec{r}} = \int_0^{2\pi} \mathrm{d}\theta \, \mathbf{\Delta}_r^{\parallel}e^{-i\vec{k}\cdot\vec{r}} = \pi\mathds{1}$,  which leads to
 \begin{multline}\label{eq::sigma_L}
   \Sigma_{L}^{\perp}(\vec{q},\omega) = \Sigma_{L}^{\parallel}(\vec{q},\omega) = \pi \rho^2t(\omega)^3 \int_0^{\infty} \mathrm{d}r\, r \left[1+h_2(r) \right]
\\\times
   \left[ \frac{G_0^{t}(r,\omega_0)^2}{1-t(\omega)^2 G_0^{t}(r,\omega_0)^2} 
      + \frac{G_0^{l}(r,\omega_0)^2}{1-t(\omega)^2 G_0^{l}(r,\omega_0)^2} \right].
\end{multline}
In addition, the contribution of the singular part of $\mathbf{G}_0(\vec{r},\omega)$ to
$\vec{\Sigma}_{L}(\mathbf{q},\omega)$  vanishes thanks to a simplification of the $\delta$-terms appearing in
Eq.~\eqref{eq:defSigmaL}, $\Sigma_{L}^{\text{LL}}(\mathbf{q},\omega)=0$.

Finally, we consider the boomerang term $\vec{\Sigma}_B$ that we write in the form
\begin{equation}
   \vec{\Sigma}_{B}(\mathbf{q},\omega) =
      \Sigma_{B}^{\perp}(\mathbf{q},\omega)\mathbf{\Delta}_q^{\perp}
      +
      \Sigma_{B}^{\parallel}(\mathbf{q},\omega)\mathbf{\Delta}_q^{\parallel}
      +
      \Sigma_{B}^{\text{LL}}(\mathbf{q},\omega)\mathds{1}.
\end{equation}
  Using Eqs.~\eqref{eq:Gtl} and \eqref{eq:int_proj}, we obtain
\begin{widetext}
   \begin{align}\label{eq::sigma_B3}
      \Sigma_{B}^{\perp}(\mathbf{q},\omega)& = 2\pi \rho^2 \int_0^{\infty} \mathrm{d}r \, r \left[1+h_2(r) \right]
            \left\{ \frac{t(\omega)^4  G_0^{t3}(r,\omega_0)}{1-t(\omega)^2 G_0^{t2}(r,\omega_0)} \left[\op{J_0}(q r) - \frac{\op{J_1}(q r)}{q r}\right]
            + \frac{t(\omega)^4 G_0^{l3}(r,\omega_0)}{1-t(\omega)^2 G_0^{l2}(r,\omega_0)} \frac{\op{J_1}(q r)}{q r} \right\},
\\
      \Sigma_{B}^{\parallel}(\mathbf{q},\omega) &= 2\pi \rho^2 \int_0^{\infty} \mathrm{d}r \, r \left[1+h_2(r) \right]
            \left\{ \frac{t(\omega)^4  G_0^{t3}(r,\omega_0)}{1-t(\omega)^2 G_0^{t2}(r,\omega_0)} \frac{\op{J_1}(q u)}{q r}
            + \frac{t(\omega)^4 G_0^{l3}(r,\omega_0)}{1-t(\omega)^2 G_0^{l2}(r,\omega_0)} \left[\op{J_0}(q r) - \frac{\op{J_1}(q u)}{q r}\right] \right\},
            \\
      \Sigma_{B}^{\text{LL}}(\vec{q},\omega) &= -\frac{\rho^2t(\omega)^2(1+h_2(0))}{2k_0^2}.
   \end{align}
\end{widetext}

It is interesting to note that the two terms coming from the singular part of the Green function simplify which gives
\begin{equation}
   \Sigma^{\text{LL}}(\vec{q},\omega) = \Sigma_{C}^{\text{LL}}(\vec{q},\omega)+\Sigma_{B}^{\text{LL}}(\vec{q},\omega)
      =-\frac{\rho^2t^2}{2k_0^2}.
\end{equation}
At first order, this contribution is identical to the one we would obtain considering point-like repulsion between
scatterers and known as Lorentz-Lorenz correction.

In the evaluation of the density of states, only the large wave-vector limit participates hence removing the
contribution of $\Sigma_B$ and the wave-vector dependent part of $\Sigma_C$. The only term which needs to be computed
numerically is $\Sigma_L$. The first integrand involving $G_0^t$ is regular at the origin, but not absolutely convergent
for large arguments. Indeed, $r G_0^{t2}(r,\omega_0)$ is equivalent to $i\exp[2ik_0r]/8\pi$. The integral can be
computed nonetheless by setting a cut-off $r_c$ in the following way~\cite{kwong_coherent_2019}:
\begin{multline}
   \int_{0}^{\infty} \mathrm{d}r\, r \left[1+h_2(r) \right]\frac{G_0^{t}(r,\omega_0)^2}{1-t(\omega)^2 G_0^{t}(r,\omega_0)^2}
\\
      =\int_{0}^{r_c} \mathrm{d}r\, r \left[1+h_2(r) \right]\frac{G_0^{t}(r,\omega_0)^2}{1-t(\omega)^2 G_0^{t}(r,\omega_0)^2}
\\
      +\int_{r_c}^{\infty} \mathrm{d}r\, r \left[1+h_2(r) \right]\frac{G_0^{t}(r,\omega_0)^2}{1-t(\omega)^2
      G_0^{t}(r,\omega_0)^2}.
      \label{eq:IntCut}
\end{multline}
For a large enough value of $r_c$, the first integral is computed numerically and the second one is approximated by
\begin{equation}
   \int_{r_c}^{\infty} \mathrm{d}r\, r G_0^{t}(r,\omega_0)^2
   \simeq \frac{i}{8\pi}\int_{r_c}^{\infty} \exp[2 i k_0r] \mathrm{d}r  = \frac{\exp[2ik_0r_c]}{16\pi k_0}.
\end{equation}
The final result~\eqref{eq:IntCut} is independent of  $r_c$.

\section{Computation of the anisotropy factor $g$ for TE waves}\label{Appendix:Anisotropy}

The transport mean free path $\ell_t$ can be linked to the scattering mean free path $\ell_s$ through the anisotropy
factor $g$. In 2D, it is given by
\begin{equation}
   \frac{1}{\ell_t} = \frac{1-g}{\ell_s} = \frac{ \bra \left(1-\hat{\vec{q}}\cdot \hat{\vec{q}}'\right) U^{\perp} (\vec{q},\vec{q}', \omega) \ket_{\hat{\vec{q}}'}}{4 k_0}, \label{eq::U_perp}
\end{equation}
where $\hat{\vec{q}}=\vec{q}/q$ with $q=k_0$,  $\bra \ldots \ket_{\hat{\vec{q}'}}$ denotes an average over the direction of
$\hat{\vec{q}}'$, and $U^{\perp}(\vec{q},\vec{q}', \omega) =\sum_{i,j,k,l}
\Delta_{\vec{q},ij}^{\perp}U_{ij,kl}(\vec{q},\vec{q}', \omega) \Delta_{\vec{q}',kl}^{\perp}$ is the transverse part of
the irreducible vertex $\vec{U}(\vec{q},\vec{q}', \omega)$ of the Bethe-Salpeter equation~\cite{sheng06}. Here, we are
specifically interested in the anisotropy factor evaluated on the shell $\vert \vec{q} \vert = k_0$, which reads
\begin{equation}
   g = \ell_s \frac{\bra \hat{\vec{q}}\cdot \hat{\vec{q}}' \; U^{\perp}  (k_0 \hat{\vec{q}}, k_0 \hat{\vec{q}'}, \omega) \ket_{\hat{\vec{q}}'}}{4k_0},
    \label{eq::g}
\end{equation}
where $\ell_s$ is given by Eq.~\eqref{EqMpf}. At the second order in density, the angular average in the numerator of
Eq.~\eqref{eq::g} has non-zero contributions from the following diagrams~\cite{vantiggelen94, cherroret16}:
\begin{widetext}
   \begin{align}\label{eq:U_1}
      \vec{U}_C \;=&\; \diag{-3}{
         \ccorreldeuxd{0}{3}{0}{-3}
         \pparticule{0}{-3}
         \pparticule{0}{3}
      }\\
      \vec{U}_L \;=&\; \diag{-3}{
         \iidentiquea{0}{12}
         \iidentique{6}{-3}{6}{3}
         \gginc{0}{6}{3}
         \gginc{6}{12}{3}
         \pparticule{6}{-3}
         \pparticule{0}{3}
         \pparticule{6}{3}
         \pparticule{12}{3}
      }+\diag{-3}{
         \iidentiquea{0}{12}
         \iidentique{6}{-3}{6}{3}
         \ccorreldeuxa{0}{6}
         \gginc{0}{6}{3}
         \gginc{6}{12}{3}
         \pparticule{6}{-3}
         \pparticule{0}{3}
         \pparticule{6}{3}
         \pparticule{12}{3}
      }+\diag{-3}{
         \iidentiquea{0}{12}
         \iidentique{6}{-3}{6}{3}
         \iidentiquea{12}{24}
         \iidentiqueaa{6}{18}
         \gginc{0}{6}{3}
         \gginc{6}{12}{3}
         \gginc{12}{18}{3}
         \gginc{18}{24}{3}
         \pparticule{6}{-3}
         \pparticule{0}{3}
         \pparticule{6}{3}
         \pparticule{12}{3}
         \pparticule{18}{3}
         \pparticule{24}{3}
      }+\diag{-3}{
         \iidentiquea{0}{12}
         \iidentique{6}{-3}{6}{3}
         \iidentiquea{12}{24}
         \iidentiqueaa{6}{18}
         \ccorreldeuxa{0}{6}
         \gginc{0}{24}{3}
         \pparticule{6}{-3}
         \pparticule{0}{3}
         \pparticule{6}{3}
         \pparticule{12}{3}
         \pparticule{18}{3}
         \pparticule{24}{3}
      } + \ldots + c.c. \notag \\[5pt]
      &+\diag{-5}{
         \iidentiquea{0}{12}
         \iidentiqueb{0}{12}
         \iidentique{0}{-3}{6}{3}
         \iidentique{6}{-3}{0}{3}
         \gginc{0}{12}{3}
         \gginc{0}{12}{-3}
         \pparticule{0}{3}
         \pparticule{6}{3}
         \pparticule{12}{3}
         \pparticule{0}{-3}
         \pparticule{6}{-3}
         \pparticule{12}{-3}
      }+\diag{-5}{
         \iidentiquea{0}{12}
         \iidentiqueb{0}{12}
         \iidentique{0}{-3}{6}{3}
         \iidentique{6}{-3}{0}{3}
         \ccorreldeuxa{0}{6}
         \gginc{0}{12}{3}
         \gginc{0}{12}{-3}
         \pparticule{0}{3}
         \pparticule{6}{3}
         \pparticule{12}{3}
         \pparticule{0}{-3}
         \pparticule{6}{-3}
         \pparticule{12}{-3}
      }+\diag{-5}{
         \iidentiquea{0}{12}
         \iidentiquea{12}{24}
         \iidentiqueaa{6}{18}
         \iidentiqueb{0}{12}
         \iidentique{0}{-3}{6}{3}
         \iidentique{6}{-3}{0}{3}
         \gginc{0}{24}{3}
         \gginc{0}{12}{-3}
         \pparticule{0}{3}
         \pparticule{6}{3}
         \pparticule{12}{3}
         \pparticule{18}{3}
         \pparticule{24}{3}
         \pparticule{0}{-3}
         \pparticule{6}{-3}
         \pparticule{12}{-3}
      }+\diag{-5}{
         \iidentiquea{0}{12}
         \iidentiquea{12}{24}
         \iidentiqueaa{6}{18}
         \iidentiqueb{0}{12}
         \iidentique{0}{-3}{6}{3}
         \iidentique{6}{-3}{0}{3}
         \ccorreldeuxa{0}{6}
         \gginc{0}{24}{3}
         \gginc{0}{12}{-3}
         \pparticule{0}{3}
         \pparticule{6}{3}
         \pparticule{12}{3}
         \pparticule{18}{3}
         \pparticule{24}{3}
         \pparticule{0}{-3}
         \pparticule{6}{-3}
         \pparticule{12}{-3}
      }+\ldots+ c.c.\\[15pt] 
       \vec{U}_B \;=&\; \diag{-3.5}{
         \iidentique{0}{-3}{6}{3}
         \iidentique{6}{-3}{0}{3}
         \gginc{0}{6}{3}
         \gginc{0}{6}{-3}
         \pparticule{6}{-3}
         \pparticule{0}{3}
         \pparticule{6}{3}
         \pparticule{0}{-3}
      }+\diag{-3.5}{
         \iidentique{0}{-3}{6}{3}
         \iidentique{6}{-3}{0}{3}
         \ccorreldeuxa{0}{6}
         \gginc{0}{6}{3}
         \gginc{0}{6}{-3}
         \pparticule{6}{-3}
         \pparticule{0}{3}
         \pparticule{6}{3}
         \pparticule{0}{-3}
      }+\diag{-3.5}{
         \iidentiquea{0}{12}
         \iidentiqueaa{6}{18}
         \iidentique{0}{-3}{6}{3}
         \iidentique{6}{-3}{0}{3}
         \gginc{0}{18}{3}
         \gginc{0}{6}{-3}
         \pparticule{6}{-3}
         \pparticule{0}{3}
         \pparticule{6}{3}
         \pparticule{12}{3}
         \pparticule{0}{-3}
         \pparticule{18}{3}
      }+\diag{-3.5}{
         \iidentiquea{0}{12}
         \iidentiqueaa{6}{18}
         \iidentique{0}{-3}{6}{3}
         \iidentique{6}{-3}{0}{3}
         \ccorreldeuxa{0}{6}
         \gginc{0}{18}{3}
         \gginc{0}{6}{-3}
         \pparticule{6}{-3}
         \pparticule{0}{3}
         \pparticule{6}{3}
         \pparticule{12}{3}
         \pparticule{0}{-3}
         \pparticule{18}{3}
      }+\diag{-3.5}{
         \iidentiquea{0}{12}
         \iidentiquea{12}{24}
         \iidentiqueaa{6}{18}
         \iidentiqueaa{18}{30}
         \iidentique{0}{-3}{6}{3}
         \iidentique{6}{-3}{0}{3}
         \gginc{0}{30}{3}
         \gginc{0}{6}{-3}
         \pparticule{6}{-3}
         \pparticule{0}{3}
         \pparticule{6}{3}
         \pparticule{12}{3}
         \pparticule{0}{-3}
         \pparticule{18}{3}
         \pparticule{24}{3}
         \pparticule{30}{3}
      }+\ldots+c.c.
   \end{align}
   where the notations are identical to those used in  Eq.~\eqref{eq:sigma_2_diag}, with the upper and lower lines
   accounting for the propagating field and its conjugate, respectively.
   Explicitly, the first term reads
   \begin{equation}
      \vec{U}_C(\vec{q},\vec{q}', \omega) = \rho^2|t(\omega)|^2 h_2(\vec{q}-\vec{q}') \mathds{1},
   \end{equation}
   so that the angular average of its transverse part can be written in the form 
   \be
   \bra \hat{\vec{q}}\cdot \hat{\vec{q}}' U_C^{\perp}(\vec{q},\vec{q}', \omega) \ket_{\hat{\vec{q}}'} =\rho|t(\omega)|^2
      \int_0^{2\pi} \frac{\mathrm{d}\theta}{2\pi} h_2\left(2q\left|\sin(\theta/2)\right| \right) \cos^3 \theta.
   \ee
   The loop and boomerang contributions, $\vec{U}_L$ and $\vec{U}_B$, are
   \begin{align}
      \vec{U}_L(\vec{q},\vec{q}', \omega) &= \rho^2|t(\omega)|^2 \int \mathrm{d}\vec{r} \left[ 1+h_2(r)\right]  e^{i(\vec{q}-\vec{q}')\cdot \vec{r}}
      \left\lbrace \frac{\mathds{1}}{\left[\mathds{1} - t(\omega)^2\vec{G_0}(\vec{r},\omega_0)^2 \right]\otimes\left[\mathds{1} - t(\omega)^2\vec{G_0}(\vec{r},\omega_0)^2  \right]^{*}} - \mathds{1} \right\rbrace,
      \\
      \vec{U}_B(\vec{q},\vec{q}', \omega) & = \rho^2|t(\omega)|^4  \int \mathrm{d}\vec{r} \left[ 1+h_2(r)\right] e^{i(\vec{q}+\vec{q}')\cdot \vec{r}}
      \frac{\vec{G_0}(\vec{r},\omega_0) \otimes \vec{G_0}^{*}(\vec{r},\omega_0)}{\left[\mathds{1} - t(\omega)^2\vec{G_0}(\vec{r},\omega_0)^2 \right]\otimes\left[\mathds{1} - t(\omega)^2\vec{G_0}(\vec{r},\omega_0)^2  \right]^{*}}.
   \end{align}
   
   The angular integrations over the directions of $\vec{r}$ and $\hat{\vec{q}}$ that appear in $\bra \hat{\vec{q}}\cdot
   \hat{\vec{q}}' \; U_L^{\perp}  (\vec{q},\vec{q}', \omega) \ket_{\hat{\vec{q}}'}$ and $\bra \hat{\vec{q}}\cdot
   \hat{\vec{q}}' \; U_B^{\perp}  (\vec{q},\vec{q}', \omega) \ket_{\hat{\vec{q}}'}$ can be performed using the
   decomposition~\eqref{eq:decompoG_real} and the relation~\eqref{eq:Gtl}. After a lengthy calculation, we find 
   \begin{align}
        \bra \hat{\vec{q}}\cdot \hat{\vec{q}}' U_L^{\perp}(\vec{q},\vec{q}', \omega) \ket_{\hat{\vec{q}}'} = -\rho^2|t(\omega)|^2\int_0^{\infty} \mathrm{d}r\; r \left[ 1+h_2(r) \right] \left\lbrace A(qr) \left| \frac{1}{1-t(\omega)^2 G_0^t(r,\omega_0)^2} \right|^2 + B(qr) \left| \frac{1}{1-t(\omega)^2 G_0^l(r,\omega_0)^2} \right|^2  \right. \notag\\
      + \left. 2C(qr) \op{Re}\left[ \left( \frac{1}{1-t(\omega)^2G_0^t(r,\omega)^2} \right)\left( \frac{1}{1-t(\omega)^2G_0^l(r,\omega)^2} \right)^{*} \right] -D(qr) \right\rbrace,
      \\
\bra \hat{\vec{q}}\cdot \hat{\vec{q}}' U_B^{\perp}(\vec{q},\vec{q}', \omega) \ket_{\hat{\vec{q}}'} = \rho^2|t(\omega)|^4 \int_0^{\infty} \mathrm{d}r\; r \left[ 1+h_2(r) \right]  \left\lbrace A(qr) \left| \frac{G_0^t(r,\omega_0)}{1-t(\omega)^2 G_0^t(r,\omega_0)^2} \right|^2 + B(qr) \left| \frac{G_0^l(r,\omega_0)}{1-t(\omega)^2 G_0^l(r,\omega_0)^2} \right|^2  \right. \notag\\
      + \left. 2C(qr) \op{Re}\left[ \left( \frac{G_0^t(r,\omega_0)}{1-t(\omega)^2G_0^t(r,\omega)^2} \right)\left( \frac{G_0^l(r,\omega_0)}{1-t(\omega)^2G_0^l(r,\omega)^2} \right)^{*} \right] \right\rbrace,
   \end{align}
\end{widetext}

\begin{figure}[b]
   \centering
   \includegraphics[width=0.8\linewidth]{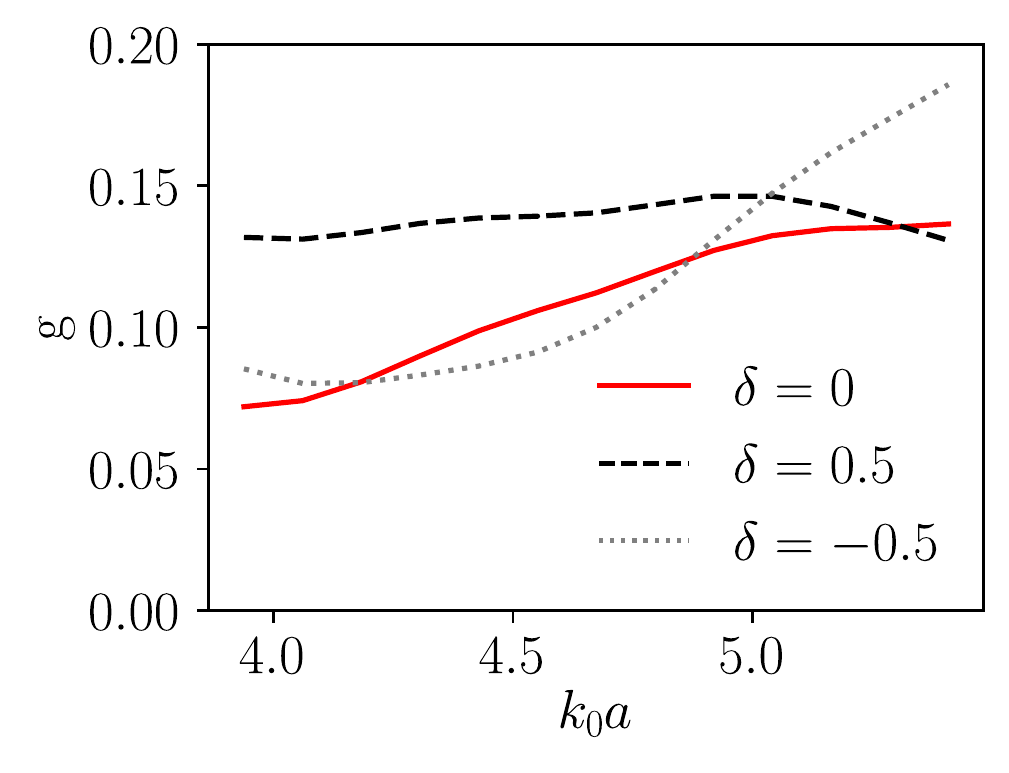}
   \caption{Anistropy factor $g$ for TE waves propagating in a highly correlated ensemble of resonators ($\chi=0.5$),
   with detuning $\delta=0$ (solid red line), $\delta=0.5$ (dashed black line), and $\delta=-0.5$ (dotted grey line).
   The anisotropy factor fluctuates between $0.05$ and $0.2$ in the density window where localization is prominent
   (see Fig.~\ref{fig:kxi_num}).}
   \label{fig:g_values}
\end{figure}
   
where $A(x) = -2\pi \left[J_2(x)-xJ_1(x)\right]^2/x^2$, $B(x)=C(x) =-2\pi J_2(x)^2/x^2$, $D(x) = A(x) + 3 B(x)$, and
$J_1(x)$ and  $J_2(x)$  are Bessel functions of the first kind. We note that the remaining integration over $r$ in not
convergent for $r \to \infty$ because of the term $rA(qr) \vert G_0^t(r,\omega_0)\vert^2 \sim \cos (kr)^2/r$. We
regularize this logarithmic divergence by replacing the free space Green's function $G_0^t(r,\omega_0)$ with the
far-field expansion of the average Green's function, $\bra G^t(r,\omega_0)  \ket \simeq G_0^t(r,\omega_0)e^{-r/2\ell_s}$
.
    
We show in Fig.~\ref{fig:g_values} the anisotropy factor $g$  evaluated from Eq.~\eqref{EqMpf} and Eq.~\eqref{eq::g}
with $U^\perp= U_C^\perp+ U_B^\perp+ U_L^\perp$, for the range of density probed in Fig.~\eqref{fig:kxi_num} and
different values of detuning $\delta$. In particular, at resonance ($\delta =0$), we find $g \sim 0.07-0.13$, which gives
$\ell_t \sim 1.07-1.15\, \ell_s$.

\section{Numerical evaluation of $\ell_s$ and $\xi$}\label{appendix:ls_xi_computation}

\begin{figure}[t]
   \centering
   \includegraphics[width=0.9\linewidth]{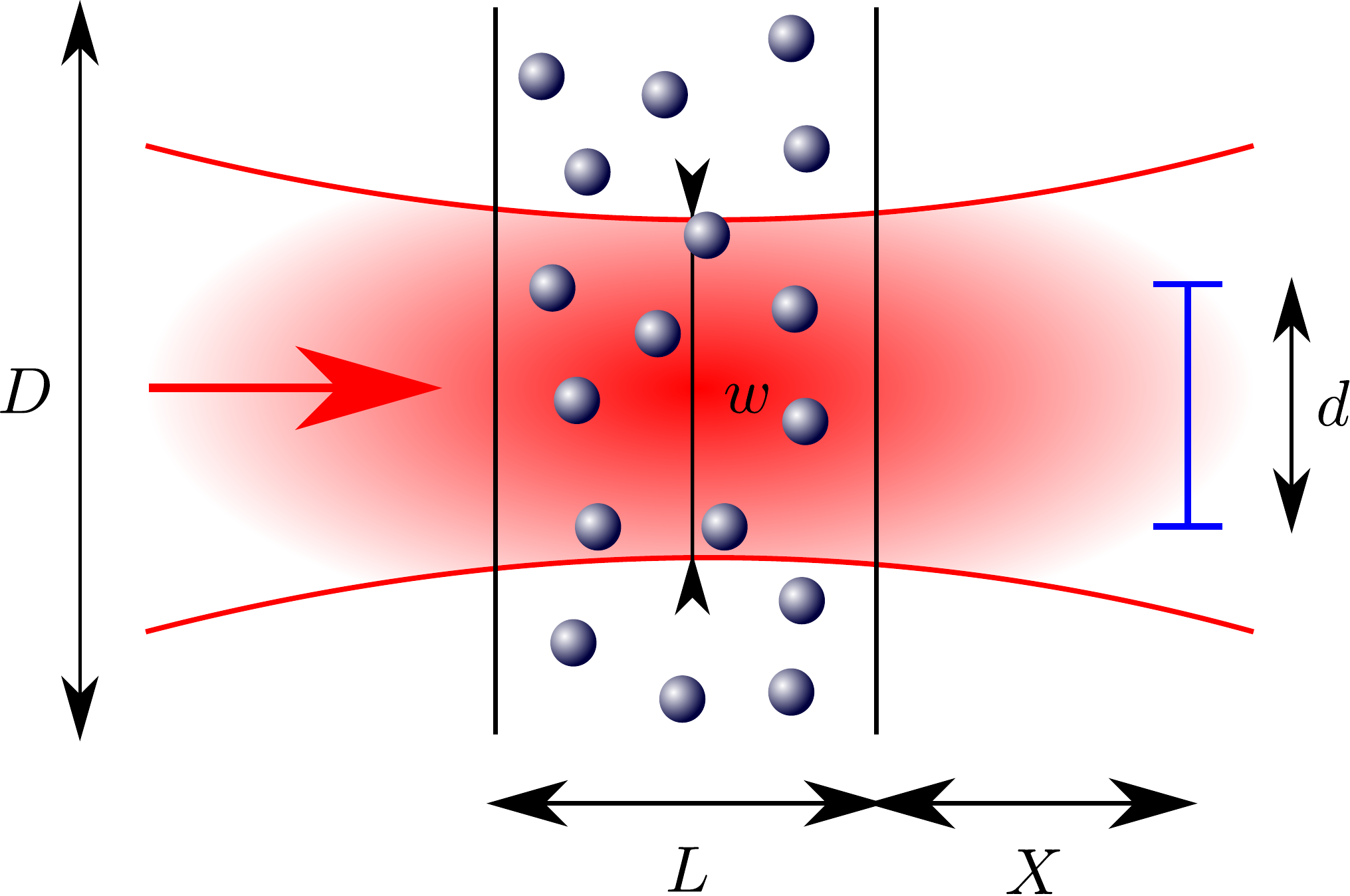}
   \caption{Numerical setup used to estimate $\ell_s$ and $\xi$. Red: gaussian beam illuminating the medium. Blue:
   screen on which the transmitted field is computed.}
   \label{fig:slab_geometry}
\end{figure}

The numerical estimate of the scattering mean-free path $\ell_s$ and of the localization length $\xi$ is performed
through \emph{ab-initio} computations using the coupled dipoles method. We place a SHU pattern into a rectangle of
length $L$ (typically $kL\in[10,60]$) and transverse size $D$. In order to mimic a slab geometry, which is the most
convenient geometry to have access to astimates of $\ell_s$ and $\xi$, we choose $D\gg L$ (typically $D=20L$, see
Fig.~\ref{fig:slab_geometry}).  This system is shined using a gaussian beam of waist $w\gg\lambda$ (typically $kw=200$)
given by
\be
   \vec{E}\left(\vec{r},\omega\right)=\frac{\vec{E}_0}{\sqrt{1+i\alpha}}
      \exp\left[ikx-\frac{y^2}{w^2\left(1+i\alpha\right)}\right]
\ee
for TE waves. $\vec{E}_0$ is the amplitude and $\alpha=2x/\left(kw^2\right)$.  This specific illumination is chosen such
that it smoothes the transverse finite-size effects. The transmitted electric field is computed for each disordered
configurations on a screen of size $d$ lying at a distance $X$ from the output interface of the slab. We take
$X>\lambda$ (typically $kX=10$) in order to avoid potential near-field effects close to the interface.

\begin{figure}[t]
   \centering
   \includegraphics[width=.45\textwidth]{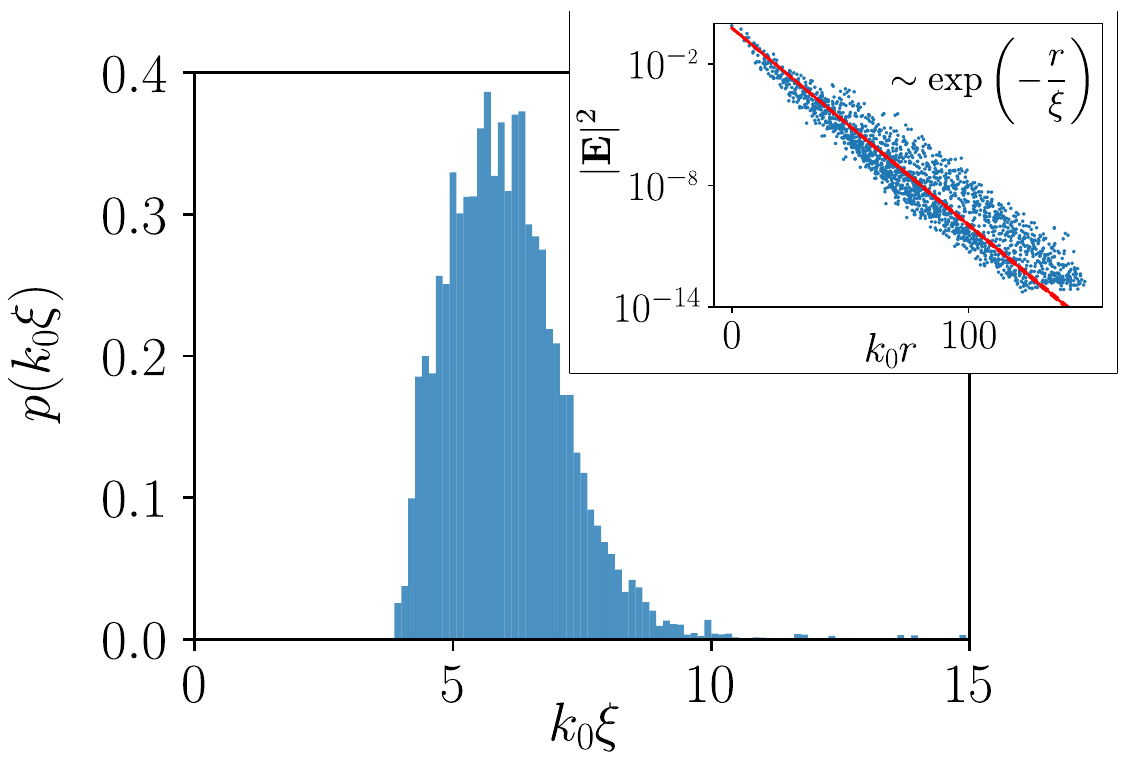}
   \caption{Histogram of the localization length $\xi$ obtained from a fit of the eigenstates of $\mathcal{H}(\omega_0)$
   in the TE polarization,  at $k_0a=4.5$ and $\chi=0.5$. Here the eigenstates are associated to detuning $\delta \in
   [-0.05, 0.05]$.  The inset shows an example of the intensity profile of an eigenstate with the best fitting estimate.
   A weighted fit with a $(1+r)^{-1}$ penalty has been used to account for the variable number of data points with the
   distance $r$.}
   \label{fig:xi_fit}
\end{figure}

In order to estimate $\ell_s$, we average the transmitted field over many SHU configurations. To accelerate the
numerical convergence, we also perform a spatial average over the observation screen assuming ergodicity and a size
$d>\lambda$ (typically $kd=10$). Then we compute the intensity of this average field and perform a fit with the
formula
\be
   \left|\bra\vec{E}\ket\right|^2=A\exp\left[-\frac{L}{\ell_s}\right]
\ee
as a function of the thickness of the slab $L$. $A$ and $\ell_s$ are the fitted parameters.

The estimate of the localization length $\xi$ is performed from the fit of the average of the intensity logarithm as a
function of $L$. The fitting formula is given by
\be
   \bra\ln |\vec{E}|^2\ket=B-L/\xi
   \label{EqXifit}
\ee
where $B$ and $\xi$ are the fitted parameters.

We also compared the value of $\xi$ found from Eq.~\eqref{EqXifit} with the length characterizing the exponential profile
of the eigenstates of $\mathcal{H}(\omega_0)$. For eigenstates associated to low collective linewidth ($\gamma\leq
10^{-4}$), the uncertainty on the fitting parameter  $\xi$  of the exponential profile is of the order of the percent,
ensuring that the eigenstates are indeed exponentially localized over a length smaller than the system size. For a given
value of $k_0a$, eigenstates corresponding to a certain detuning $\delta$ do not have necessarily the same localization
length $\xi$.  However, in the regime of density and detuning where we expect strong localization, the distribution
$p(\xi)$ becomes noticeably narrow. An example of $p(\xi)$ obtained in this way is shown in Fig.~\ref{fig:xi_fit}.   The
values of $\xi$ represented in Fig.~\ref{fig:kxi_num} correspond to the most probable value of $p(\xi)$, computed for
different $k_0a$ at $\delta=0$, using $32$ SHU configurations of $N=2000$ resonators distributed in a disk. For all
$k_0a$, the disk radius is  significantly  larger than $\xi$ ($k_0R\in [100-150]$).

\bibliography{biblio.bib}

\begin{thebibliography}{66}%
\makeatletter
\providecommand \@ifxundefined [1]{%
 \@ifx{#1\undefined}
}%
\providecommand \@ifnum [1]{%
 \ifnum #1\expandafter \@firstoftwo
 \else \expandafter \@secondoftwo
 \fi
}%
\providecommand \@ifx [1]{%
 \ifx #1\expandafter \@firstoftwo
 \else \expandafter \@secondoftwo
 \fi
}%
\providecommand \natexlab [1]{#1}%
\providecommand \enquote  [1]{``#1''}%
\providecommand \bibnamefont  [1]{#1}%
\providecommand \bibfnamefont [1]{#1}%
\providecommand \citenamefont [1]{#1}%
\providecommand \href@noop [0]{\@secondoftwo}%
\providecommand \href [0]{\begingroup \@sanitize@url \@href}%
\providecommand \@href[1]{\@@startlink{#1}\@@href}%
\providecommand \@@href[1]{\endgroup#1\@@endlink}%
\providecommand \@sanitize@url [0]{\catcode `\\12\catcode `\$12\catcode
  `\&12\catcode `\#12\catcode `\^12\catcode `\_12\catcode `\%12\relax}%
\providecommand \@@startlink[1]{}%
\providecommand \@@endlink[0]{}%
\providecommand \url  [0]{\begingroup\@sanitize@url \@url }%
\providecommand \@url [1]{\endgroup\@href {#1}{\urlprefix }}%
\providecommand \urlprefix  [0]{URL }%
\providecommand \Eprint [0]{\href }%
\providecommand \doibase [0]{http://dx.doi.org/}%
\providecommand \selectlanguage [0]{\@gobble}%
\providecommand \bibinfo  [0]{\@secondoftwo}%
\providecommand \bibfield  [0]{\@secondoftwo}%
\providecommand \translation [1]{[#1]}%
\providecommand \BibitemOpen [0]{}%
\providecommand \bibitemStop [0]{}%
\providecommand \bibitemNoStop [0]{.\EOS\space}%
\providecommand \EOS [0]{\spacefactor3000\relax}%
\providecommand \BibitemShut  [1]{\csname bibitem#1\endcsname}%
\let\auto@bib@innerbib\@empty
\bibitem [{\citenamefont {{V}ynck}\ \emph {et~al.}(2021)\citenamefont
  {{V}ynck}, \citenamefont {{P}ierrat}, \citenamefont {{C}arminati},
  \citenamefont {{F}roufe {P}{\'e}rez}, \citenamefont {{S}cheffold},
  \citenamefont {{S}apienza}, \citenamefont {{V}ignolini},\ and\ \citenamefont
  {{S}{\'a}enz}}]{vynck21}%
  \BibitemOpen
  \bibfield  {author} {\bibinfo {author} {\bibfnamefont {K.}~\bibnamefont
  {{V}ynck}}, \bibinfo {author} {\bibfnamefont {R.}~\bibnamefont {{P}ierrat}},
  \bibinfo {author} {\bibfnamefont {R.}~\bibnamefont {{C}arminati}}, \bibinfo
  {author} {\bibfnamefont {L.~S.}\ \bibnamefont {{F}roufe {P}{\'e}rez}},
  \bibinfo {author} {\bibfnamefont {F.}~\bibnamefont {{S}cheffold}}, \bibinfo
  {author} {\bibfnamefont {R.}~\bibnamefont {{S}apienza}}, \bibinfo {author}
  {\bibfnamefont {S.}~\bibnamefont {{V}ignolini}}, \ and\ \bibinfo {author}
  {\bibfnamefont {J.~J.}\ \bibnamefont {{S}{\'a}enz}},\ }\href@noop {} {\
  (\bibinfo {year} {2021})},\ \Eprint {http://arxiv.org/abs/2106.13892}
  {ar{X}iv:2106.13892} \BibitemShut {NoStop}%
\bibitem [{\citenamefont {Benedek}(1971)}]{benedek71}%
  \BibitemOpen
  \bibfield  {author} {\bibinfo {author} {\bibfnamefont {G.~B.}\ \bibnamefont
  {Benedek}},\ }\href {\doibase 10.1364/AO.10.000459} {\bibfield  {journal}
  {\bibinfo  {journal} {Appl. Opt.}\ }\textbf {\bibinfo {volume} {10}},\
  \bibinfo {pages} {459} (\bibinfo {year} {1971})}\BibitemShut {NoStop}%
\bibitem [{\citenamefont {Jacucci}\ \emph {et~al.}(2020)\citenamefont
  {Jacucci}, \citenamefont {Vignolini},\ and\ \citenamefont
  {Schertel}}]{jacucci20}%
  \BibitemOpen
  \bibfield  {author} {\bibinfo {author} {\bibfnamefont {G.}~\bibnamefont
  {Jacucci}}, \bibinfo {author} {\bibfnamefont {S.}~\bibnamefont {Vignolini}},
  \ and\ \bibinfo {author} {\bibfnamefont {L.}~\bibnamefont {Schertel}},\
  }\href {\doibase 10.1073/pnas.2010486117} {\bibfield  {journal} {\bibinfo
  {journal} {Proc. Natl. Acad. Sci. U.S.A.}\ }\textbf {\bibinfo {volume}
  {117}},\ \bibinfo {pages} {23345} (\bibinfo {year} {2020})}\BibitemShut
  {NoStop}%
\bibitem [{\citenamefont {Rojas-Ochoa}\ \emph {et~al.}(2004)\citenamefont
  {Rojas-Ochoa}, \citenamefont {Mendez-Alcaraz}, \citenamefont {S\'aenz},
  \citenamefont {Schurtenberger},\ and\ \citenamefont {Scheffold}}]{ochoa04}%
  \BibitemOpen
  \bibfield  {author} {\bibinfo {author} {\bibfnamefont {L.~F.}\ \bibnamefont
  {Rojas-Ochoa}}, \bibinfo {author} {\bibfnamefont {J.~M.}\ \bibnamefont
  {Mendez-Alcaraz}}, \bibinfo {author} {\bibfnamefont {J.~J.}\ \bibnamefont
  {S\'aenz}}, \bibinfo {author} {\bibfnamefont {P.}~\bibnamefont
  {Schurtenberger}}, \ and\ \bibinfo {author} {\bibfnamefont {F.}~\bibnamefont
  {Scheffold}},\ }\href {\doibase 10.1103/PhysRevLett.93.073903} {\bibfield
  {journal} {\bibinfo  {journal} {Phys. Rev. Lett.}\ }\textbf {\bibinfo
  {volume} {93}},\ \bibinfo {pages} {073903} (\bibinfo {year}
  {2004})}\BibitemShut {NoStop}%
\bibitem [{\citenamefont {Fraden}\ and\ \citenamefont
  {Maret}(1990)}]{fraden90}%
  \BibitemOpen
  \bibfield  {author} {\bibinfo {author} {\bibfnamefont {S.}~\bibnamefont
  {Fraden}}\ and\ \bibinfo {author} {\bibfnamefont {G.}~\bibnamefont {Maret}},\
  }\href {\doibase 10.1103/PhysRevLett.65.512} {\bibfield  {journal} {\bibinfo
  {journal} {Phys. Rev. Lett.}\ }\textbf {\bibinfo {volume} {65}},\ \bibinfo
  {pages} {512} (\bibinfo {year} {1990})}\BibitemShut {NoStop}%
\bibitem [{\citenamefont {John}(1987)}]{john87}%
  \BibitemOpen
  \bibfield  {author} {\bibinfo {author} {\bibfnamefont {S.}~\bibnamefont
  {John}},\ }\href {\doibase 10.1103/PhysRevLett.58.2486} {\bibfield  {journal}
  {\bibinfo  {journal} {Phys. Rev. Lett.}\ }\textbf {\bibinfo {volume} {58}},\
  \bibinfo {pages} {2486} (\bibinfo {year} {1987})}\BibitemShut {NoStop}%
\bibitem [{\citenamefont {Garc\'{\i}a}\ \emph {et~al.}(2011)\citenamefont
  {Garc\'{\i}a}, \citenamefont {Sapienza}, \citenamefont {Toninelli},
  \citenamefont {L\'opez},\ and\ \citenamefont {Wiersma}}]{garcia11}%
  \BibitemOpen
  \bibfield  {author} {\bibinfo {author} {\bibfnamefont {P.~D.}\ \bibnamefont
  {Garc\'{\i}a}}, \bibinfo {author} {\bibfnamefont {R.}~\bibnamefont
  {Sapienza}}, \bibinfo {author} {\bibfnamefont {C.}~\bibnamefont {Toninelli}},
  \bibinfo {author} {\bibfnamefont {C.}~\bibnamefont {L\'opez}}, \ and\
  \bibinfo {author} {\bibfnamefont {D.~S.}\ \bibnamefont {Wiersma}},\ }\href
  {\doibase 10.1103/PhysRevA.84.023813} {\bibfield  {journal} {\bibinfo
  {journal} {Phys. Rev. A}\ }\textbf {\bibinfo {volume} {84}},\ \bibinfo
  {pages} {023813} (\bibinfo {year} {2011})}\BibitemShut {NoStop}%
\bibitem [{\citenamefont {Jin}\ \emph {et~al.}(2001)\citenamefont {Jin},
  \citenamefont {Meng}, \citenamefont {Cheng}, \citenamefont {Li},\ and\
  \citenamefont {Zhang}}]{jin01}%
  \BibitemOpen
  \bibfield  {author} {\bibinfo {author} {\bibfnamefont {C.}~\bibnamefont
  {Jin}}, \bibinfo {author} {\bibfnamefont {X.}~\bibnamefont {Meng}}, \bibinfo
  {author} {\bibfnamefont {B.}~\bibnamefont {Cheng}}, \bibinfo {author}
  {\bibfnamefont {Z.}~\bibnamefont {Li}}, \ and\ \bibinfo {author}
  {\bibfnamefont {D.}~\bibnamefont {Zhang}},\ }\href {\doibase
  10.1103/PhysRevB.63.195107} {\bibfield  {journal} {\bibinfo  {journal} {Phys.
  Rev. B}\ }\textbf {\bibinfo {volume} {63}},\ \bibinfo {pages} {195107}
  (\bibinfo {year} {2001})}\BibitemShut {NoStop}%
\bibitem [{\citenamefont {Edagawa}\ \emph {et~al.}(2008)\citenamefont
  {Edagawa}, \citenamefont {Kanoko},\ and\ \citenamefont {Notomi}}]{edagawa08}%
  \BibitemOpen
  \bibfield  {author} {\bibinfo {author} {\bibfnamefont {K.}~\bibnamefont
  {Edagawa}}, \bibinfo {author} {\bibfnamefont {S.}~\bibnamefont {Kanoko}}, \
  and\ \bibinfo {author} {\bibfnamefont {M.}~\bibnamefont {Notomi}},\ }\href
  {\doibase 10.1103/PhysRevLett.100.013901} {\bibfield  {journal} {\bibinfo
  {journal} {Phys. Rev. Lett.}\ }\textbf {\bibinfo {volume} {100}},\ \bibinfo
  {pages} {013901} (\bibinfo {year} {2008})}\BibitemShut {NoStop}%
\bibitem [{\citenamefont {Liew}\ \emph {et~al.}(2011)\citenamefont {Liew},
  \citenamefont {Yang}, \citenamefont {Noh}, \citenamefont {Schreck},
  \citenamefont {Dufresne}, \citenamefont {O'Hern},\ and\ \citenamefont
  {Cao}}]{liew11}%
  \BibitemOpen
  \bibfield  {author} {\bibinfo {author} {\bibfnamefont {S.~F.}\ \bibnamefont
  {Liew}}, \bibinfo {author} {\bibfnamefont {J.-K.}\ \bibnamefont {Yang}},
  \bibinfo {author} {\bibfnamefont {H.}~\bibnamefont {Noh}}, \bibinfo {author}
  {\bibfnamefont {C.~F.}\ \bibnamefont {Schreck}}, \bibinfo {author}
  {\bibfnamefont {E.~R.}\ \bibnamefont {Dufresne}}, \bibinfo {author}
  {\bibfnamefont {C.~S.}\ \bibnamefont {O'Hern}}, \ and\ \bibinfo {author}
  {\bibfnamefont {H.}~\bibnamefont {Cao}},\ }\href {\doibase
  10.1103/PhysRevA.84.063818} {\bibfield  {journal} {\bibinfo  {journal} {Phys.
  Rev. A}\ }\textbf {\bibinfo {volume} {84}},\ \bibinfo {pages} {063818}
  (\bibinfo {year} {2011})}\BibitemShut {NoStop}%
\bibitem [{\citenamefont {Froufe-P{\'e}rez}\ \emph {et~al.}(2017)\citenamefont
  {Froufe-P{\'e}rez}, \citenamefont {Engel}, \citenamefont {S{\'a}enz},\ and\
  \citenamefont {Scheffold}}]{froufe17}%
  \BibitemOpen
  \bibfield  {author} {\bibinfo {author} {\bibfnamefont {L.~S.}\ \bibnamefont
  {Froufe-P{\'e}rez}}, \bibinfo {author} {\bibfnamefont {M.}~\bibnamefont
  {Engel}}, \bibinfo {author} {\bibfnamefont {J.~J.}\ \bibnamefont
  {S{\'a}enz}}, \ and\ \bibinfo {author} {\bibfnamefont {F.}~\bibnamefont
  {Scheffold}},\ }\href {\doibase 10.1073/pnas.1705130114} {\bibfield
  {journal} {\bibinfo  {journal} {Proc. Natl. Acad. Sci. U.S.A.}\ }\textbf
  {\bibinfo {volume} {114}},\ \bibinfo {pages} {9570} (\bibinfo {year}
  {2017})}\BibitemShut {NoStop}%
\bibitem [{\citenamefont {Florescu}\ \emph {et~al.}(2009)\citenamefont
  {Florescu}, \citenamefont {Torquato},\ and\ \citenamefont
  {Steinhardt}}]{florescu09}%
  \BibitemOpen
  \bibfield  {author} {\bibinfo {author} {\bibfnamefont {M.}~\bibnamefont
  {Florescu}}, \bibinfo {author} {\bibfnamefont {S.}~\bibnamefont {Torquato}},
  \ and\ \bibinfo {author} {\bibfnamefont {P.~J.}\ \bibnamefont {Steinhardt}},\
  }\href {\doibase 10.1073/pnas.0907744106} {\bibfield  {journal} {\bibinfo
  {journal} {Proc. Natl. Acad. Sci. U.S.A.}\ }\textbf {\bibinfo {volume}
  {106}},\ \bibinfo {pages} {20658} (\bibinfo {year} {2009})}\BibitemShut
  {NoStop}%
\bibitem [{\citenamefont {Froufe-P\'erez}\ \emph {et~al.}(2016)\citenamefont
  {Froufe-P\'erez}, \citenamefont {Engel}, \citenamefont {Damasceno},
  \citenamefont {Muller}, \citenamefont {Haberko}, \citenamefont {Glotzer},\
  and\ \citenamefont {Scheffold}}]{froufe16}%
  \BibitemOpen
  \bibfield  {author} {\bibinfo {author} {\bibfnamefont {L.~S.}\ \bibnamefont
  {Froufe-P\'erez}}, \bibinfo {author} {\bibfnamefont {M.}~\bibnamefont
  {Engel}}, \bibinfo {author} {\bibfnamefont {P.~F.}\ \bibnamefont
  {Damasceno}}, \bibinfo {author} {\bibfnamefont {N.}~\bibnamefont {Muller}},
  \bibinfo {author} {\bibfnamefont {J.}~\bibnamefont {Haberko}}, \bibinfo
  {author} {\bibfnamefont {S.~C.}\ \bibnamefont {Glotzer}}, \ and\ \bibinfo
  {author} {\bibfnamefont {F.}~\bibnamefont {Scheffold}},\ }\href {\doibase
  10.1103/PhysRevLett.117.053902} {\bibfield  {journal} {\bibinfo  {journal}
  {Phys. Rev. Lett.}\ }\textbf {\bibinfo {volume} {117}},\ \bibinfo {pages}
  {053902} (\bibinfo {year} {2016})}\BibitemShut {NoStop}%
\bibitem [{\citenamefont {Weaire}(1971)}]{weaire71}%
  \BibitemOpen
  \bibfield  {author} {\bibinfo {author} {\bibfnamefont {D.}~\bibnamefont
  {Weaire}},\ }\href {\doibase 10.1103/PhysRevLett.26.1541} {\bibfield
  {journal} {\bibinfo  {journal} {Phys. Rev. Lett.}\ }\textbf {\bibinfo
  {volume} {26}},\ \bibinfo {pages} {1541} (\bibinfo {year}
  {1971})}\BibitemShut {NoStop}%
\bibitem [{\citenamefont {Thorpe}\ and\ \citenamefont
  {Weaire}(1971)}]{thorpe71}%
  \BibitemOpen
  \bibfield  {author} {\bibinfo {author} {\bibfnamefont {M.~F.}\ \bibnamefont
  {Thorpe}}\ and\ \bibinfo {author} {\bibfnamefont {D.}~\bibnamefont
  {Weaire}},\ }\href {\doibase 10.1103/PhysRevLett.27.1581} {\bibfield
  {journal} {\bibinfo  {journal} {Phys. Rev. Lett.}\ }\textbf {\bibinfo
  {volume} {27}},\ \bibinfo {pages} {1581} (\bibinfo {year}
  {1971})}\BibitemShut {NoStop}%
\bibitem [{\citenamefont {Lidorikis}\ \emph {et~al.}(1998)\citenamefont
  {Lidorikis}, \citenamefont {Sigalas}, \citenamefont {Economou},\ and\
  \citenamefont {Soukoulis}}]{lidorikis98}%
  \BibitemOpen
  \bibfield  {author} {\bibinfo {author} {\bibfnamefont {E.}~\bibnamefont
  {Lidorikis}}, \bibinfo {author} {\bibfnamefont {M.~M.}\ \bibnamefont
  {Sigalas}}, \bibinfo {author} {\bibfnamefont {E.~N.}\ \bibnamefont
  {Economou}}, \ and\ \bibinfo {author} {\bibfnamefont {C.~M.}\ \bibnamefont
  {Soukoulis}},\ }\href {\doibase 10.1103/PhysRevLett.81.1405} {\bibfield
  {journal} {\bibinfo  {journal} {Phys. Rev. Lett.}\ }\textbf {\bibinfo
  {volume} {81}},\ \bibinfo {pages} {1405} (\bibinfo {year}
  {1998})}\BibitemShut {NoStop}%
\bibitem [{\citenamefont {Vynck}\ \emph {et~al.}(2009)\citenamefont {Vynck},
  \citenamefont {Felbacq}, \citenamefont {Centeno}, \citenamefont
  {C\ifmmode~\u{a}\else \u{a}\fi{}buz}, \citenamefont {Cassagne},\ and\
  \citenamefont {Guizal}}]{vynck09}%
  \BibitemOpen
  \bibfield  {author} {\bibinfo {author} {\bibfnamefont {K.}~\bibnamefont
  {Vynck}}, \bibinfo {author} {\bibfnamefont {D.}~\bibnamefont {Felbacq}},
  \bibinfo {author} {\bibfnamefont {E.}~\bibnamefont {Centeno}}, \bibinfo
  {author} {\bibfnamefont {A.~I.}\ \bibnamefont {C\ifmmode~\u{a}\else
  \u{a}\fi{}buz}}, \bibinfo {author} {\bibfnamefont {D.}~\bibnamefont
  {Cassagne}}, \ and\ \bibinfo {author} {\bibfnamefont {B.}~\bibnamefont
  {Guizal}},\ }\href {\doibase 10.1103/PhysRevLett.102.133901} {\bibfield
  {journal} {\bibinfo  {journal} {Phys. Rev. Lett.}\ }\textbf {\bibinfo
  {volume} {102}},\ \bibinfo {pages} {133901} (\bibinfo {year}
  {2009})}\BibitemShut {NoStop}%
\bibitem [{\citenamefont {Rockstuhl}\ \emph {et~al.}(2006)\citenamefont
  {Rockstuhl}, \citenamefont {Peschel},\ and\ \citenamefont
  {Lederer}}]{rockstuhl06}%
  \BibitemOpen
  \bibfield  {author} {\bibinfo {author} {\bibfnamefont {C.}~\bibnamefont
  {Rockstuhl}}, \bibinfo {author} {\bibfnamefont {U.}~\bibnamefont {Peschel}},
  \ and\ \bibinfo {author} {\bibfnamefont {F.}~\bibnamefont {Lederer}},\
  }\href@noop {} {\bibfield  {journal} {\bibinfo  {journal} {Opt. Lett.}\
  }\textbf {\bibinfo {volume} {31}},\ \bibinfo {pages} {1741} (\bibinfo {year}
  {2006})}\BibitemShut {NoStop}%
\bibitem [{\citenamefont {Sellers}\ \emph {et~al.}(2017)\citenamefont
  {Sellers}, \citenamefont {Man}, \citenamefont {Sahba},\ and\ \citenamefont
  {Florescu}}]{sellers17}%
  \BibitemOpen
  \bibfield  {author} {\bibinfo {author} {\bibfnamefont {S.~R.}\ \bibnamefont
  {Sellers}}, \bibinfo {author} {\bibfnamefont {W.}~\bibnamefont {Man}},
  \bibinfo {author} {\bibfnamefont {S.}~\bibnamefont {Sahba}}, \ and\ \bibinfo
  {author} {\bibfnamefont {M.}~\bibnamefont {Florescu}},\ }\href {\doibase
  10.1038/ncomms14439} {\bibfield  {journal} {\bibinfo  {journal} {Nat.
  Commun.}\ }\textbf {\bibinfo {volume} {8}},\ \bibinfo {pages} {14439}
  (\bibinfo {year} {2017})}\BibitemShut {NoStop}%
\bibitem [{\citenamefont {Lagendijk}\ and\ \citenamefont {van
  Tiggelen}(1996)}]{lagendijk96}%
  \BibitemOpen
  \bibfield  {author} {\bibinfo {author} {\bibfnamefont {A.}~\bibnamefont
  {Lagendijk}}\ and\ \bibinfo {author} {\bibfnamefont {B.~A.}\ \bibnamefont
  {van Tiggelen}},\ }\href@noop {} {\bibfield  {journal} {\bibinfo  {journal}
  {Phys. Rep.}\ }\textbf {\bibinfo {volume} {270}},\ \bibinfo {pages} {143}
  (\bibinfo {year} {1996})}\BibitemShut {NoStop}%
\bibitem [{\citenamefont {Imagawa}\ \emph {et~al.}(2010)\citenamefont
  {Imagawa}, \citenamefont {Edagawa}, \citenamefont {Morita}, \citenamefont
  {Niino}, \citenamefont {Kagawa},\ and\ \citenamefont {Notomi}}]{imagawa10}%
  \BibitemOpen
  \bibfield  {author} {\bibinfo {author} {\bibfnamefont {S.}~\bibnamefont
  {Imagawa}}, \bibinfo {author} {\bibfnamefont {K.}~\bibnamefont {Edagawa}},
  \bibinfo {author} {\bibfnamefont {K.}~\bibnamefont {Morita}}, \bibinfo
  {author} {\bibfnamefont {T.}~\bibnamefont {Niino}}, \bibinfo {author}
  {\bibfnamefont {Y.}~\bibnamefont {Kagawa}}, \ and\ \bibinfo {author}
  {\bibfnamefont {M.}~\bibnamefont {Notomi}},\ }\href {\doibase
  10.1103/PhysRevB.82.115116} {\bibfield  {journal} {\bibinfo  {journal} {Phys.
  Rev. B}\ }\textbf {\bibinfo {volume} {82}},\ \bibinfo {pages} {115116}
  (\bibinfo {year} {2010})}\BibitemShut {NoStop}%
\bibitem [{\citenamefont {Rechtsman}\ \emph {et~al.}(2011)\citenamefont
  {Rechtsman}, \citenamefont {Szameit}, \citenamefont {Dreisow}, \citenamefont
  {Heinrich}, \citenamefont {Keil}, \citenamefont {Nolte},\ and\ \citenamefont
  {Segev}}]{rechtsman11}%
  \BibitemOpen
  \bibfield  {author} {\bibinfo {author} {\bibfnamefont {M.}~\bibnamefont
  {Rechtsman}}, \bibinfo {author} {\bibfnamefont {A.}~\bibnamefont {Szameit}},
  \bibinfo {author} {\bibfnamefont {F.}~\bibnamefont {Dreisow}}, \bibinfo
  {author} {\bibfnamefont {M.}~\bibnamefont {Heinrich}}, \bibinfo {author}
  {\bibfnamefont {R.}~\bibnamefont {Keil}}, \bibinfo {author} {\bibfnamefont
  {S.}~\bibnamefont {Nolte}}, \ and\ \bibinfo {author} {\bibfnamefont
  {M.}~\bibnamefont {Segev}},\ }\href {\doibase 10.1103/PhysRevLett.106.193904}
  {\bibfield  {journal} {\bibinfo  {journal} {Phys. Rev. Lett.}\ }\textbf
  {\bibinfo {volume} {106}},\ \bibinfo {pages} {193904} (\bibinfo {year}
  {2011})}\BibitemShut {NoStop}%
\bibitem [{\citenamefont {Conley}\ \emph {et~al.}(2014)\citenamefont {Conley},
  \citenamefont {Burresi}, \citenamefont {Pratesi}, \citenamefont {Vynck},\
  and\ \citenamefont {Wiersma}}]{conley14}%
  \BibitemOpen
  \bibfield  {author} {\bibinfo {author} {\bibfnamefont {G.~M.}\ \bibnamefont
  {Conley}}, \bibinfo {author} {\bibfnamefont {M.}~\bibnamefont {Burresi}},
  \bibinfo {author} {\bibfnamefont {F.}~\bibnamefont {Pratesi}}, \bibinfo
  {author} {\bibfnamefont {K.}~\bibnamefont {Vynck}}, \ and\ \bibinfo {author}
  {\bibfnamefont {D.~S.}\ \bibnamefont {Wiersma}},\ }\href {\doibase
  10.1103/PhysRevLett.112.143901} {\bibfield  {journal} {\bibinfo  {journal}
  {Phys. Rev. Lett.}\ }\textbf {\bibinfo {volume} {112}},\ \bibinfo {pages}
  {143901} (\bibinfo {year} {2014})}\BibitemShut {NoStop}%
\bibitem [{\citenamefont {Haberko}\ \emph {et~al.}(2020)\citenamefont
  {Haberko}, \citenamefont {Froufe-P\'erez},\ and\ \citenamefont
  {Scheffold}}]{haberko20}%
  \BibitemOpen
  \bibfield  {author} {\bibinfo {author} {\bibfnamefont {J.}~\bibnamefont
  {Haberko}}, \bibinfo {author} {\bibfnamefont {L.~S.}\ \bibnamefont
  {Froufe-P\'erez}}, \ and\ \bibinfo {author} {\bibfnamefont {F.}~\bibnamefont
  {Scheffold}},\ }\href@noop {} {\bibfield  {journal} {\bibinfo  {journal}
  {Nat. Commun.}\ }\textbf {\bibinfo {volume} {11}},\ \bibinfo {pages} {4867}
  (\bibinfo {year} {2020})}\BibitemShut {NoStop}%
\bibitem [{\citenamefont {Aubry}\ \emph {et~al.}(2020)\citenamefont {Aubry},
  \citenamefont {Froufe-P\'erez}, \citenamefont {Kuhl}, \citenamefont
  {Legrand}, \citenamefont {Scheffold},\ and\ \citenamefont
  {Mortessagne}}]{aubry20}%
  \BibitemOpen
  \bibfield  {author} {\bibinfo {author} {\bibfnamefont {G.~J.}\ \bibnamefont
  {Aubry}}, \bibinfo {author} {\bibfnamefont {L.~S.}\ \bibnamefont
  {Froufe-P\'erez}}, \bibinfo {author} {\bibfnamefont {U.}~\bibnamefont
  {Kuhl}}, \bibinfo {author} {\bibfnamefont {O.}~\bibnamefont {Legrand}},
  \bibinfo {author} {\bibfnamefont {F.}~\bibnamefont {Scheffold}}, \ and\
  \bibinfo {author} {\bibfnamefont {F.}~\bibnamefont {Mortessagne}},\ }\href
  {\doibase 10.1103/PhysRevLett.125.127402} {\bibfield  {journal} {\bibinfo
  {journal} {Phys. Rev. Lett.}\ }\textbf {\bibinfo {volume} {125}},\ \bibinfo
  {pages} {127402} (\bibinfo {year} {2020})}\BibitemShut {NoStop}%
\bibitem [{\citenamefont {Sheng}(2006)}]{sheng06}%
  \BibitemOpen
  \bibfield  {author} {\bibinfo {author} {\bibfnamefont {P.}~\bibnamefont
  {Sheng}},\ }\href@noop {} {\emph {\bibinfo {title} {Introduction to Wave
  Scattering, Localization and Mesoscopic Phenomena}}}\ (\bibinfo  {publisher}
  {Springer-Verlag Berlin Heidelberg},\ \bibinfo {year} {2006})\BibitemShut
  {NoStop}%
\bibitem [{\citenamefont {Hu}\ \emph {et~al.}(2008)\citenamefont {Hu},
  \citenamefont {Strybulevych}, \citenamefont {Page}, \citenamefont
  {Skipetrov},\ and\ \citenamefont {van Tiggelen}}]{hu08}%
  \BibitemOpen
  \bibfield  {author} {\bibinfo {author} {\bibfnamefont {H.}~\bibnamefont
  {Hu}}, \bibinfo {author} {\bibfnamefont {A.}~\bibnamefont {Strybulevych}},
  \bibinfo {author} {\bibfnamefont {J.~H.}\ \bibnamefont {Page}}, \bibinfo
  {author} {\bibfnamefont {S.~E.}\ \bibnamefont {Skipetrov}}, \ and\ \bibinfo
  {author} {\bibfnamefont {B.~A.}\ \bibnamefont {van Tiggelen}},\ }\href
  {\doibase 10.1038/nphys1101} {\bibfield  {journal} {\bibinfo  {journal}
  {Nature Phys.}\ }\textbf {\bibinfo {volume} {4}},\ \bibinfo {pages} {945}
  (\bibinfo {year} {2008})}\BibitemShut {NoStop}%
\bibitem [{\citenamefont {John}(1991)}]{john91}%
  \BibitemOpen
  \bibfield  {author} {\bibinfo {author} {\bibfnamefont {S.}~\bibnamefont
  {John}},\ }\href {\doibase 10.1063/1.881300} {\bibfield  {journal} {\bibinfo
  {journal} {Physics Today}\ }\textbf {\bibinfo {volume} {44}},\ \bibinfo
  {pages} {32} (\bibinfo {year} {1991})}\BibitemShut {NoStop}%
\bibitem [{\citenamefont {John}(1996)}]{john96}%
  \BibitemOpen
  \bibfield  {author} {\bibinfo {author} {\bibfnamefont {S.}~\bibnamefont
  {John}},\ }\enquote {\bibinfo {title} {Localization of light: Theory of
  photonic band gap materials},}\ in\ \href@noop {} {\emph {\bibinfo
  {booktitle} {Photonic Band Gap Materials}}},\ \bibinfo {editor} {edited by\
  \bibinfo {editor} {\bibfnamefont {C.~M.}\ \bibnamefont {Soukoulis}}}\
  (\bibinfo  {publisher} {Springer Netherlands},\ \bibinfo {address}
  {Dordrecht},\ \bibinfo {year} {1996})\ pp.\ \bibinfo {pages}
  {563--665}\BibitemShut {NoStop}%
\bibitem [{\citenamefont {Abrahams}\ \emph {et~al.}(1979)\citenamefont
  {Abrahams}, \citenamefont {Anderson}, \citenamefont {Licciardello},\ and\
  \citenamefont {Ramakrishnan}}]{abrahams79}%
  \BibitemOpen
  \bibfield  {author} {\bibinfo {author} {\bibfnamefont {E.}~\bibnamefont
  {Abrahams}}, \bibinfo {author} {\bibfnamefont {P.~W.}\ \bibnamefont
  {Anderson}}, \bibinfo {author} {\bibfnamefont {D.~C.}\ \bibnamefont
  {Licciardello}}, \ and\ \bibinfo {author} {\bibfnamefont {T.~V.}\
  \bibnamefont {Ramakrishnan}},\ }\href {\doibase 10.1103/PhysRevLett.42.673}
  {\bibfield  {journal} {\bibinfo  {journal} {Phys. Rev. Lett.}\ }\textbf
  {\bibinfo {volume} {42}},\ \bibinfo {pages} {673} (\bibinfo {year}
  {1979})}\BibitemShut {NoStop}%
\bibitem [{\citenamefont {Vollhardt}\ and\ \citenamefont
  {W\"{o}lfle}(1992)}]{vollhardt92}%
  \BibitemOpen
  \bibfield  {author} {\bibinfo {author} {\bibfnamefont {D.}~\bibnamefont
  {Vollhardt}}\ and\ \bibinfo {author} {\bibfnamefont {P.}~\bibnamefont
  {W\"{o}lfle}},\ }\enquote {\bibinfo {title} {Self-consistent theory of
  anderson localization},}\ in\ \href@noop {} {\emph {\bibinfo {booktitle}
  {Electronic Phase Transitions}}},\ \bibinfo {editor} {edited by\ \bibinfo
  {editor} {\bibfnamefont {W.}~\bibnamefont {Hanke}}\ and\ \bibinfo {editor}
  {\bibfnamefont {Y.~V.}\ \bibnamefont {Kopaev}}}\ (\bibinfo  {publisher}
  {Elsevier Science Publishers},\ \bibinfo {year} {1992})\ pp.\ \bibinfo
  {pages} {1--78}\BibitemShut {NoStop}%
\bibitem [{\citenamefont {Antezza}\ and\ \citenamefont
  {Castin}(2009{\natexlab{a}})}]{antezza09_prl}%
  \BibitemOpen
  \bibfield  {author} {\bibinfo {author} {\bibfnamefont {M.}~\bibnamefont
  {Antezza}}\ and\ \bibinfo {author} {\bibfnamefont {Y.}~\bibnamefont
  {Castin}},\ }\href {\doibase 10.1103/PhysRevLett.103.123903} {\bibfield
  {journal} {\bibinfo  {journal} {Phys. Rev. Lett.}\ }\textbf {\bibinfo
  {volume} {103}},\ \bibinfo {pages} {123903} (\bibinfo {year}
  {2009}{\natexlab{a}})}\BibitemShut {NoStop}%
\bibitem [{\citenamefont {Perczel}\ \emph {et~al.}(2017)\citenamefont
  {Perczel}, \citenamefont {Borregaard}, \citenamefont {Chang}, \citenamefont
  {Pichler}, \citenamefont {Yelin}, \citenamefont {Zoller},\ and\ \citenamefont
  {Lukin}}]{perczel17}%
  \BibitemOpen
  \bibfield  {author} {\bibinfo {author} {\bibfnamefont {J.}~\bibnamefont
  {Perczel}}, \bibinfo {author} {\bibfnamefont {J.}~\bibnamefont {Borregaard}},
  \bibinfo {author} {\bibfnamefont {D.~E.}\ \bibnamefont {Chang}}, \bibinfo
  {author} {\bibfnamefont {H.}~\bibnamefont {Pichler}}, \bibinfo {author}
  {\bibfnamefont {S.~F.}\ \bibnamefont {Yelin}}, \bibinfo {author}
  {\bibfnamefont {P.}~\bibnamefont {Zoller}}, \ and\ \bibinfo {author}
  {\bibfnamefont {M.~D.}\ \bibnamefont {Lukin}},\ }\href {\doibase
  10.1103/PhysRevA.96.063801} {\bibfield  {journal} {\bibinfo  {journal} {Phys.
  Rev. A}\ }\textbf {\bibinfo {volume} {96}},\ \bibinfo {pages} {063801}
  (\bibinfo {year} {2017})}\BibitemShut {NoStop}%
\bibitem [{\citenamefont {Skipetrov}(2020{\natexlab{a}})}]{skipetrov20_epjb}%
  \BibitemOpen
  \bibfield  {author} {\bibinfo {author} {\bibfnamefont {S.~E.}\ \bibnamefont
  {Skipetrov}},\ }\href@noop {} {\bibfield  {journal} {\bibinfo  {journal}
  {Eur. Phys. J. B}\ }\textbf {\bibinfo {volume} {93}},\ \bibinfo {pages} {70}
  (\bibinfo {year} {2020}{\natexlab{a}})}\BibitemShut {NoStop}%
\bibitem [{\citenamefont {Goetschy}\ and\ \citenamefont
  {Skipetrov}(2011{\natexlab{a}})}]{goetschy11_pre}%
  \BibitemOpen
  \bibfield  {author} {\bibinfo {author} {\bibfnamefont {A.}~\bibnamefont
  {Goetschy}}\ and\ \bibinfo {author} {\bibfnamefont {S.}~\bibnamefont
  {Skipetrov}},\ }\href@noop {} {\bibfield  {journal} {\bibinfo  {journal}
  {Phys. Rev. E}\ }\textbf {\bibinfo {volume} {84}},\ \bibinfo {pages} {011150}
  (\bibinfo {year} {2011}{\natexlab{a}})}\BibitemShut {NoStop}%
\bibitem [{\citenamefont {Skipetrov}\ and\ \citenamefont
  {Sokolov}(2014)}]{skipetrov14}%
  \BibitemOpen
  \bibfield  {author} {\bibinfo {author} {\bibfnamefont {S.~E.}\ \bibnamefont
  {Skipetrov}}\ and\ \bibinfo {author} {\bibfnamefont {I.~M.}\ \bibnamefont
  {Sokolov}},\ }\href {\doibase 10.1103/PhysRevLett.112.023905} {\bibfield
  {journal} {\bibinfo  {journal} {Phys. Rev. Lett.}\ }\textbf {\bibinfo
  {volume} {112}},\ \bibinfo {pages} {023905} (\bibinfo {year}
  {2014})}\BibitemShut {NoStop}%
\bibitem [{\citenamefont {Bellando}\ \emph {et~al.}(2014)\citenamefont
  {Bellando}, \citenamefont {Gero}, \citenamefont {Akkermans},\ and\
  \citenamefont {Kaiser}}]{bellando14}%
  \BibitemOpen
  \bibfield  {author} {\bibinfo {author} {\bibfnamefont {L.}~\bibnamefont
  {Bellando}}, \bibinfo {author} {\bibfnamefont {A.}~\bibnamefont {Gero}},
  \bibinfo {author} {\bibfnamefont {E.}~\bibnamefont {Akkermans}}, \ and\
  \bibinfo {author} {\bibfnamefont {R.}~\bibnamefont {Kaiser}},\ }\href@noop {}
  {\bibfield  {journal} {\bibinfo  {journal} {Phys. Rev. A}\ }\textbf {\bibinfo
  {volume} {90}},\ \bibinfo {pages} {063822} (\bibinfo {year}
  {2014})}\BibitemShut {NoStop}%
\bibitem [{\citenamefont {M\'aximo}\ \emph {et~al.}(2015)\citenamefont
  {M\'aximo}, \citenamefont {Piovella}, \citenamefont {Courteille},
  \citenamefont {Kaiser},\ and\ \citenamefont {Bachelard}}]{maximo15}%
  \BibitemOpen
  \bibfield  {author} {\bibinfo {author} {\bibfnamefont {C.~E.}\ \bibnamefont
  {M\'aximo}}, \bibinfo {author} {\bibfnamefont {N.}~\bibnamefont {Piovella}},
  \bibinfo {author} {\bibfnamefont {P.~W.}\ \bibnamefont {Courteille}},
  \bibinfo {author} {\bibfnamefont {R.}~\bibnamefont {Kaiser}}, \ and\ \bibinfo
  {author} {\bibfnamefont {R.}~\bibnamefont {Bachelard}},\ }\href {\doibase
  10.1103/PhysRevA.92.062702} {\bibfield  {journal} {\bibinfo  {journal} {Phys.
  Rev. A}\ }\textbf {\bibinfo {volume} {92}},\ \bibinfo {pages} {062702}
  (\bibinfo {year} {2015})}\BibitemShut {NoStop}%
\bibitem [{\citenamefont {Skipetrov}(2016)}]{skipetrov16}%
  \BibitemOpen
  \bibfield  {author} {\bibinfo {author} {\bibfnamefont {S.~E.}\ \bibnamefont
  {Skipetrov}},\ }\href {\doibase 10.1103/PhysRevB.94.064202} {\bibfield
  {journal} {\bibinfo  {journal} {Phys. Rev. B}\ }\textbf {\bibinfo {volume}
  {94}},\ \bibinfo {pages} {064202} (\bibinfo {year} {2016})}\BibitemShut
  {NoStop}%
\bibitem [{\citenamefont {Nieuwenhuizen}\ \emph {et~al.}(1994)\citenamefont
  {Nieuwenhuizen}, \citenamefont {Burin}, \citenamefont {Kagan},\ and\
  \citenamefont {Schlyapnikov}}]{nieuwenhuizen94}%
  \BibitemOpen
  \bibfield  {author} {\bibinfo {author} {\bibfnamefont {T.~M.}\ \bibnamefont
  {Nieuwenhuizen}}, \bibinfo {author} {\bibfnamefont {A.}~\bibnamefont
  {Burin}}, \bibinfo {author} {\bibfnamefont {Y.}~\bibnamefont {Kagan}}, \ and\
  \bibinfo {author} {\bibfnamefont {G.}~\bibnamefont {Schlyapnikov}},\
  }\href@noop {} {\bibfield  {journal} {\bibinfo  {journal} {Phys. Lett. A}\
  }\textbf {\bibinfo {volume} {184}},\ \bibinfo {pages} {360} (\bibinfo {year}
  {1994})}\BibitemShut {NoStop}%
\bibitem [{\citenamefont {Skipetrov}\ and\ \citenamefont
  {Page}(2016)}]{skipetrov16_njp}%
  \BibitemOpen
  \bibfield  {author} {\bibinfo {author} {\bibfnamefont {S.~E.}\ \bibnamefont
  {Skipetrov}}\ and\ \bibinfo {author} {\bibfnamefont {J.~H.}\ \bibnamefont
  {Page}},\ }\href@noop {} {\bibfield  {journal} {\bibinfo  {journal} {New J.
  Phys.}\ }\textbf {\bibinfo {volume} {18}},\ \bibinfo {pages} {021001}
  (\bibinfo {year} {2016})}\BibitemShut {NoStop}%
\bibitem [{\citenamefont {van Tiggelen}\ and\ \citenamefont
  {Skipetrov}(2021)}]{vantiggelen21}%
  \BibitemOpen
  \bibfield  {author} {\bibinfo {author} {\bibfnamefont {B.~A.}\ \bibnamefont
  {van Tiggelen}}\ and\ \bibinfo {author} {\bibfnamefont {S.~E.}\ \bibnamefont
  {Skipetrov}},\ }\href {\doibase 10.1103/PhysRevB.103.174204} {\bibfield
  {journal} {\bibinfo  {journal} {Phys. Rev. B}\ }\textbf {\bibinfo {volume}
  {103}},\ \bibinfo {pages} {174204} (\bibinfo {year} {2021})}\BibitemShut
  {NoStop}%
\bibitem [{\citenamefont {Foldy}(1945)}]{foldy45}%
  \BibitemOpen
  \bibfield  {author} {\bibinfo {author} {\bibfnamefont {L.~L.}\ \bibnamefont
  {Foldy}},\ }\href {\doibase 10.1103/PhysRev.67.107} {\bibfield  {journal}
  {\bibinfo  {journal} {Phys. Rev.}\ }\textbf {\bibinfo {volume} {67}},\
  \bibinfo {pages} {107} (\bibinfo {year} {1945})}\BibitemShut {NoStop}%
\bibitem [{\citenamefont {Lax}(1951)}]{lax51}%
  \BibitemOpen
  \bibfield  {author} {\bibinfo {author} {\bibfnamefont {M.}~\bibnamefont
  {Lax}},\ }\href {\doibase 10.1103/RevModPhys.23.287} {\bibfield  {journal}
  {\bibinfo  {journal} {Rev. Mod. Phys.}\ }\textbf {\bibinfo {volume} {23}},\
  \bibinfo {pages} {287} (\bibinfo {year} {1951})}\BibitemShut {NoStop}%
\bibitem [{\citenamefont {Lehmberg}(1970)}]{lehmberg70}%
  \BibitemOpen
  \bibfield  {author} {\bibinfo {author} {\bibfnamefont {R.~H.}\ \bibnamefont
  {Lehmberg}},\ }\href {\doibase 10.1103/PhysRevA.2.883} {\bibfield  {journal}
  {\bibinfo  {journal} {Phys. Rev. A}\ }\textbf {\bibinfo {volume} {2}},\
  \bibinfo {pages} {883} (\bibinfo {year} {1970})}\BibitemShut {NoStop}%
\bibitem [{\citenamefont {Goetschy}(2011)}]{goetschy11_thesis}%
  \BibitemOpen
  \bibfield  {author} {\bibinfo {author} {\bibfnamefont {A.}~\bibnamefont
  {Goetschy}},\ }\href@noop {} {\emph {\bibinfo {title} {Light in disordered
  atomic systems: Euclidean matrix theory of random lasing. PhD thesis}}}\
  (\bibinfo  {publisher} {J. Fourier University--Grenoble 1, France},\ \bibinfo
  {year} {2011})\BibitemShut {NoStop}%
\bibitem [{\citenamefont {Klugkist}\ \emph {et~al.}(2006)\citenamefont
  {Klugkist}, \citenamefont {Mostovoy},\ and\ \citenamefont
  {Knoester}}]{klugkist06}%
  \BibitemOpen
  \bibfield  {author} {\bibinfo {author} {\bibfnamefont {J.~A.}\ \bibnamefont
  {Klugkist}}, \bibinfo {author} {\bibfnamefont {M.}~\bibnamefont {Mostovoy}},
  \ and\ \bibinfo {author} {\bibfnamefont {J.}~\bibnamefont {Knoester}},\
  }\href {\doibase 10.1103/PhysRevLett.96.163903} {\bibfield  {journal}
  {\bibinfo  {journal} {Phys. Rev. Lett.}\ }\textbf {\bibinfo {volume} {96}},\
  \bibinfo {pages} {163903} (\bibinfo {year} {2006})}\BibitemShut {NoStop}%
\bibitem [{\citenamefont {Antezza}\ and\ \citenamefont
  {Castin}(2009{\natexlab{b}})}]{antezza09_pra}%
  \BibitemOpen
  \bibfield  {author} {\bibinfo {author} {\bibfnamefont {M.}~\bibnamefont
  {Antezza}}\ and\ \bibinfo {author} {\bibfnamefont {Y.}~\bibnamefont
  {Castin}},\ }\href@noop {} {\bibfield  {journal} {\bibinfo  {journal} {PRA}\
  }\textbf {\bibinfo {volume} {80}},\ \bibinfo {pages} {013816} (\bibinfo
  {year} {2009}{\natexlab{b}})}\BibitemShut {NoStop}%
\bibitem [{\citenamefont {Akkermans}\ \emph {et~al.}(2008)\citenamefont
  {Akkermans}, \citenamefont {Gero},\ and\ \citenamefont
  {Kaiser}}]{akkermans08}%
  \BibitemOpen
  \bibfield  {author} {\bibinfo {author} {\bibfnamefont {E.}~\bibnamefont
  {Akkermans}}, \bibinfo {author} {\bibfnamefont {A.}~\bibnamefont {Gero}}, \
  and\ \bibinfo {author} {\bibfnamefont {R.}~\bibnamefont {Kaiser}},\ }\href
  {\doibase 10.1103/PhysRevLett.101.103602} {\bibfield  {journal} {\bibinfo
  {journal} {Phys. Rev. Lett.}\ }\textbf {\bibinfo {volume} {101}},\ \bibinfo
  {pages} {103602} (\bibinfo {year} {2008})}\BibitemShut {NoStop}%
\bibitem [{\citenamefont {Skipetrov}\ and\ \citenamefont
  {Goetschy}(2011)}]{skipetrov11}%
  \BibitemOpen
  \bibfield  {author} {\bibinfo {author} {\bibfnamefont {S.~E.}\ \bibnamefont
  {Skipetrov}}\ and\ \bibinfo {author} {\bibfnamefont {A.}~\bibnamefont
  {Goetschy}},\ }\href@noop {} {\bibfield  {journal} {\bibinfo  {journal} {J.
  Phys. A}\ }\textbf {\bibinfo {volume} {44}},\ \bibinfo {pages} {065102}
  (\bibinfo {year} {2011})}\BibitemShut {NoStop}%
\bibitem [{\citenamefont {Goetschy}\ and\ \citenamefont
  {Skipetrov}(2011{\natexlab{b}})}]{goetschy11_epl}%
  \BibitemOpen
  \bibfield  {author} {\bibinfo {author} {\bibfnamefont {A.}~\bibnamefont
  {Goetschy}}\ and\ \bibinfo {author} {\bibfnamefont {S.~E.}\ \bibnamefont
  {Skipetrov}},\ }\href@noop {} {\bibfield  {journal} {\bibinfo  {journal}
  {Europhys. Lett.}\ }\textbf {\bibinfo {volume} {96}},\ \bibinfo {pages}
  {34005} (\bibinfo {year} {2011}{\natexlab{b}})}\BibitemShut {NoStop}%
\bibitem [{\citenamefont {Torquato}(2018)}]{torquato18}%
  \BibitemOpen
  \bibfield  {author} {\bibinfo {author} {\bibfnamefont {S.}~\bibnamefont
  {Torquato}},\ }\href {\doibase https://doi.org/10.1016/j.physrep.2018.03.001}
  {\bibfield  {journal} {\bibinfo  {journal} {Physics Reports}\ }\textbf
  {\bibinfo {volume} {745}},\ \bibinfo {pages} {1} (\bibinfo {year}
  {2018})}\BibitemShut {NoStop}%
\bibitem [{\citenamefont {Torquato}\ \emph {et~al.}(2015)\citenamefont
  {Torquato}, \citenamefont {Zhang},\ and\ \citenamefont
  {Stillinger}}]{torquato15}%
  \BibitemOpen
  \bibfield  {author} {\bibinfo {author} {\bibfnamefont {S.}~\bibnamefont
  {Torquato}}, \bibinfo {author} {\bibfnamefont {G.}~\bibnamefont {Zhang}}, \
  and\ \bibinfo {author} {\bibfnamefont {F.~H.}\ \bibnamefont {Stillinger}},\
  }\href {\doibase 10.1103/PhysRevX.5.021020} {\bibfield  {journal} {\bibinfo
  {journal} {Phys. Rev. X}\ }\textbf {\bibinfo {volume} {5}},\ \bibinfo {pages}
  {021020} (\bibinfo {year} {2015})}\BibitemShut {NoStop}%
\bibitem [{\citenamefont {Skipetrov}\ and\ \citenamefont
  {Sokolov}(2018)}]{skipetrov18}%
  \BibitemOpen
  \bibfield  {author} {\bibinfo {author} {\bibfnamefont {S.~E.}\ \bibnamefont
  {Skipetrov}}\ and\ \bibinfo {author} {\bibfnamefont {I.~M.}\ \bibnamefont
  {Sokolov}},\ }\href {\doibase 10.1103/PhysRevB.98.064207} {\bibfield
  {journal} {\bibinfo  {journal} {Phys. Rev. B}\ }\textbf {\bibinfo {volume}
  {98}},\ \bibinfo {pages} {064207} (\bibinfo {year} {2018})}\BibitemShut
  {NoStop}%
\bibitem [{\citenamefont {Skipetrov}(2020{\natexlab{b}})}]{skipetrov20_prb}%
  \BibitemOpen
  \bibfield  {author} {\bibinfo {author} {\bibfnamefont {S.~E.}\ \bibnamefont
  {Skipetrov}},\ }\href {\doibase 10.1103/PhysRevB.102.134206} {\bibfield
  {journal} {\bibinfo  {journal} {Phys. Rev. B}\ }\textbf {\bibinfo {volume}
  {102}},\ \bibinfo {pages} {134206} (\bibinfo {year}
  {2020}{\natexlab{b}})}\BibitemShut {NoStop}%
\bibitem [{\citenamefont {Yang}\ \emph {et~al.}(2010)\citenamefont {Yang},
  \citenamefont {Schreck}, \citenamefont {Noh}, \citenamefont {Liew},
  \citenamefont {Guy}, \citenamefont {O'Hern},\ and\ \citenamefont
  {Cao}}]{yang10}%
  \BibitemOpen
  \bibfield  {author} {\bibinfo {author} {\bibfnamefont {J.-K.}\ \bibnamefont
  {Yang}}, \bibinfo {author} {\bibfnamefont {C.}~\bibnamefont {Schreck}},
  \bibinfo {author} {\bibfnamefont {H.}~\bibnamefont {Noh}}, \bibinfo {author}
  {\bibfnamefont {S.-F.}\ \bibnamefont {Liew}}, \bibinfo {author}
  {\bibfnamefont {M.~I.}\ \bibnamefont {Guy}}, \bibinfo {author} {\bibfnamefont
  {C.~S.}\ \bibnamefont {O'Hern}}, \ and\ \bibinfo {author} {\bibfnamefont
  {H.}~\bibnamefont {Cao}},\ }\href {\doibase 10.1103/PhysRevA.82.053838}
  {\bibfield  {journal} {\bibinfo  {journal} {Phys. Rev. A}\ }\textbf {\bibinfo
  {volume} {82}},\ \bibinfo {pages} {053838} (\bibinfo {year}
  {2010})}\BibitemShut {NoStop}%
\bibitem [{\citenamefont {Goetschy}\ and\ \citenamefont
  {Skipetrov}(2013)}]{goetschy13_review}%
  \BibitemOpen
  \bibfield  {author} {\bibinfo {author} {\bibfnamefont {A.}~\bibnamefont
  {Goetschy}}\ and\ \bibinfo {author} {\bibfnamefont {S.}~\bibnamefont
  {Skipetrov}},\ }\href@noop {} {\bibfield  {journal} {\bibinfo  {journal}
  {arXiv: 1303.2880}\ } (\bibinfo {year} {2013})}\BibitemShut {NoStop}%
\bibitem [{\citenamefont {Akkermans}\ and\ \citenamefont
  {Montambeaux}(2007)}]{akkermans07}%
  \BibitemOpen
  \bibfield  {author} {\bibinfo {author} {\bibfnamefont {E.}~\bibnamefont
  {Akkermans}}\ and\ \bibinfo {author} {\bibfnamefont {G.}~\bibnamefont
  {Montambeaux}},\ }\href@noop {} {\emph {\bibinfo {title} {Mesoscopic physics
  of electrons and photons}}}\ (\bibinfo  {publisher} {Cambridge University
  Press},\ \bibinfo {year} {2007})\BibitemShut {NoStop}%
\bibitem [{\citenamefont {van Tiggelen}\ and\ \citenamefont
  {Lagendijk}(1994)}]{vantiggelen94}%
  \BibitemOpen
  \bibfield  {author} {\bibinfo {author} {\bibfnamefont {B.~A.}\ \bibnamefont
  {van Tiggelen}}\ and\ \bibinfo {author} {\bibfnamefont {A.}~\bibnamefont
  {Lagendijk}},\ }\href {\doibase 10.1103/PhysRevB.50.16729} {\bibfield
  {journal} {\bibinfo  {journal} {Phys. Rev. B}\ }\textbf {\bibinfo {volume}
  {50}},\ \bibinfo {pages} {16729} (\bibinfo {year} {1994})}\BibitemShut
  {NoStop}%
\bibitem [{\citenamefont {Morice}\ \emph {et~al.}(1995)\citenamefont {Morice},
  \citenamefont {Castin},\ and\ \citenamefont {Dalibard}}]{morice95}%
  \BibitemOpen
  \bibfield  {author} {\bibinfo {author} {\bibfnamefont {O.}~\bibnamefont
  {Morice}}, \bibinfo {author} {\bibfnamefont {Y.}~\bibnamefont {Castin}}, \
  and\ \bibinfo {author} {\bibfnamefont {J.}~\bibnamefont {Dalibard}},\ }\href
  {\doibase 10.1103/PhysRevA.51.3896} {\bibfield  {journal} {\bibinfo
  {journal} {Phys. Rev. A}\ }\textbf {\bibinfo {volume} {51}},\ \bibinfo
  {pages} {3896} (\bibinfo {year} {1995})}\BibitemShut {NoStop}%
\bibitem [{\citenamefont {Cherroret}\ \emph {et~al.}(2016)\citenamefont
  {Cherroret}, \citenamefont {Delande},\ and\ \citenamefont {van
  Tiggelen}}]{cherroret16}%
  \BibitemOpen
  \bibfield  {author} {\bibinfo {author} {\bibfnamefont {N.}~\bibnamefont
  {Cherroret}}, \bibinfo {author} {\bibfnamefont {D.}~\bibnamefont {Delande}},
  \ and\ \bibinfo {author} {\bibfnamefont {B.~A.}\ \bibnamefont {van
  Tiggelen}},\ }\href {\doibase 10.1103/PhysRevA.94.012702} {\bibfield
  {journal} {\bibinfo  {journal} {Phys. Rev. A}\ }\textbf {\bibinfo {volume}
  {94}},\ \bibinfo {pages} {012702} (\bibinfo {year} {2016})}\BibitemShut
  {NoStop}%
\bibitem [{\citenamefont {{U}che}\ \emph {et~al.}(2004)\citenamefont {{U}che},
  \citenamefont {{S}tillinger},\ and\ \citenamefont {{T}orquato}}]{uche04}%
  \BibitemOpen
  \bibfield  {author} {\bibinfo {author} {\bibfnamefont {O.~U.}\ \bibnamefont
  {{U}che}}, \bibinfo {author} {\bibfnamefont {F.~H.}\ \bibnamefont
  {{S}tillinger}}, \ and\ \bibinfo {author} {\bibfnamefont {S.}~\bibnamefont
  {{T}orquato}},\ }\href {\doibase 10.1103/{P}hys{R}ev{E}.70.046122} {\bibfield
   {journal} {\bibinfo  {journal} {{P}hys. {R}ev. {E}}\ }\textbf {\bibinfo
  {volume} {70}},\ \bibinfo {pages} {046122} (\bibinfo {year}
  {2004})}\BibitemShut {NoStop}%
\bibitem [{\citenamefont {{U}che}\ \emph {et~al.}(2006)\citenamefont {{U}che},
  \citenamefont {{T}orquato},\ and\ \citenamefont {{S}tillinger}}]{uche06}%
  \BibitemOpen
  \bibfield  {author} {\bibinfo {author} {\bibfnamefont {O.~U.}\ \bibnamefont
  {{U}che}}, \bibinfo {author} {\bibfnamefont {S.}~\bibnamefont {{T}orquato}},
  \ and\ \bibinfo {author} {\bibfnamefont {F.~H.}\ \bibnamefont
  {{S}tillinger}},\ }\href {\doibase 10.1103/{P}hys{R}ev{E}.74.031104}
  {\bibfield  {journal} {\bibinfo  {journal} {{P}hys. {R}ev. {E}}\ }\textbf
  {\bibinfo {volume} {74}},\ \bibinfo {pages} {031104} (\bibinfo {year}
  {2006})}\BibitemShut {NoStop}%
\bibitem [{\citenamefont {{B}atten}\ \emph {et~al.}(2008)\citenamefont
  {{B}atten}, \citenamefont {{S}tillinger},\ and\ \citenamefont
  {{T}orquato}}]{batten08}%
  \BibitemOpen
  \bibfield  {author} {\bibinfo {author} {\bibfnamefont {R.~D.}\ \bibnamefont
  {{B}atten}}, \bibinfo {author} {\bibfnamefont {F.~H.}\ \bibnamefont
  {{S}tillinger}}, \ and\ \bibinfo {author} {\bibfnamefont {S.}~\bibnamefont
  {{T}orquato}},\ }\href@noop {} {\bibfield  {journal} {\bibinfo  {journal}
  {{J}. {A}ppl. {P}hys.}\ }\textbf {\bibinfo {volume} {104}},\ \bibinfo {pages}
  {033504} (\bibinfo {year} {2008})}\BibitemShut {NoStop}%
\bibitem [{\citenamefont {{L}eseur}\ \emph {et~al.}(2016)\citenamefont
  {{L}eseur}, \citenamefont {{P}ierrat},\ and\ \citenamefont
  {{C}arminati}}]{leseur16}%
  \BibitemOpen
  \bibfield  {author} {\bibinfo {author} {\bibfnamefont {O.}~\bibnamefont
  {{L}eseur}}, \bibinfo {author} {\bibfnamefont {R.}~\bibnamefont {{P}ierrat}},
  \ and\ \bibinfo {author} {\bibfnamefont {R.}~\bibnamefont {{C}arminati}},\
  }\href {\doibase 10.1364/{O}{P}{T}{I}{C}{A}.3.000763} {\bibfield  {journal}
  {\bibinfo  {journal} {{O}ptica}\ }\textbf {\bibinfo {volume} {3}},\ \bibinfo
  {pages} {763} (\bibinfo {year} {2016})}\BibitemShut {NoStop}%
\bibitem [{\citenamefont {{K}wong}\ \emph {et~al.}(2019)\citenamefont
  {{K}wong}, \citenamefont {{W}ilkowski}, \citenamefont {{D}elande},\ and\
  \citenamefont {{P}ierrat}}]{kwong_coherent_2019}%
  \BibitemOpen
  \bibfield  {author} {\bibinfo {author} {\bibfnamefont {C.~C.}\ \bibnamefont
  {{K}wong}}, \bibinfo {author} {\bibfnamefont {D.}~\bibnamefont
  {{W}ilkowski}}, \bibinfo {author} {\bibfnamefont {D.}~\bibnamefont
  {{D}elande}}, \ and\ \bibinfo {author} {\bibfnamefont {R.}~\bibnamefont
  {{P}ierrat}},\ }\href {\doibase 10.1103/{P}hys{R}ev{A}.99.043806} {\bibfield
  {journal} {\bibinfo  {journal} {{P}hys. {R}ev. {A}}\ }\textbf {\bibinfo
  {volume} {99}},\ \bibinfo {pages} {043806} (\bibinfo {year}
  {2019})}\BibitemShut {NoStop}%
\end{thebibliography}%

\end{document}